\documentclass[reqno,12pt]{amsart}

\pdfoutput=1

\usepackage{amsmath,amssymb,amsfonts}
\usepackage{amsaddr}  
\usepackage{graphicx}
\usepackage{bm}
\usepackage[labelfont=rm]{subcaption}
\usepackage{color}
\usepackage[numbers,sort&compress]{natbib}
\usepackage{hyperref}
\usepackage[letterpaper,hmarginratio=1:1]{geometry}
\usepackage[capitalize]{cleveref}
\graphicspath{{figs/}}

\usepackage{ifthen}
\def\arXiv{}  

\numberwithin{equation}{section}

\usepackage{amsthm,thmtools}
\declaretheoremstyle[
  spaceabove=6pt, spacebelow=6pt,
  headfont=\normalfont\bfseries,
  notefont=\mdseries, notebraces={(}{)},
  bodyfont=\normalfont,
  postheadspace=1em,
  qed={\lower-0.3ex\hbox{$\triangle$}}
]{examplestyle}
\declaretheorem[style=examplestyle]{Example}
\crefname{Example}{Example}{Examples}
\Crefname{Example}{Example}{Examples}

\newcommand{\ie}{\textit{i.e.}}
\newcommand{\etal}{\textit{et~al.}}
\newcommand{\mathnotation}[2]{\newcommand{#1}{\ensuremath{#2}}}

\newcommand{\nofrac}[2]{#1/#2}
\newcommand{\Order}[1]{\mathrm{O}(#1)}      

\DeclareMathOperator{\sgn}{sgn}             

\renewcommand{\l}{\mathopen{}\mathclose\bgroup\left}
\renewcommand{\r}{\aftergroup\egroup\right}
\mathnotation{\pd}{\partial}
\mathnotation{\ldef}{\mathrel{\raisebox{.069ex}{:}\!\!=}}
\mathnotation{\rdef}{\mathrel{=\!\!\raisebox{.069ex}{:}}}
\mathnotation{\dint}{\,{\mathrm{d}}}        
\mathnotation{\ee}{\mathrm{e}}              
\mathnotation{\imi}{\mathrm{i}}             
\mathnotation{\grad}{\nabla}                
\renewcommand{\div}{\nabla\cdot}          

\mathnotation{\bndry}{\mathrm{b}}           
\mathnotation{\Xcb}{X_\bndry}               
\mathnotation{\Ycb}{Y_\bndry}               
\mathnotation{\Rvb}{\bm{\R}_\bndry}         
\mathnotation{\xcb}{x_\bndry}               
\mathnotation{\ycb}{y_\bndry}               
\mathnotation{\rvb}{\bm{r}_\bndry}          
\mathnotation{\tvb}{\bm{t}_\bndry}          
\mathnotation{\Xc}{X}                       %
\mathnotation{\Yc}{Y}                       %
\mathnotation{\R}{R}                        
\mathnotation{\Rv}{\bm{\R}}                 
\mathnotation{\xc}{x}                       
\mathnotation{\yc}{y}                       
\mathnotation{\rv} {\bm{r}}                 
\mathnotation{\Qrot}{\mathbb{Q}}            
\mathnotation{\Xcrot}{\Xc_{\mathrm{rot}}}   
\mathnotation{\Ycrot}{\Yc_{\mathrm{rot}}}   
\mathnotation{\Rvrot}{\Rv_{\mathrm{rot}}}   
\mathnotation{\aaa}{a}                      
\mathnotation{\bbb}{b}                      
\mathnotation{\ecc}{e}                      

\mathnotation{\ycw}{\zeta}
\mathnotation{\thetaL}{\theta^{\mathrm{L}}}
\mathnotation{\thetaR}{\theta^{\mathrm{R}}}

\renewcommand{\t}{t}                        
\renewcommand{\d}{\mathrm{d}}              
\mathnotation{\dt}{\d{\t}}                  
\mathnotation{\dXc}{\d{\Xc}}                
\mathnotation{\dYc}{\d{\Yc}}                
\mathnotation{\dtheta}{\d{\theta}}          
\mathnotation{\dW}{\d{W}}                   
\mathnotation{\dxc}{\d{\xc}}                
\mathnotation{\dyc}{\d{\yc}}                
\mathnotation{\U}{U}                        
\mathnotation{\Uv}{\bm{\U}}                 
\mathnotation{\Uxc}{u}                      
\mathnotation{\Uyc}{v}                      
\mathnotation{\p}{p}                        
\mathnotation{\Drot}{D_\theta}              
\mathnotation{\DX}{D_X}                     
\mathnotation{\DY}{D_Y}                     
\mathnotation{\Dt}{\mathbb{D}}              
\mathnotation{\Dxx}{D_{\xc\xc}}             
\mathnotation{\Dxy}{D_{\xc\yc}}             
\mathnotation{\Dyy}{D_{\yc\yc}}             
\mathnotation{\fv}{\bm{f}}                  
\mathnotation{\nv}{\bm{n}}                  
\mathnotation{\thetauv}{\hat{\bm{\theta}}}  
\mathnotation{\xuv}{\hat{\bm{x}}}           
\mathnotation{\yuv}{\hat{\bm{y}}}           
\mathnotation{\Vol}{V}                      
\mathnotation{\dV}{\dint\Vol}               
\mathnotation{\dSv}{\dint{\bm{S}}}          

\mathnotation{\bp}{{\bar\p}}                
\mathnotation{\bfv}{\bar{\fv}}              
\mathnotation{\bnv}{\bar{\bm{n}}}           

\mathnotation{\eps}{\varepsilon}            
\mathnotation{\iPer}{\eps}                  
\mathnotation{\T}{T}                        
\mathnotation{\w}{w}                        
\mathnotation{\vth}{\sigma}
\mathnotation{\Q}{Q}
\renewcommand{\P}{P}                        
\mathnotation{\Dyc}{\Delta}

\mathnotation{\thetaLint}{\thetaL} 
\mathnotation{\thetaRint}{\thetaR} 
\mathnotation{\cc}{c}                       
\mathnotation{\Pinv}{\mathcal{\P}}          
\mathnotation{\Qinv}{\mathcal{\Q}}          
\mathnotation{\E}{\mathbb{E}}               
\mathnotation{\mub}{\bar{\mu}}              

\mathnotation{\wtau}{\mathrm{T}}
\mathnotation{\Pinvt}{\widetilde{\Pinv}}
\mathnotation{\cct}{\tilde{\cc}}
\mathnotation{\Phit}{\tilde{\Phi}}
\mathnotation{\Gt}{\widetilde{G}}
\mathnotation{\Cc}{\mathcal{C}}
\mathnotation{\taurev}{\tau_{\mathrm{rev}}} 
\mathnotation{\EE}{E}                       
\mathnotation{\EK}{K}                       
\mathnotation{\EPi}{\Pi}                    

\mathnotation{\Lc}{\mathcal{L}}
\newcommand{\avg}[1]{\l\langle #1 \r\rangle}
\mathnotation{\Deff}{D_{\mathrm{eff}}}
\mathnotation{\Denh}{D_{\mathrm{enh}}}
\mathnotation{\Pinvz}{\rho}
\mathnotation{\thom}{\t}
\mathnotation{\xchom}{\xc}
\mathnotation{\phom}{\mathrm{P}} 
\mathnotation{\Chi}{\mathrm{X}}
\mathnotation{\dd}{d}                       

\begin{document}

\title[Shape matters: A Brownian microswimmer in a channel]%
  {Shape matters:\\[4pt] A Brownian microswimmer in a channel}

\author{Hongfei Chen and Jean-Luc Thiffeault}
\address{Department of Mathematics, University of Wisconsin -- Madison \\
  480 Lincoln Dr., Madison, WI 53706, USA}
\email{hchen475@wisc.edu, jeanluc@math.wisc.edu}

\date{}

\begin{abstract}
  We consider the active Brownian particle (ABP) model for a two-dimensional
  microswimmer with fixed speed, whose direction of swimming changes according
  to a Brownian process.  The probability density for the swimmer evolves
  according to a Fokker--Planck equation defined on the configuration space,
  whose structure depends on the swimmer's shape, center of rotation and
  domain of swimming.  We enforce zero probability flux at the boundaries of
  configuration space.  We derive a reduced equation for a swimmer in an
  infinite channel, in the limit of small rotational diffusivity, and find
  that the invariant density depends strongly on the swimmer's precise shape
  and center of rotation.  We also give a formula for the \emph{mean reversal
    time}: the expected time taken for a swimmer to completely reverse
  direction in the channel.  Using homogenization theory, we find an
  expression for the effective longitudinal diffusivity of a swimmer in the
  channel, and show that it is bounded by the mean reversal time.
\end{abstract}

\maketitle

[\emph{Note to \ifthenelse{\isundefined{\arXiv}}{referees}{reader}: we have
  included a table of notation in \cref{apx:notation}.}]

\section{Introduction}
\label{sec:intro}

Microswimmers are common in nature --- they include bacteria, spermatozoa,
some algae, and synthetic swimmers.  In almost all contexts these swimmers
interact with boundaries, either biological (e.g., the gut, cell walls) or
man-made (e.g., tubes, filters).  These interactions have been studied
experimentally, numerically, and theoretically by many groups.  The two main
aspects of interaction are hydrodynamic (mediated by the fluid) and steric
(direct contact with the boundary); they can have different relative
importance depending on the context, but it is widely accepted that both can
play a crucial role \cite{Drescher2011, Kantsler2013, Contino2015,
  Bianchi2017}.

In the present paper we will be concerned with modeling the steric interaction
of a microswimmer with solid surfaces, with an emphasis on the role of the
swimmer's shape.  For simplicity, the swimmer will be two dimensional with a
fixed shape, though in principle the theory could be extended to include a
deformable body or flagella.

\subsection{Previous work}

Many models have been proposed to mimic the behavior of microswimmers, with
the simplest being the \emph{active Brownian particle} (ABP) model
\cite{Ai2013, Solon2015, Zottl2016, Wagner2017, Redner2013,Stenhammar2014}
where a particle moves with constant speed and both its swimming direction and
spatial position are subject to independent diffusion processes.  A more
complicated model has the organism moving in a straight line for a random time
(run), followed by a random change in direction (tumble); such run-and-tumble
models have been investigated both theoretically and numerically
\cite{Elgeti2015, Tailleur2009, Martens2012,Lambert2010, Koumakis2014,
  Razin2017, Elgeti2016, Lushi2014, Chen2018, Sepulveda2017, Molaei2014,
  Costanzo2012, Ezhilan2012, Ezhilan2015b,Nash2010, Lee2019, Cates2013,
  Lushi2016, Lushi2012}.  There are also more complex models that incorporate
hydrodynamic effects \cite{Crowdy2010, Crowdy2011, Obuse2012, Takatori2014,
  Li2014, tenHagen2011, Spagnolie2015, Lushi2014, Evans2010, Theillard2017,
  Wagner2019, Costanzo2012, DaddiMoussaIder2018, Shum2010, Theers2018,
  Mathijssen2016, Fauci1995, Wioland2016, Rusconi2010, Lushi2015,
  Saintillan2006ii, Alonso-Matilla2019, Saintillan2007, Saintillan2008,
  tenHagen2015, Lushi2018, Saintillan2010, Yeo2015, Saintillan2006,
  Saintillan2013, Bricard2015}.  In the present paper we will limit ourselves
to the ABP model.

Experiments with microswimmers near boundaries are also plentiful
\cite{Woolley2003, Li2008, Lauga2006, Hill2007, Volpe2011, Drescher2011,
  Kantsler2013, Contino2015, DiLuzio2005, Berke2008, Rothschild1963,Kim2014,
  Frymier1995}.  As early as 1963, Rothschild \cite{Rothschild1963} measured
the density of bull spermatozoa between two glass plates and found
accumulation near the plates.  Accumulation as well as local alignment and
preferred tail rotation were also observed in later experiments
\cite{Woolley2003, Li2008, Lauga2006, Hill2007, Volpe2011, Drescher2011,
  Kantsler2013, Contino2015, Li2011, Berke2008, Lefauve2014, Rothschild1963,
  Denissenko2012}. Simulations have shown that either hydrodynamic
interactions or steric interactions with thermal fluctuations can lead to
accumulation.  Later work found that steric effects dominate at walls, while
hydrodynamic interactions can play an important role depending on the shape of
the swimmer and its orientation with respect to the wall \cite{Drescher2011,
  Kantsler2013, Contino2015, Bianchi2017}.

In principle, the interaction of microswimmers with boundaries requires
modeling both hydrodynamic and steric interactions.  Some groups include only
hydrodynamic interactions \cite{Crowdy2010, Mirzakhanloo2018, Crowdy2011,
  Kaynan2017, Obuse2012, Sipos2015, HernandezOrtiz2005, Spagnolie2012,
  tenHagen2015, Schaar2015, Spagnolie2015, Swan2007, Lopez2014, Wagner2019,
  Elgeti2009,Saintillan2006ii, Katz1975,Katz1974, Katz1975ii}, which can be
done in several ways: either explicitly with a solution of the Stokes
equation, or implicitly through a resistance or mobility matrix, with various
approximations in both cases.  Spagnolie \& Lauga \cite{Spagnolie2012} used a
multipole expansion which in principle can be applied to any swimmer shape,
and Takagi \etal\ \cite{Takagi2014} solved the Stokes equation in the
lubrication limit.  Zargar \etal\ \cite{Zargar2009} approximated the mobility
matrix by restricting the swimmer to planar motion near a wall.  Despite the
simplifications in these models, they are fairly accurate away from boundaries
and reproduce observed behavior \cite{Zottl2016, Marchetti2013, Bechinger2016,
  Lauga2009}.  However, such models remain in general fairly complicated and
either ignore the details of swimmers such as shape, or are swimmer-dependent.
Most importantly, they ignore steric interactions and hence are not accurate
near walls.

Steric interactions are often included phenomenologically, such as by using
wall potential functions \cite{Costanzo2012, Tian2017,
  vanTeeffelen2008,Wensink2008, Caprini2018, Sepulveda2018,
  HernandezOrtiz2009, Mathijssen2016,Wagner2019, DaddiMoussaIder2018,
  Kaiser2012, Chilukuri2014}. This approach makes it easy to add boundaries to
free-space simulations and has low simulation cost.  Sep\'ulveda \etal\
\cite{Sepulveda2018} used Weeks--Chandler--Anderson functions and a Gaussian
potential function for deformable swimmers; they treat the effective wall
force as a smooth repulsive force.  Hernandez-Ortiz \etal\
\cite{HernandezOrtiz2009} used Gay--Berne functions for steric exclusion of
rod-shape swimmers.  These models are usually a smooth approximation to the
true dynamics, whereas for rigid swimmers the steric effect is by volume
exclusion.  Moreover, the complicated potential functions make it hard to
obtain predictive formulas, although some authors used a simple harmonic
potential to represent an elastic boundary, which made the problem tractable
\cite{Drescher2011,Caprini2018,Nikola2016}.

Another way to model steric interactions is to assign specific dynamics near
the boundary such as reflecting \cite{Volpe2014, Chen2018, Li2009,
  Koumakis2014} or vanishing velocity \cite{Sepulveda2018} boundary
conditions. It is worth noting that neither boundary condition is
realistic. Ezhilan \etal\ \cite{Ezhilan2015} observed that reflecting boundary
conditions do not recover the behavior observed in experiments.  Other groups
did not try to prescribe any specific law for swimmers at a boundary and
focused on statistics such as the invariant density of swimmers
\cite{Yariv2014,Yan2015,Ezhilan2012} and swim pressure \cite{Chen2018,
  Caprini2018, Yan2015, Speck2020}.

When modeling a collection of stochastic swimmers or the statistics of a
single swimmer, an ideal approach to include steric interactions is to use
no-flux boundary conditions that prevent the organism's body from entering the
wall, as described by Nitsche \& Brenner~\cite{Nitsche1990} for passive
particles.  Most work in the literature using these boundary conditions
assumes that swimmers have negligible size or are of spherical shape, so that
they can rotate freely at a wall \cite{Alonso-Matilla2019, Alonso-Matilla2018,
  Yan2015, Yariv2014, Ezhilan2015, Bearon2011}.  Lee \etal\ \cite{Lee2013}
attempted to solve for the invariant density without spatial diffusion for
point swimmers in a channel, and managed to find a solution when swimmers had
only six swimming directions.  Wagner \etal\ \cite{Wagner2017} continued Lee's
work and found the invariant density for continuous swimming directions by
introducing a wall density function.  Schaar \etal\ \cite{Schaar2015} later
calculated the trapping time at a wall.  Ai \etal\ \cite{Ai2013} predicted for
point swimmers the optimal swimming speed, the spatial diffusion, and the
strength of wall potentials for maximal effective diffusion.  For spherical or
point swimmers, Elgeti \& Gompper \cite{Elgeti2013} and Ezhilan \& Saintillan
\cite{Ezhilan2015} investigated asymptotic solutions to the Fokker--Planck
equation associated with the ABP model.  Elgeti \& Gompper \cite{Elgeti2016}
subsequently found the invariant density for run-and-tumble spherical
swimmers.

\subsection{The role of shape}
The admissible positions and orientations of a nonspherical swimmer are
constrained by the presence of walls; the set of admissible values of the
degrees of freedom is the \emph{configuration space} \cite{Nitsche1990} (see
\cref{sec:conf}).  A natural way to include the shape of a swimmer into a
model is thus to impose no-flux boundary conditions on the configuration space
itself.  Krochak \etal\ \cite{Krochak2010} and Ezhilan \etal\
\cite{Ezhilan2015} used steric exclusion with no-flux boundary conditions for
rigid fibers and rod-like swimmers, respectively, and simulated the invariant
density in a channel in the presence of flow.

In the present paper we propose a framework to incorporate the shape of a
swimmer into the boundary conditions of a partial differential equation
describing its dynamics.  The PDE is the Fokker--Planck equation derived from
the two-dimensional ABP model with no-flux boundary conditions in
configuration space.  The configuration space is determined by the swimmer's
shape and the domain of swimming, which we take to be an infinite channel.  We
solve explicitly for the invariant density in the limit of small rotational
diffusivity.  In the same limit, we also solve for the mean reversal time and
the longitudinal effective diffusivity of the swimmer.  All these quantities
are greatly influenced by the shape of the swimmer.  In particular the
effective diffusivity can become very large when the swimmer tends to align
parallel to the channel walls, which can surprisingly occur even for circular
swimmers when their center of rotation does not coincide with the geometric
center (see below).

\begin{figure}
  \begin{center}
    \includegraphics[height=.23\textheight]{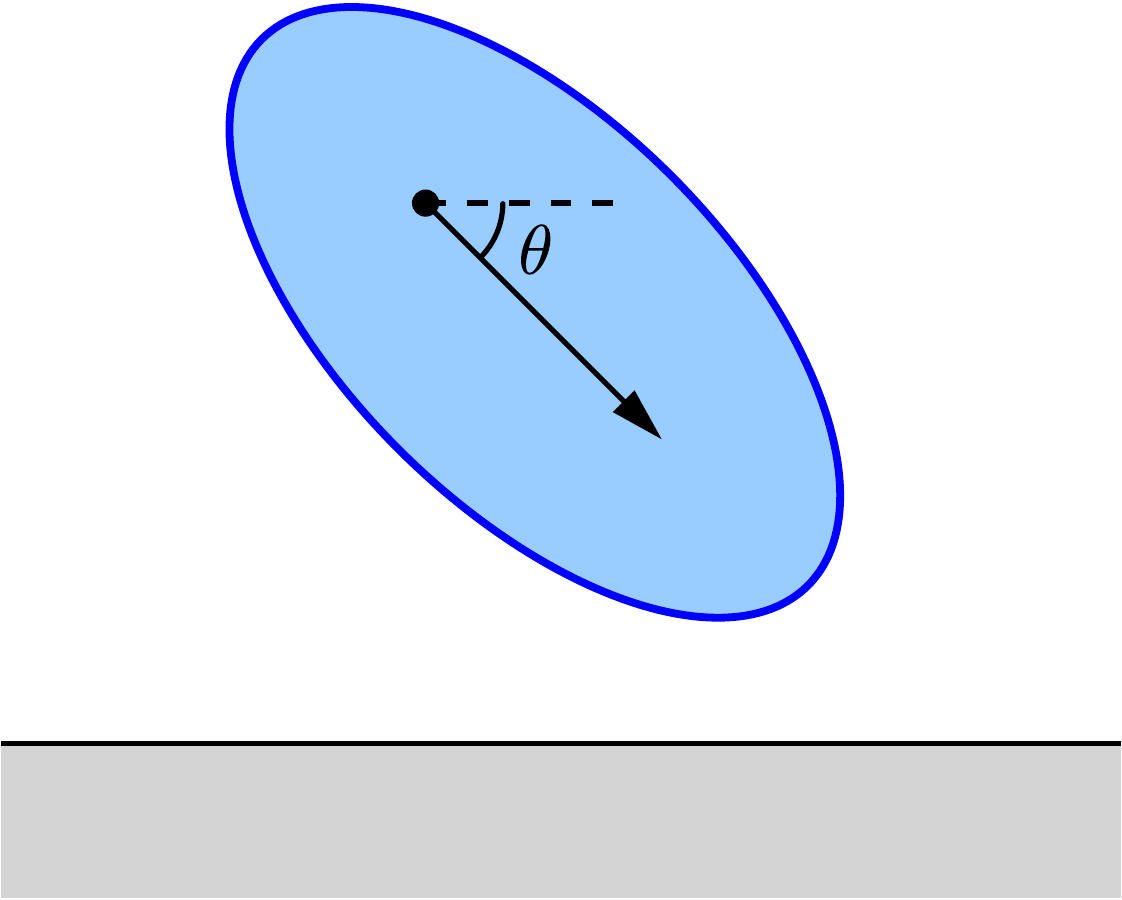}
  \end{center}
  \caption{An elliptical swimmer approaching a wall at a
    direction~$\theta = -\pi/4$.  The center of rotation is not necessarily
    the geometric center of the ellipse.}
  \label{fig:swimmer_ellipse}
\end{figure}

An important observation is in order regarding the ABP model for a finite-size
swimmer.  The ABP model implicitly assumes that the swimmer rotates about some
distinguished fixed point in a co-moving frame.  The precise location of this
point becomes important when the swimmer has finite size and boundaries are
present.  For example, \Cref{fig:swimmer_ellipse} shows an elliptical swimmer
approaching a boundary: the direction of swimming is given by the
angle~$\theta$, which is measured counterclockwise from the horizontal, so
that $\theta<0$ corresponds to swimming towards the wall.  The angle $\theta$
is measured around a point we call the \emph{center of rotation} of the
swimmer.  For a free particle, this corresponds to the \emph{center of
  hydrodynamic reaction} defined by Happel \& Brenner
\cite[p.~174]{HappelBrenner}.  For an ellipse it would coincides with the
geometrical center.  However, the swimmer's propulsion mechanism (e.g.,
flagella), which is abstracted here since we consider fixed shapes, can
displace this center.  Hence, we treat the center of rotation as a parameter
that may be adjusted to model a particular swimmer.  To parallel terminology
based on the type of propulsion used by a microorganism
\cite{HernandezOrtiz2005,HernandezOrtiz2009,Saintillan2007,Saintillan2011},
when the center of rotation is ahead of the geometric center we will call the
swimmer \emph{puller-like}; when it is behind we call it \emph{pusher-like}.
If the swimmer doesn't interact with boundaries, then the center of rotation
is not particularly important to the dynamics; but with external boundaries it
can influence the tendency of the swimmer to align parallel or perpendicular
to a wall, depending on its shape~\cite{Lushi2017}.  In fact, we will see that
even a circular particle can align with a wall if its center of rotation is
behind the geometric center, despite the absence of hydrodynamic interactions.

\subsection{Outline}
In this paper we focus exclusively on a two-dimensional swimmer undergoing
steric interactions.  In \cref{sec:conf} we describe the configuration space
for a swimmer with an arbitrary fixed convex shape, in particular when the
swimmer is confined to a channel consisting of two infinite parallel walls.  A
crucial quantity is the \emph{wall distance function}, which describes the
swimmer's closest point of approach to a wall as a function of the swimmer's
orientation.  We give explicit examples for needle (rod-like), elliptical, and
teardrop-shaped swimmers.  The wall distance function can then be used twice
to determine the full configuration space in a channel.  This configuration
space is \emph{open} if the swimmer can reverse direction in the channel, or
\emph{closed} if the channel is too narrow to do so.  We also describe
symmetries of the configuration space that follow from symmetries of the
swimmer and channel, the most important being the case where a swimmer is
\emph{left-right symmetric}.

In \cref{sec:stochmod} we describe the stochastic ABP model for our swimmer,
and give its corresponding Fokker--Planck equation.  This leads to the natural
no-flux boundary conditions that we impose at the solid walls.  For the
infinite channel geometry we average over the lengthwise coordinate.  In
\cref{sec:reduced} we simplify the model by assuming a small rotational
diffusivity.  This leads to a \emph{reduced equation}, which is a partial
differential equation in one time variable and one angle.  The configuration
geometry is completely encoded into a single effective angular drift function.

In \cref{sec:inv} we solve for the steady state of the reduced equation, which
gives us the \emph{invariant probability density} of the swimmer, or invariant
density for short.  The invariant density is strongly dependent on the shape
and center of rotation, as we show with some explicit examples, typically in
the limit of rapid swimming.  In particular, we show that circular swimmers
can align either parallel or perpendicular to the walls, depending on whether
they are pusher- or puller-like, respectively.  When a swimmer has broken
left-right symmetry, it can undergo a net rotation due to repeated biased
interactions with the walls.

We introduce the \emph{mean reversal time} (MRT) of a swimmer in
\cref{sec:MERT}: the expected time for a swimmer to fully reverse direction in
an open channel.  This is a generalization of the turnaround time of Holcman
\& Schuss \cite{Holcman2014}, who described the expected time for a Brownian
needle to reverse direction when its length is slightly shorter than the
channel width.  For a left-right symmetric swimmer we give a simple integral
formula for the MRT.  We explicitly compute the MRT in some limits, in
particular for a fast swimmer.  The MRT in this fast case is exponentially
long, since the swimmer sticks to a wall for a very long time before
undergoing a large enough random fluctuation that causes reversal.

In \cref{sec:effdiff} we use a homogenization theory approach to find the
longitudinal \emph{effective diffusivity} $\Deff$ for the swimmer in an open
channel.  In the same reduced limit as above (small rotational diffusion) we
give an integral formula for the diffusivity.  For a fast swimmer we might
expect that the effective diffusivity is related to the MRT: the swimmer makes
large excursions and sometimes reverses direction, thereby undergoing an
effective random walk for long times.  Indeed, we obtain the rigorous bound
\begin{equation*}
  \Deff
  \le
  \DX
  +
  \tfrac12\taurev\,U^2,
\end{equation*}
where~$\DX$ is the diffusivity of the swimmer along the direction of swimming,
$\taurev$ is the MRT, and $U$ is the swimming speed.  The
term~$\tfrac12(\taurev\,U)^2/\taurev$ is equal to the diffusivity for an
unbiased random walk with step size~$\taurev U$ and step time~$\taurev$.
Finally, we offer some concluding remarks in \cref{sec:discussion}.

\section{Configuration space}
\label{sec:conf}

In this section we describe how the swimmer's shape interacts with boundaries
to create \emph{configuration space}.  We first establish coordinate systems
for a convex swimmer (\cref{sec:walldist}): the fixed lab frame and a frame
rotating with the swimmer.  These are both necessary since the direction of
swimming and the magnitude of diffusion are tied to the swimmer's shape, and
may be different along different axes (\cref{sec:stochmod}).  We use the term
``swimmer'' throughout, but our entire formalism applies to passive particles
as well, for which $\U=0$.  We consider an arbitrary contact point between a
swimmer's body and a single wall, and show how to derive the \emph{wall
  distance function} for the swimmer.  We present a few examples: a needle
(one-dimensional segment), an ellipse, and a `teardrop' shape.  These are all
swimmers with a left-right axis of symmetry, but our formalism applies to more
general swimmers as well.

In \cref{sec:channelgeom} we use the wall distance function to obtain the
configuration space for a swimmer confined between two infinite, parallel
walls.  Two very different cases emerges: in the \emph{open} channel
configuration the channel is wide enough to allow the swimmer to completely
reverse direction, whereas in the \emph{closed} configuration the swimmer is
unable to do so.

\subsection{The wall distance function}
\label{sec:walldist}

\begin{figure}
  \begin{center}
    \subcaptionbox{\label{fig:swimmer_shape_body}}{
      \includegraphics[height=.28\textheight]{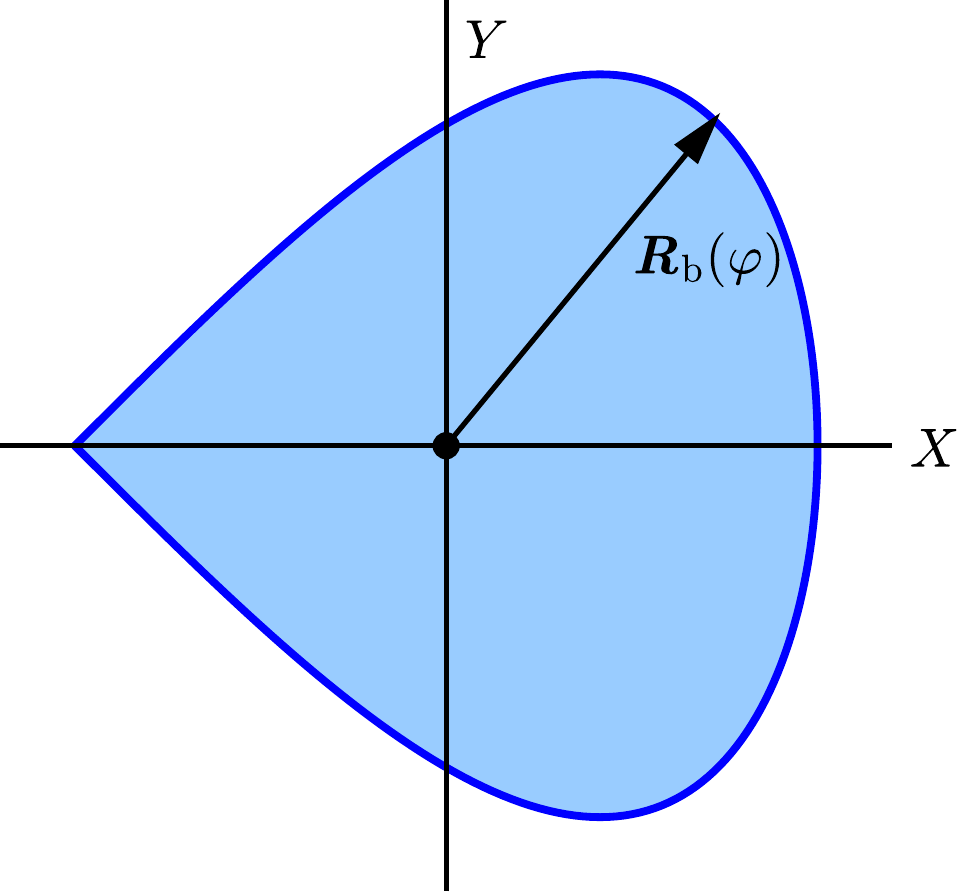}}
    \hspace{.05\textwidth}
    \subcaptionbox{\label{fig:swimmer_shape_lab}}{
      \includegraphics[height=.28\textheight]{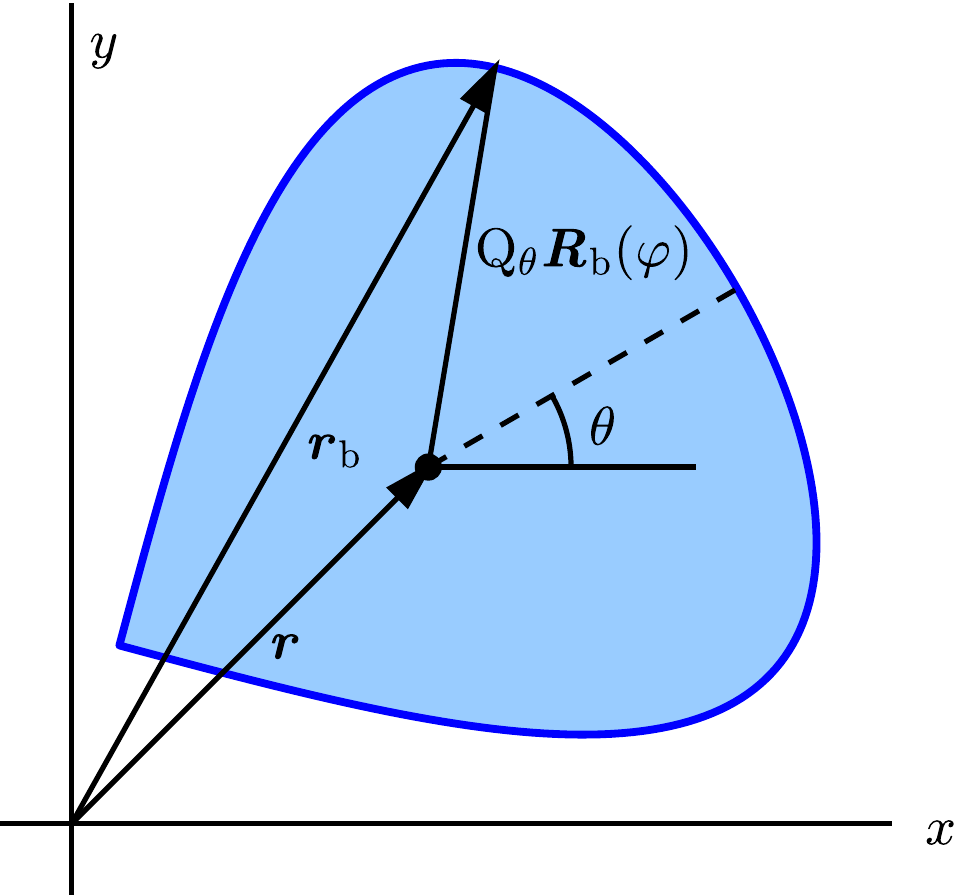}}
  \end{center}
  \caption{Boundary of a convex swimmer in (a) the swimmer's frame, with the
    swimming direction along the positive~$\Xc$ axis, and (b) the fixed lab
    frame, where the swimming direction makes an angle~$\theta$ with the~$\xc$
    axis.}
  \label{fig:swimmer_shape}
\end{figure}

The shape of a swimmer is expressed by giving its boundary in parametric form
\begin{equation}
  \Rv = \Rvb(\varphi) = (\Xcb(\varphi),\Ycb(\varphi)),
  \qquad
  -\pi < \varphi \le \pi,
  \label{eq:shape}
\end{equation}
where~$\Rvb(\varphi)$ is a $2\pi$-periodic, piecewise-smooth function
(\cref{fig:swimmer_shape_body}).  By convention, the swimming
direction~$\Rvb(0)$ is along the positive~$\Xc$ axis in the swimmer's
co-moving and co-rotating frame.  The origin of the $\Rv=(\Xc,\Yc)$ coordinate
system is the \emph{center of rotation} of the swimmer.  Note that we do not
require~$\tan\varphi = \Ycb(\varphi)/\Xcb(\varphi)$, that is, $\varphi$ does
not necessarily correspond to the polar angle of~$\Rvb(\varphi)$.

In a fixed (lab) frame, the boundary of the swimmer is located at
(\cref{fig:swimmer_shape_lab})
\begin{equation}
  (\xcb,\ycb) = \rvb(\theta,\varphi)
  = \rv + \Qrot_\theta\cdot\Rvb(\varphi)\,,
  \qquad
  -\pi < \varphi,\theta \le \pi,
\end{equation}
where~$\rv = (\xc,\yc)$ denotes the center of rotation of the swimmer
and
\begin{equation}
  \Qrot_\theta = \begin{pmatrix}
    \cos\theta & -\sin\theta \\
    \sin\theta & \phantom{-}\cos\theta
  \end{pmatrix}
\end{equation}
is a rotation matrix.

\begin{figure}
  \begin{center}
  \subcaptionbox{\label{fig:swimmer_corner_touch}}{
    \includegraphics[height=.18\textheight]{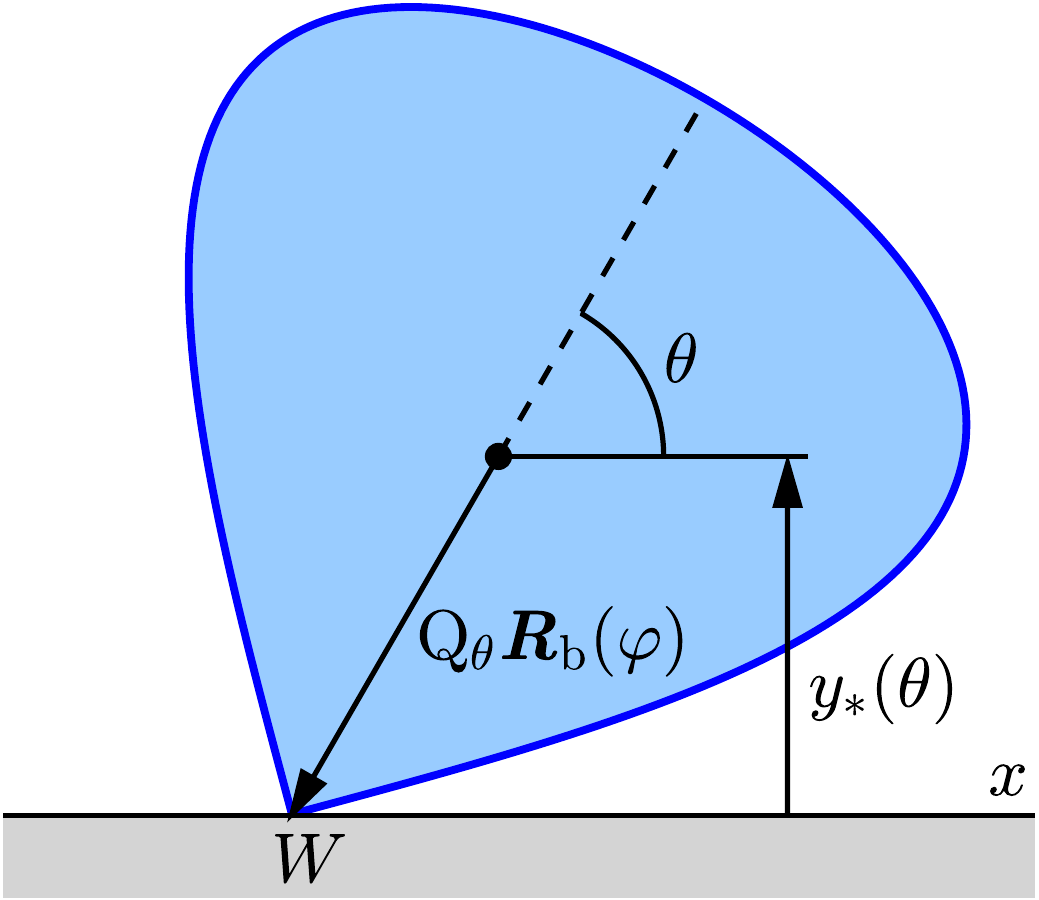}}
  \hspace{.005\textwidth}
  \subcaptionbox{\label{fig:swimmer_corner_touch_theta_minus}}{
    \includegraphics[height=.18\textheight]{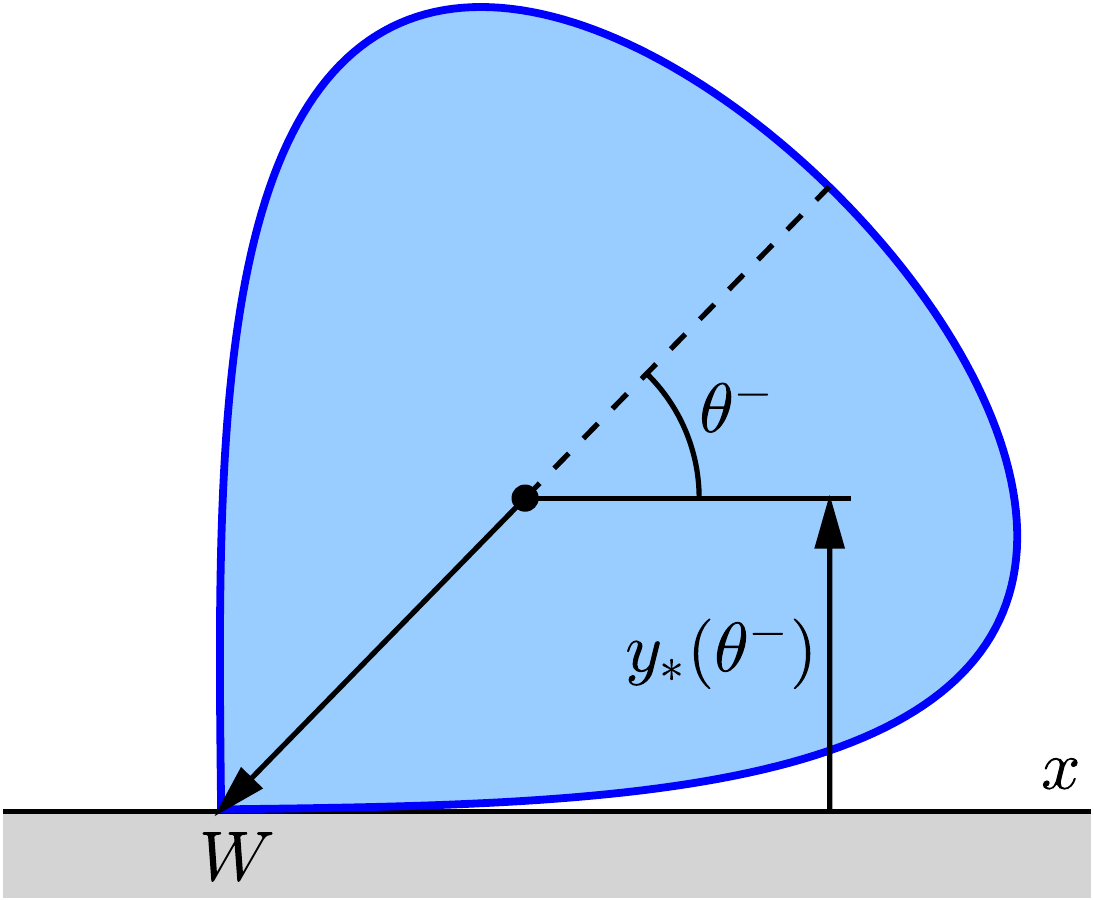}}
  \hspace{.005\textwidth}
  \subcaptionbox{\label{fig:swimmer_corner_touch_theta_plus}}{
    \includegraphics[height=.18\textheight]{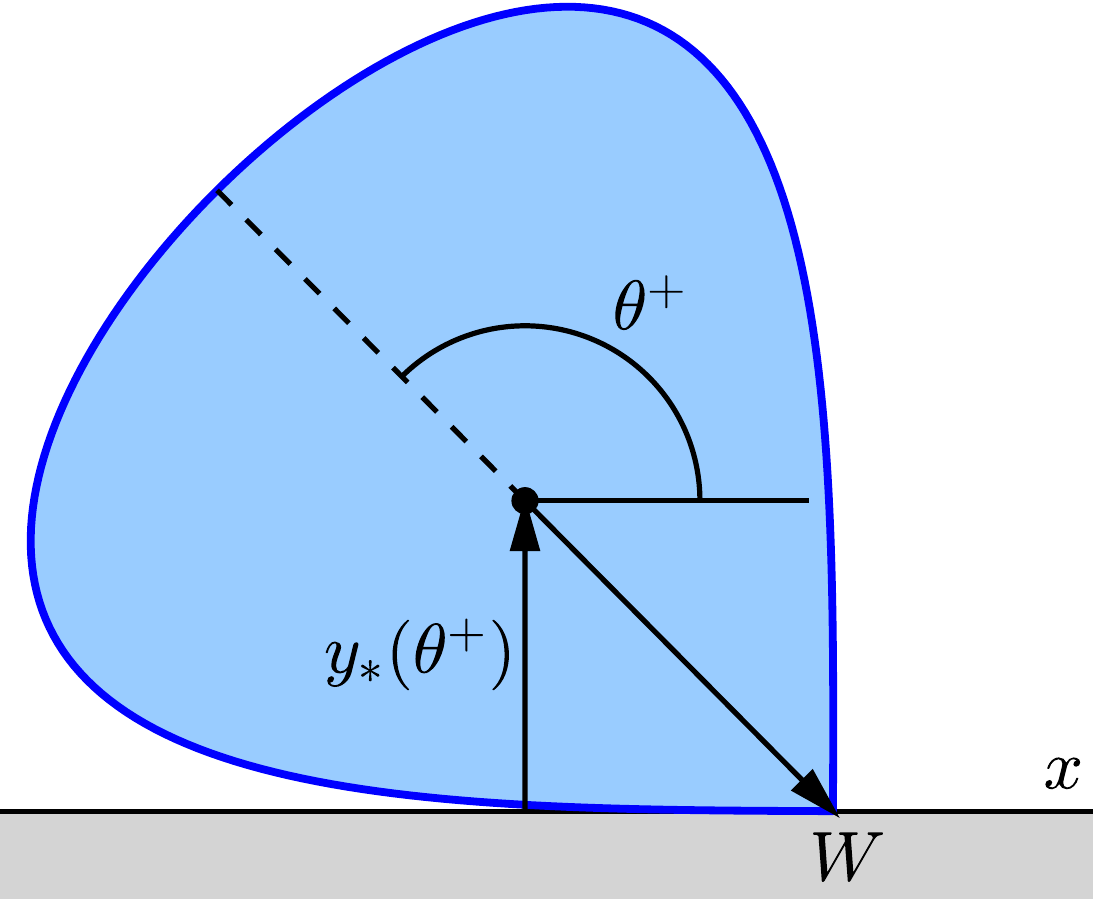}}
  \end{center}
  \caption{(a) Convex swimmer touching a horizontal wall at a corner
    point~$W$.  (b)--(c) Holding $W$ fixed, the angle $\theta$ can vary from
    the right-tangency angle~$\theta^-$ to the left-tangency angle
    $\theta^+$.}
  \label{fig:swimmer_corner_touch_all}
\end{figure}

Now take the swimmer to be touching an infinite wall along~$\yc=0$, as shown
in \cref{fig:swimmer_corner_touch}.  The contact point~$W$ between the swimmer
and the wall has coordinates
\begin{equation}
  (\xcb,0) = \rv + \Qrot_\theta\cdot\Rvb(\varphi).
  \label{eq:W}
\end{equation}
We wish to solve for the swimmer's center of rotation~$\rv = (\xc,\yc)$, which
depends only on the convex hull of the swimmer; hence the swimmer's shape may
be assumed convex without loss of generality.  We proceed differently
depending on whether the contact point~$W$ is a corner or a smooth boundary
point.  (Note that the analysis below can be couched in the language of
Legendre transformations and convex analysis, but we opt here for a direct
treatment.)

\subsubsection{Corner}
\label{sec:corner}

Consider first the case where the contact point~$W$ corresponds to a corner of
the piecewise-smooth boundary, as in \cref{fig:swimmer_corner_touch}.  The
parameter~$\varphi$ has a fixed value for corner~$W$.  The allowable range
of~$\theta$ is then determined by the right- and left-tangency values
of~$\theta$:
\begin{equation}
  \theta^- \le \theta \le \theta^+,
  \qquad
  \tan\theta^\pm
  =
  -\Ycb'(\varphi^\pm)/\Xcb'(\varphi^\pm),
\end{equation}
as depicted in \cref{%
  fig:swimmer_corner_touch_theta_minus,%
  fig:swimmer_corner_touch_theta_plus%
}.  For this range of~$\theta$, we can then use~\cref{eq:W} to deduce the
range of~$\yc$ values:
\begin{equation}
  \yc_*(\theta)
  =
  -\sin\theta\,\Xcb(\varphi)
  -\cos\theta\,\Ycb(\varphi),
  \qquad
  \theta^- \le \theta \le \theta^+.
  \label{eq:ycstar_corner}
\end{equation}
We call~$\yc_*$ the \emph{wall distance function}.  It characterizes the
minimum distance from the swimmer's center of rotation to a wall at~$\ycb=0$,
as a function of the swimmer's orientation.  Observe that a given corner
corresponds to a single~$\varphi$ value, but a range of~$\theta$ values.

\begin{figure}
  \begin{center}
  \subcaptionbox{\label{fig:swimmer_walldist_needle}}{
    \includegraphics[height=.22\textheight]{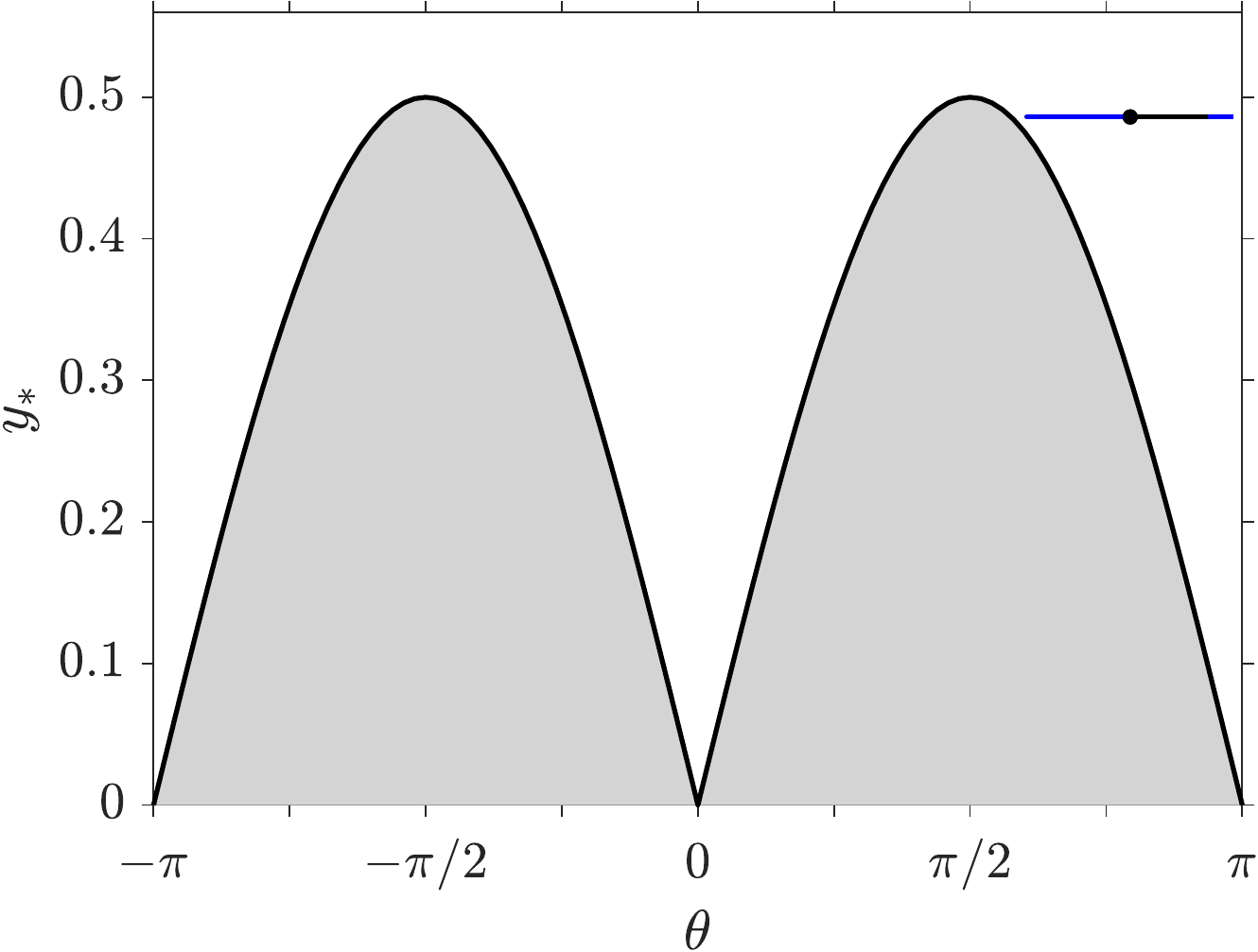}}
  \hspace{.005\textwidth}
  \subcaptionbox{\label{fig:swimmer_walldist_needle_Xrot}}{
    \includegraphics[height=.22\textheight]{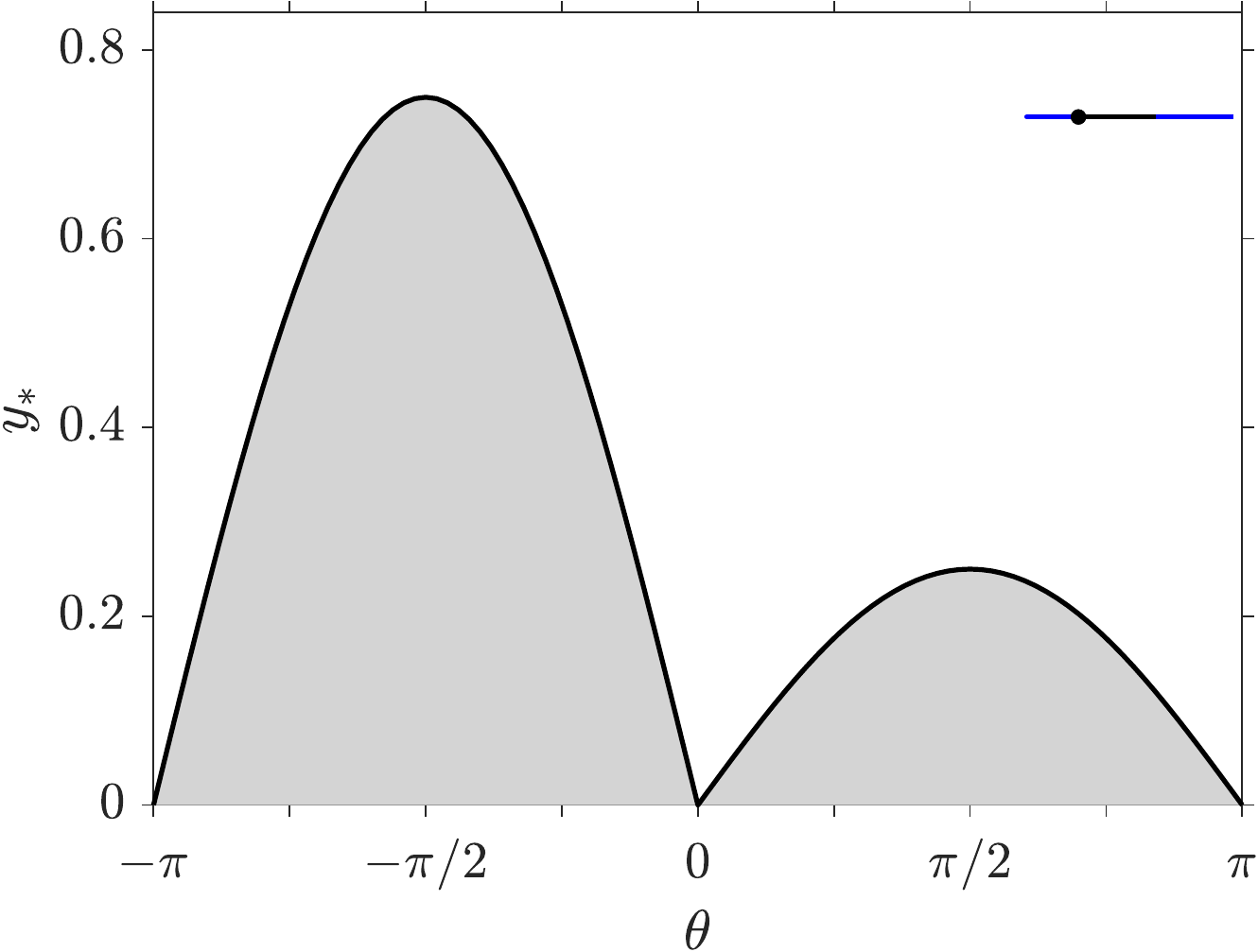}}

  \vspace{.025\textheight}

  \subcaptionbox{\label{fig:swimmer_walldist_ellipse}}{
    \includegraphics[height=.22\textheight]{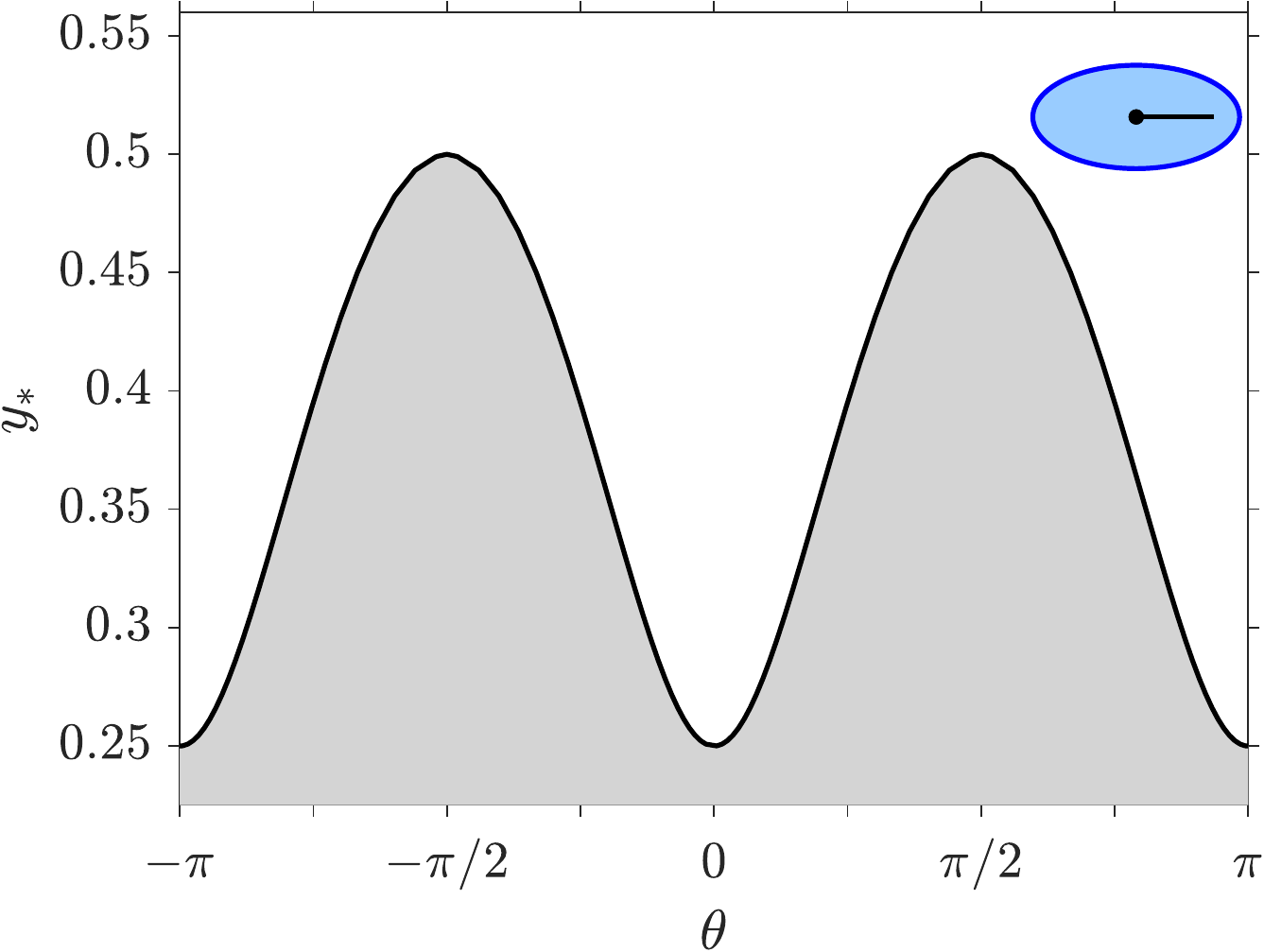}}
  \hspace{.005\textwidth}
  \subcaptionbox{\label{fig:swimmer_walldist_ellipse_Xrot}}{
    \includegraphics[height=.22\textheight]{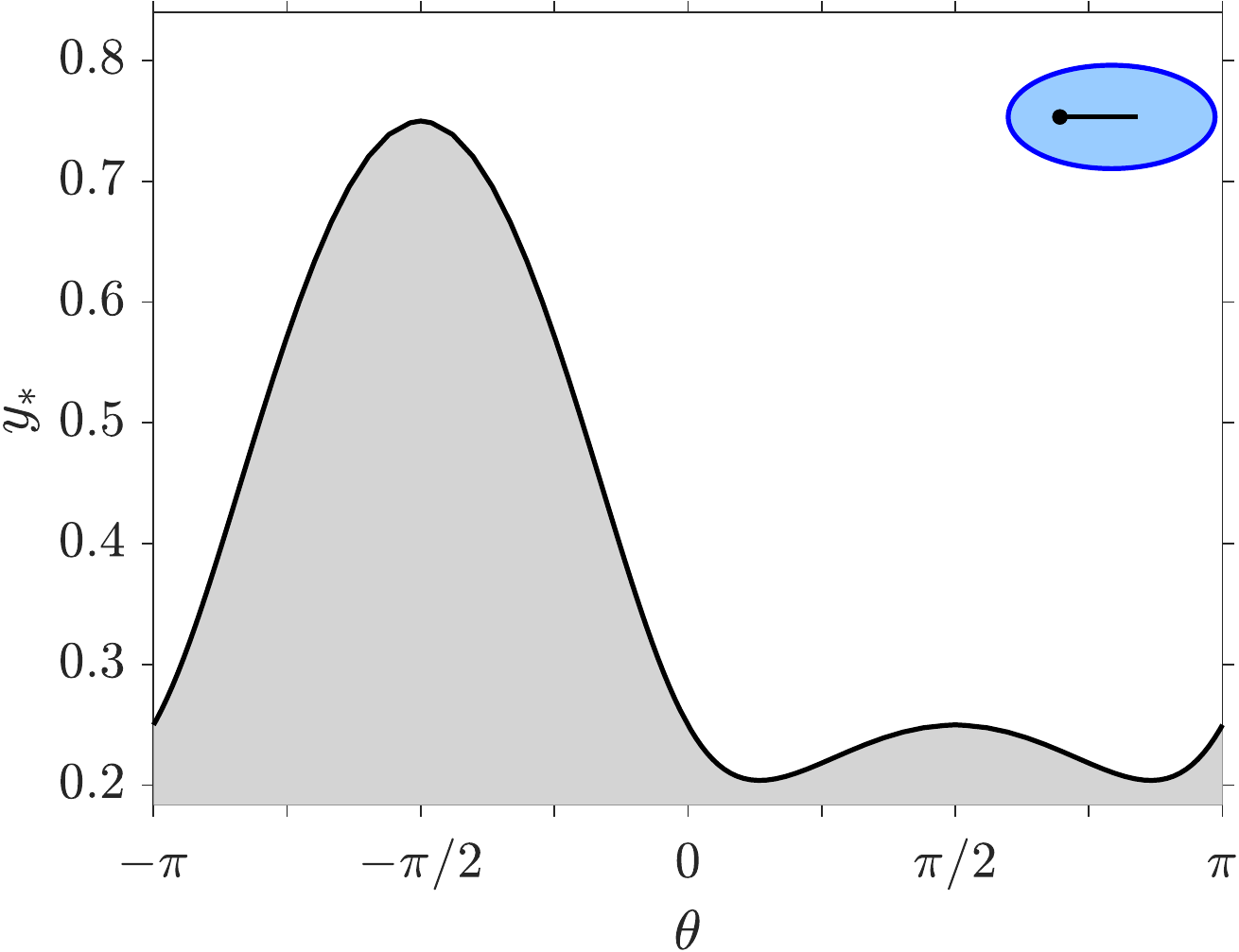}}

  \vspace{.025\textheight}

  \subcaptionbox{\label{fig:swimmer_walldist_tear}}{
    \includegraphics[height=.22\textheight]{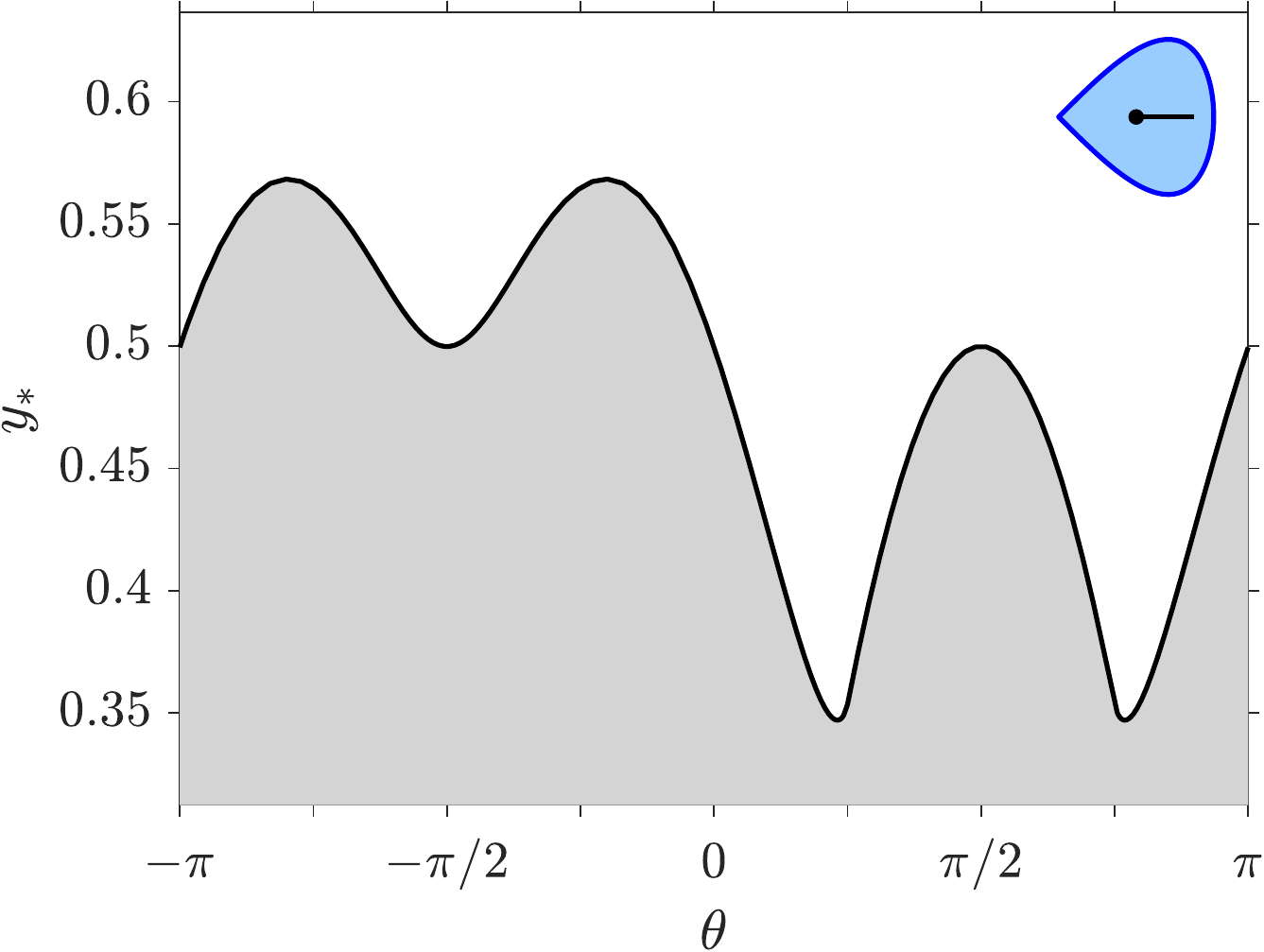}}
  \hspace{.005\textwidth}
  \subcaptionbox{\label{fig:swimmer_walldist_tear_Xrot}}{
    \includegraphics[height=.22\textheight]{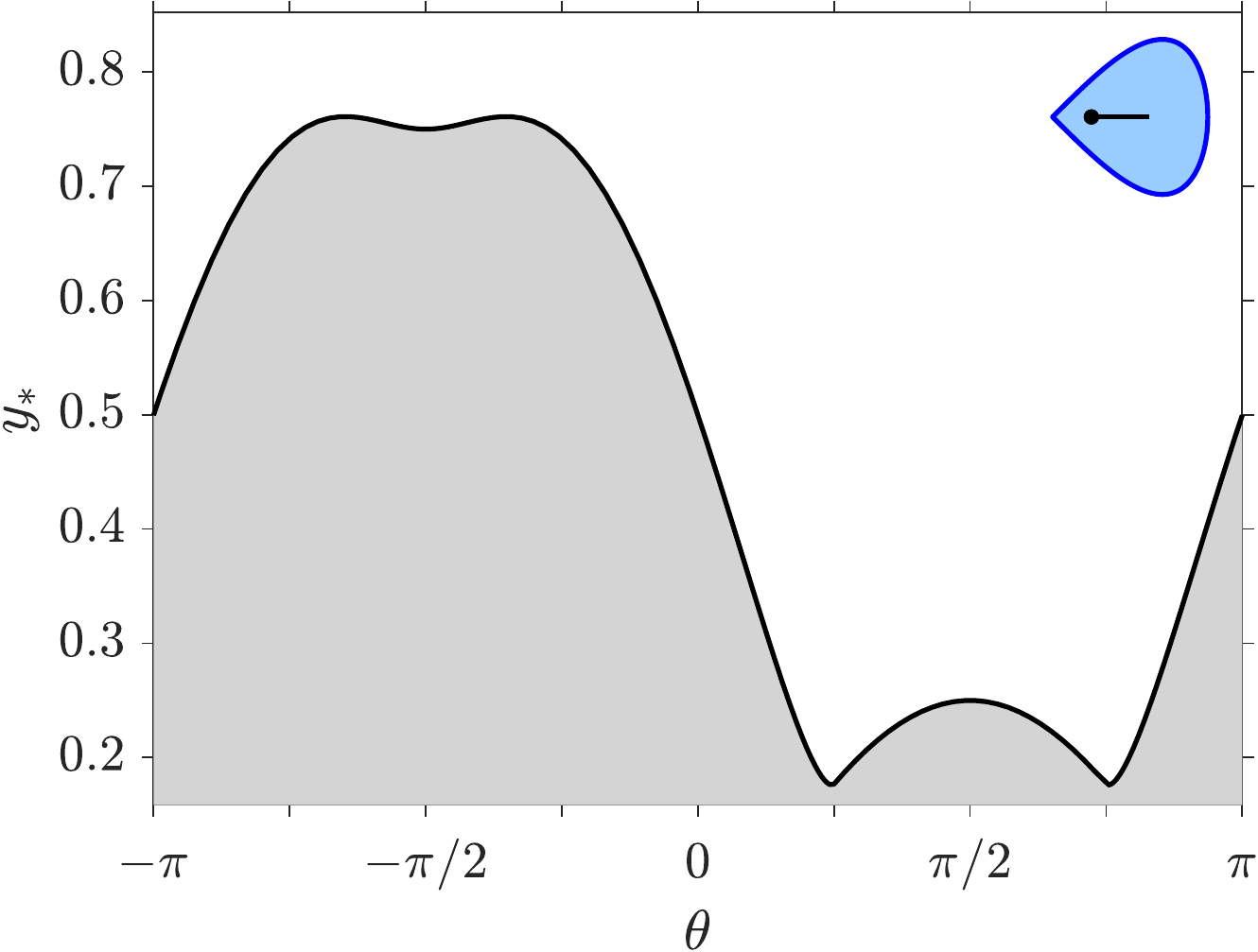}}
  \end{center}
  \caption{The wall distance function $\yc_*(\theta)$ for three different
    swimmers: (a,b) needle of length $\ell=2\aaa=1$; (c,d) ellipse of length
    $\ell=2\aaa=1$ and width~$2\bbb=1/2$; (e,f) teardrop-shaped swimmer of
    size~$1$ by~$1$.  The inset shows the swimmer shape and center of
    rotation, with swimming direction to the right.  The right column has
    center of rotation displaced to~$\Xcrot=-1/4$.}
  \label{fig:swimmer_walldist}
\end{figure}

\begin{Example}[needle swimmer]\label{ex:needle}
  As a simple example, take
  \begin{equation}
    \Xcb(\varphi) = \tfrac12\ell\,\cos\varphi - \Xcrot,
    \qquad
    \Ycb(\varphi) = 0.
  \end{equation}
  This is the \emph{needle swimmer} with center of rotation at~$\Xc = \Xcrot$,
  with~$\lvert\Xcrot\rvert \le \ell/2$.  It consists of a one-dimensional
  segment of length~$\ell$, with degenerate ``corners'' at~$\varphi_1=0$
  and~$\varphi_2=\pi$.  At~$\varphi=\varphi_1=0$, we
  have~$-\pi \le \theta \le 0$, so from \cref{eq:ycstar_corner}
  $\yc_*(\theta) = -\sin\theta\,\Xcb(\varphi_1) = -\sin\theta(\tfrac12\ell -
  \Xcrot)$. At~$\varphi=\varphi_2=\pi$, we have~$0 \le \theta \le \pi$, so
  from \cref{eq:ycstar_corner}
  $\yc_*(\theta) = -\sin\theta\,\Xcb(\varphi_2) = -\sin\theta(-\tfrac12\ell -
  \Xcrot)$.  We can combine these cases by writing
  \begin{equation}
    \yc_*(\theta)
    =
    \tfrac12\ell\lvert\sin\theta\rvert + \Xcrot\sin\theta,
    \qquad
    -\pi < \theta \le \pi.
    \label{eq:ycstar_needle}
  \end{equation}
  Only needle positions with~$\yc \ge \yc_*(\theta)$ are allowed; see
  \cref{fig:swimmer_walldist_needle,fig:swimmer_walldist_needle_Xrot} for a
  plot.  For the case~$\Xcrot=0$, this type of swimmer and configuration space
  was investigated by Ezhilan \& Saintillan~\cite{Ezhilan2015}.
\end{Example}

We will typically use~$\ell$ to denote the maximum diameter of a swimmer,
which controls whether or not it can reverse direction for a given channel
width (\cref{sec:channelgeom}).  The parameter~$\Xcrot$ controls the position
of the center of rotation: for~$\Xcrot > 0$ it is closer to the front, and
for~$\Xcrot < 0$ it is towards the rear.  We often refer to these cases as
\emph{puller-like} and \emph{pusher-like}, respectively, by analogy with the
classification based on the type of propulsion used by a microorganism
\cite{HernandezOrtiz2005,HernandezOrtiz2009,Saintillan2007,Saintillan2011}.

\subsubsection{Smooth boundary point}
\label{sec:smooth}

When the contact point~$W$ is at a smooth boundary point, given~$\theta$ we
wish to solve for~$(\xc,\yc)$ and~$\varphi$.  Two equations come from
\eqref{eq:W}, but we need a third, which stems from requiring that the tangent
to the swimmer,
\begin{equation}
  \tvb = \pd_\varphi\rvb
  =
  \Qrot_\theta\cdot\Rvb'(\varphi),
  \label{eq:tangent}
\end{equation}
is horizontal at~$W$:
\begin{equation}
  \Xcb'(\varphi)\, \sin\theta
  +
  \Ycb'(\varphi)\, \cos\theta
  = 0
\end{equation}
or
\begin{equation}
  \nofrac{\Ycb'(\varphi)}{\Xcb'(\varphi)}
  =
  -
  \tan\theta.
  \label{eq:tangenthoriz}
\end{equation}
We can solve \cref{eq:tangenthoriz} for~$\varphi = \varphi_*(\theta)$, which
we then use in~\cref{eq:W} to obtain~$\rv = \rv_*(\theta)$ at the contact
point:
\begin{equation}
  \rv_*(\theta)
  =
  (\xc_*(\theta),\yc_*(\theta))
  =
  -\Qrot_\theta\cdot\Rvb(\varphi_*(\theta)).
\end{equation}
\Cref{eq:tangenthoriz} can have more than one solution, but we keep the one
that leads to a non-negative wall distance function,
\begin{equation}
  \yc_*(\theta)
  =
  -\sin\theta\,\Xcb(\varphi_*(\theta))
  -\cos\theta\,\Ycb(\varphi_*(\theta)) \ge 0.
  \label{eq:ycstar}
\end{equation}

\begin{Example}[elliptical swimmer]\label{ex:ellipse}
  An ellipse-shaped swimmer with semi-axes~$\aaa$ and~$\bbb$ can be
  parameterized as
  \begin{equation}
    \Xcb(\varphi) = \aaa\cos\varphi - \Xcrot,
    \qquad
    \Ycb(\varphi) = \bbb\sin\varphi 
  \end{equation}
  with~$\lvert\Xcrot\rvert \le \aaa$.  Here~$\aaa$ and~$\bbb$ are the
  semi-axes along and perpendicular to the swimming direction, respectively.
  The tangency condition~\eqref{eq:tangenthoriz} is
  then~$\cot\varphi_*(\theta) = (\aaa/\bbb)\tan\theta$.
  After inserting in \cref{eq:ycstar} and selecting the non-negative solution
  we obtain
  \begin{equation}
    \yc_*(\theta)
    =
    \sqrt{\aaa^2\sin^2\theta + \bbb^2\cos^2\theta}
    +
    \Xcrot\,\sin\theta.
    \label{eq:ycstar_ellipse}
  \end{equation}
  This wall distance function is plotted in
  \cref{fig:swimmer_walldist_ellipse,fig:swimmer_walldist_ellipse_Xrot}.

  For~$\aaa\ge\bbb$, it is convenient to rewrite~\eqref{eq:ycstar_ellipse} as
  \begin{equation}
    \yc_*(\theta)
    =
    \aaa\,\sqrt{1 - \ecc^2\cos^2\theta} + \Xcrot\,\sin\theta
    ,
    \qquad
    \ecc \ldef \sqrt{1 - \bbb^2/\aaa^2} < 1,
    \label{eq:ycstar_ellipse_ecc}
  \end{equation}
  where~$\ecc$ is the eccentricity.  For~$\bbb=0$ ($\ecc=1$) we recover the
  needle case~\cref{eq:ycstar_needle}, with~$\aaa=\ell/2$.  The case~$\ecc=0$
  is a circular swimmer, which for~$\Xcrot=0$ has the same dynamics in our
  model as a point swimmer \cite{Elgeti2013,Lee2013}.  Note however that for
  $\Xcrot\ne0$ even a circular swimmer can exhibit alignment with the walls
  (see \cref{ex:circ_invdens}).
\end{Example}

\subsubsection{General shapes}

The convex hull for a general swimmer will consist of a combination of smooth
parts separated by corners, as for the `teardrop' swimmer depicted in
\cref{fig:swimmer_shape}.  The wall distance function $\yc_*(\theta)$ cannot
be found analytically in general, but is easy to compute numerically.  The
simplest approach is to discretize the convex hull as a polygon, and then
apply the formulas in \cref{sec:corner} to every corner.
\begin{Example}[teardrop swimmer]
  The `teardrop' swimmer depicted in \cref{fig:swimmer_shape} is parameterized
  by
  \begin{equation}
    \Xcb(\varphi) = \aaa\l(2\lvert\cos(\varphi/2)\rvert - 1\r) - \Xcrot,
    \qquad
    \Ycb(\varphi) = \bbb\sin\varphi
  \end{equation}
  with~$\lvert\Xcrot\rvert \le \aaa$.  This shape has a smooth boundary except
  for one corner at~$\varphi = \varphi_1 = \pi$.  The wall distance function
  can be obtained analytically but is a bit cumbersome; we plot it in
  \cref{fig:swimmer_walldist_tear,fig:swimmer_walldist_tear_Xrot}.  Unlike the
  previous examples, the wall distance function for the teardrop swimmer has a
  local \emph{minimum} at~$\theta=-\pi/2$, rather than a maximum.  This value
  of~$\theta$ corresponds to swimming towards the wall, and the minimum
  suggests that this shape has a tendency to align perpendicular to the wall,
  rather than parallel.  (This is similar to the triangular swimmer in Lushi
  \etal~\cite{Lushi2017}.)  In the presence of diffusion, the depth of the
  local minimum is a measure of how long a swimmer gets stuck in that position
  before fluctuating out.  See also \cref{ex:circ_invdens} for another,
  simpler model swimmer that aligns perpendicular to the wall.
\end{Example}

All the examples discussed thus far involve \emph{left-right symmetric
  swimmers}, which
satisfy~$(\Xcb(\varphi),\Ycb(\varphi)) = (\Xcb(-\varphi),-\Ycb(-\varphi))$.
For this class of swimmers, the wall distance function has the symmetry
\begin{equation}
  \yc_*(\theta) = \yc_*(\pi - \theta)
  \label{eq:LRsym}
\end{equation}
which is evident in \cref{fig:swimmer_walldist}.

\subsection{Channel geometry}
\label{sec:channelgeom}

So far we have considered a two-dimensional swimmer above a single infinite
horizontal wall.  In a channel geometry, the swimmer is confined between two
parallel infinite walls, at~$\yc=\pm L/2$.  Luckily, we do not need to derive
a separate wall distance function for the top wall: we can deduce it by
symmetry.  The center of rotation of a swimmer with wall distance
function~$\yc_*(\theta)$ will have its~$\yc$ coordinate in the range
\begin{equation}
  \ycw_-(\theta) \le \yc \le \ycw_+(\theta)
  \label{eq:ycw_range}
\end{equation}
where
\begin{equation}
  \ycw_-(\theta) = \yc_*(\theta) - L/2,
  \qquad
  \ycw_+(\theta) = -\yc_*(\theta + \pi) + L/2.
  \label{eq:ycw_mp}
\end{equation}
This means that~$\ycw_\pm$ are related by the \emph{channel symmetry}
\begin{equation}
  \ycw_+(\theta) = -\ycw_-(\theta + \pi).
  \label{eq:channelsym}
\end{equation}

The~$\xc$ coordinate of the center of rotation is unconstrained and can be any
real number, but the domain for the swimming angle~$\theta$ can either
be~$[-\pi,\pi]$ or a union of disjoint intervals.  This depends on
whether~$\ycw_-(\theta) < \ycw_+(\theta)$ for all~$\theta \in [-\pi,\pi]$,
or~$\ycw_+(\theta) = \ycw_-(\theta)$ for some $\theta$.  We call these two
cases the \emph{open channel} and the \emph{closed channel}, respectively.

\subsubsection{Open channel}
\label{sec:open}

In the simplest case, we have
\begin{equation}
  \theta \in [-\pi,\pi],
  \qquad
  \ycw_-(\theta) < \ycw_+(\theta).
\end{equation}
In this case the swimmer can fully reverse direction in the channel.  The full
configuration space for the swimmer's center of rotation is then
\begin{equation}
  \Omega = \{
  (\xc,\yc,\theta) \,:\,
  \xc \in \mathbb{R},\
  \ycw_-(\theta) \le \yc \le \ycw_+(\theta),\
  -\pi \le \theta \le \pi
  \}
\end{equation}
periodic in the~$\theta$ direction.  This configuration space is depicted in
\cref{fig:channel_conf_open}.

\subsubsection{Closed channel}
\label{sec:closed}

\begin{figure}
  \begin{center}
    \subcaptionbox{\label{fig:channel_conf_open}}{
      \includegraphics[height=.26\textheight]{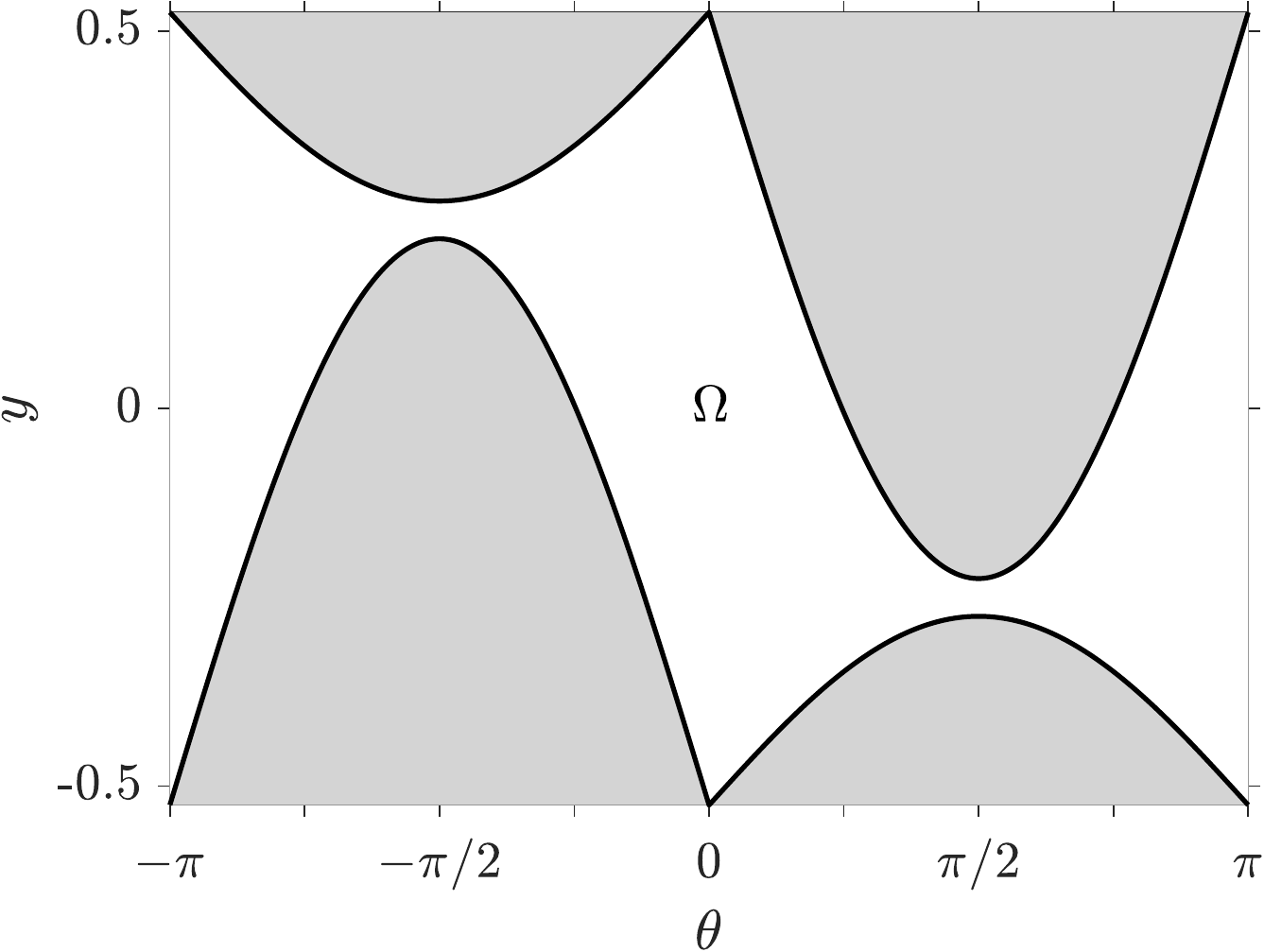}}
    \hspace{.05\textwidth}
    \subcaptionbox{\label{fig:channel_conf_closed}}{
      \includegraphics[height=.26\textheight]{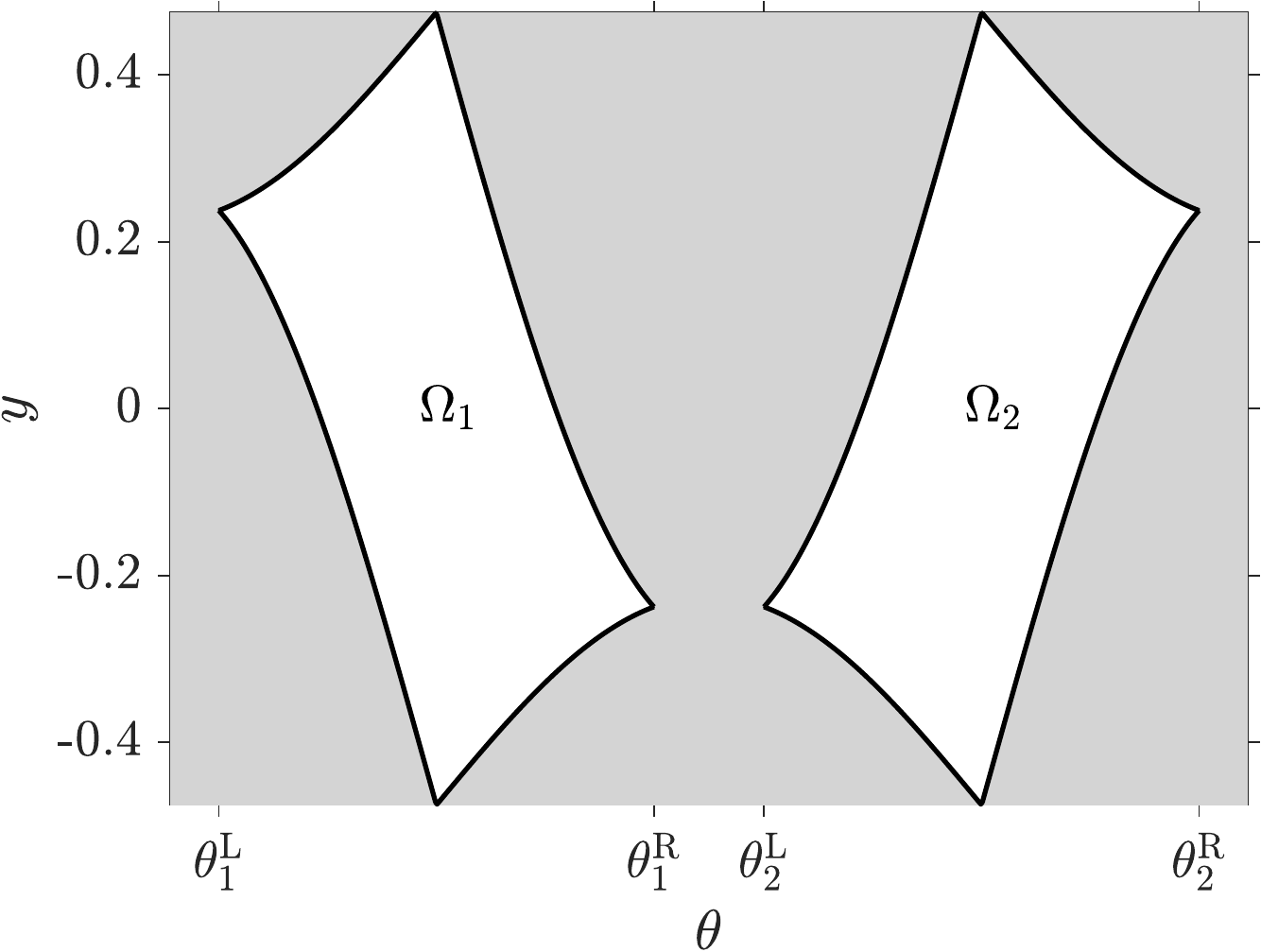}}
  \end{center}
  \caption{Configuration space for the needle in
    \cref{fig:swimmer_walldist_needle_Xrot} of length~$\ell=2\aaa=1$ in (a) an
    open channel of width~$L = 1.05$; (b) a closed channel of
    width~$L = 0.95$.  ($\xc$ direction not shown.)}
  \label{fig:channel_conf}
\end{figure}

Another possibility is that~$\ycw_+(\theta_i)=\ycw_-(\theta_i)$ for some set
of points~$\{\theta_i\}$.  This breaks up~$[-\pi,\pi]$ into inadmissible
intervals where~$\ycw_-(\theta) > \ycw_+(\theta)$, and~$N$ disjoint admissible
intervals
\begin{equation}
  \theta \in (\thetaL_i,\thetaR_i),
  \quad\text{with}\quad
  \ycw_-(\theta) < \ycw_+(\theta),
  \qquad
  i=1,\ldots,N.
\end{equation}
The relevant interval is determined by the initial orientation of the swimmer.
The motion of the swimmer then takes place in the configuration space
\begin{equation}
  \Omega_i = \{
  (\xc,\yc,\theta) \,:\,
  \xc \in \mathbb{R},\
  \ycw_-(\theta) \le \yc \le \ycw_+(\theta),\
  \thetaL_i \le \theta \le \thetaR_i
  \}
  \label{eq:Omega_i}
\end{equation}
which is \emph{not} periodic in the~$\theta$ direction.  This configuration
space is depicted in \cref{fig:channel_conf_closed}.  Note that the
condition~$\ycw_+(\theta_i)=\ycw_-(\theta_i)$ together with the channel
symmetry~\eqref{eq:channelsym} implies
that~$\ycw_+(\theta_i + \pi)=\ycw_-(\theta_i + \pi)$.

\section{Stochastic model}
\label{sec:stochmod}

Now that we've established that the domain of motion for our swimmer is
described by the configuration space of \cref{sec:conf}, we now describe the
stochastic model for the swimmer's motion, the active Brownian particle model
(ABP).

\subsection{Derivation from the SDE}
\label{sec:SDEderiv}

In the ABP model, the Brownian swimmer obeys the stochastic equation
\begin{subequations}
  \begin{align}
    \dXc &= \U\dt + \sqrt{2\DX}\,\dW_1\,; \\
    \dYc &= \sqrt{2\DY}\,\dW_2\,; \\
    \dtheta &= \sqrt{2\Drot}\,\dW_3,
\end{align}
\end{subequations}
in its own rotating reference frame.  (We omitted any intrinsic swimmer
rotation for simplicity, though this would not change the derivation
appreciably.  We will see that a net rotation can still emerge when the
swimmer is not left-right symmetric.)  In terms of absolute~$\xc$ and~$\yc$
coordinates, this becomes an It\^o stochastic equation
\begin{subequations}
  \begin{align}
    \dxc &= \bigl(\U\dt + \sqrt{2\DX}\,\dW_1\bigr)\cos\theta
    - \sin\theta\,\sqrt{2\DY}\,\dW_2\,; \\
    \dyc &= \bigl(\U\dt + \sqrt{2\DX}\,\dW_1\bigr)\sin\theta
    + \cos\theta\,\sqrt{2\DY}\,\dW_2\,; \\
    \dtheta &= \sqrt{2\Drot}\,\dW_3\,.
  \end{align}
  \label{eq:swimsde}%
\end{subequations}
For now we take~$\U$, $\DX$, $\DY$, $\Drot$ to be general functions
of~$(\xc,\yc,\theta,\t)$.  The corresponding Fokker--Planck equation for the
probability density~$\p(\xc,\yc,\theta,\t)$ is then
\begin{equation}
  \pd_\t \p
  =
  -\div\l(\Uv\p - \div\l(\Dt\,\p\r)\r) + \pd_\theta^2(\Drot\,\p)
  \label{eq:FP}
\end{equation}
where~$\grad \ldef \xuv\,\pd_\xc + \yuv\,\pd_\yc$, and the drift vector and
diffusion tensor are respectively
\begin{equation}
  \Uv
  =
  \begin{pmatrix}
    \U\cos\theta \\
    \U\sin\theta
  \end{pmatrix},
  \qquad
  \Dt
  =
  \begin{pmatrix}
    \DX\cos^2\theta + \DY\sin^2\theta & \tfrac12(\DX-\DY)\sin2\theta \\
    \tfrac12(\DX-\DY)\sin2\theta & \DX\sin^2\theta + \DY\cos^2\theta \\
  \end{pmatrix}.
  \label{eq:uD}
\end{equation}
See Kurzthaler \etal\ \cite{Kurzthaler2016, Kurzthaler2017} for the
intermediate scattering function for \cref{eq:FP} in the absence of
boundaries and with constant parameters.

For any fixed volume~$\Vol$ we have
\begin{align}
  \pd_\t \int_\Vol \p \,\dV
  &=
  -\int_\Vol\bigl(
  \div\l(\Uv\p - \grad\cdot\l(\Dt\,\p\r)\r) - \pd_\theta^2(\Drot\,\p)
  \bigl)\dV \nonumber\\
  &=
  -\int_{\pd\Vol}\fv\cdot \dSv\,,
\end{align}
where~$\pd\Vol$ is the boundary of~$V$, and the flux vector is
\begin{equation}
  \fv
  =
  \Uv\p
  -
  \div\l(\Dt\,\p\r) - \thetauv\,\pd_\theta(\Drot\,\p).
  \label{eq:flux}
\end{equation}
Thus, on the reflecting parts of the boundary we require the no-flux condition
\begin{equation}
  \fv \cdot \nv = 0,
  \quad
  \text{on\quad $\pd \Vol$,}
  \label{eq:noflux}
\end{equation}
where~$\nv$ is normal to the boundary.

\subsection{Infinite channel geometry}
\label{sec:channel}

The previous section applies to any geometry and general~$\Uv$, $\Dt$,
and~$\Drot$, which can be functions of~$(\xc,\yc,\theta,\t)$.  For our
problem, these only depend on~$\theta$.  In an infinite channel geometry
(\cref{sec:channelgeom}), which we consider in this paper, we can eliminate
the along-channel direction~$\xc$ by defining the marginal probability density
\begin{equation}
  \bp(\yc,\theta,\t)
  =
  \int_{-\infty}^\infty p(\xc,\yc,\theta,\t)\dint\xc.
  \label{eq:bp}
\end{equation}
In order for~$\bp$ to be finite,~$\p$ has to decay fast enough
as~$\lvert\xc\rvert\rightarrow\infty$; we use this assumption to discard some
terms after we integrate \cref{eq:FP} from~$\xc=-\infty$ to~$\infty$, and find
an equation for~$\bp$:
\begin{equation}
  \pd_\t \bp
  =
  -\pd_\yc\l(\U\sin\theta\,\bp\r)
  +
  \pd_\yc^2\l(\Dyy\,\bp\r) + \pd_\theta^2\l(\Drot\,\bp\r)
  \label{eq:FPbp0}
\end{equation}
where from
\eqref{eq:uD}~$\Dyy = [\Dt]_{22} = \DX\sin^2\theta + \DY\cos^2\theta$.  For
the rest of the paper we take~$\U$, $\DX$, $\DY$, and~$\Drot$ to be constants,
so that \cref{eq:FPbp0} simplifies to
\begin{equation}
  \pd_\t \bp
  =
  - \U\sin\theta\,\pd_\yc\bp
  +
  \Dyy(\theta)\,\pd_\yc^2\bp + \Drot\,\pd_\theta^2\bp.
  \label{eq:FPbp}
\end{equation}
(Note that instead of defining~$\bp$ as in~\eqref{eq:bp} we could assume
that~$\p$ is independent of~$\xc$, in which case~$\p$ is a density per unit
length that satisfies~\eqref{eq:FPbp}.)  \Cref{eq:FPbp} is our main focus.
The corresponding flux vector~\eqref{eq:flux} reduces to
\begin{equation}
  \bfv
  =
  (\U\sin\theta\,\bp - \Dyy(\theta)\,\pd_\yc\bp)\,\yuv
  - \Drot\,\pd_\theta\bp\,\thetauv.
  \label{eq:bflux}
\end{equation}
For the channel geometry, the domain can be characterized
by~$\ycw_-(\theta) < \yc < \ycw_+(\theta)$, so the normal vector is
\begin{equation}
  \bnv = \ycw_\pm'(\theta)\,\thetauv - \yuv\,.
  \label{eq:bnv}
\end{equation}
The no-flux boundary conditions on \cref{eq:FPbp} comes
from~\eqref{eq:noflux}:
\begin{equation}
  \bfv\cdot\bnv
  =
  -(\U\sin\theta\,\bp - \Dyy(\theta)\,\pd_\yc\bp)
  - \ycw_\pm'(\theta)\,\Drot\,\pd_\theta\bp
  =
  0,
  \qquad
  \yc = \ycw_\pm(\theta).
  \label{eq:nofluxbp}
\end{equation}
For convenience, we gather together the main \cref{eq:FPbp} and its no-flux
boundary condition~\eqref{eq:nofluxbp} for an infinite channel geometry:
\begin{subequations}
\begin{align}
  \pd_\t \bp
  + \U\sin\theta\,\pd_\yc\bp
  -
  \Dyy(\theta)\,\pd_\yc^2\bp - \Drot\,\pd_\theta^2\bp &= 0,
  \qquad
  \ycw_-(\theta) < \yc < \ycw_+(\theta);
  \label{eq:FPbp_both}
  \\
  \U\sin\theta\,\bp - \Dyy(\theta)\,\pd_\yc\bp
  + \ycw_\pm'(\theta)\,\Drot\,\pd_\theta\bp
  &= 0,
  \qquad
  \yc = \ycw_\pm(\theta).
  \label{eq:nofluxbp_both}
\end{align}
\label{eq:PDEbp}%
\end{subequations}
As discussed in \cref{sec:conf}, the domain in~$\theta$ is~$[-\pi,\pi]$
(periodic) for~$\ycw_-(\theta) < \ycw_+(\theta)$, which means the swimmer can
fully reverse direction in the channel (open channel configuration,
\cref{fig:channel_conf_open}).  If~$\ycw_-(\theta) \le \ycw_+(\theta)$, the
domain `pinches off' whenever~$\ycw_-(\theta) = \ycw_+(\theta)$, and consists
of two or more disconnected pieces (closed channel configuration,
\cref{fig:channel_conf_closed}).

\section{Reduced equation}
\label{sec:reduced}

\Cref{eq:PDEbp} is a challenging equation to solve, in particular because of
the complicated boundary shape.  We can dramatically simplify the problem by
assuming that the diffusivity~$\Drot$ is small, and carrying out an expansion
in powers of~$\iPer=\Drot$.  We call this the small-$\Drot$ or \emph{reduced}
limit.  The reduced form of~\cref{eq:PDEbp}, given by \cref{eq:Peq}, will
enable us to solve for the invariant density for a swimmer in \cref{sec:inv},
as well as many other quantities of interest such as a swimmer's mean reversal
time (\cref{sec:MERT}) and its effective diffusivity along the channel
(\cref{sec:effdiff}).

Take \cref{eq:PDEbp} and write~$\Drot=\iPer$:
\begin{subequations}
\begin{alignat}{2}
  \U\sin\theta\,\pd_\yc\bp
  - \Dyy(\theta)\,\pd_\yc^2\bp
  &= \iPer\,(\pd_\theta^2\bp - \pd_T\bp),
  \qquad
  && \ycw_-(\theta) < \yc < \ycw_+(\theta);
  \label{eq:steady_invdens} \\
  \U\sin\theta\,\bp
  - \Dyy(\theta)\,\pd_\yc\bp
  &= -\iPer\,\ycw_\pm'(\theta)\,\pd_\theta\bp
  ,
  \qquad
  &&\yc = \ycw_\pm(\theta),
  \label{eq:BCs_invdens}
\end{alignat}
\label{eq:pde_invdens}%
\end{subequations}
where we also defined a slow time~$\T = \iPer\t$,
$\pd_\t \rightarrow \iPer\,\pd_\T$.  We write the regular expansion
\begin{equation}
  \bp(\theta,\yc,\T)
  =
  \bp_0(\theta,\yc,\T)
  +
  \iPer\,\bp_1(\theta,\yc,\T)
  +
  \iPer^2\,\bp_2(\theta,\yc,\T)
  +
  \ldots
\end{equation}
and proceed to solve for~$\bp_i$ order-by-order.

At order~$\iPer^0$, \cref{eq:pde_invdens} is
\begin{equation}
  \U\sin\theta\,\pd_\yc\bp_0
  - \Dyy(\theta)\,\pd_\yc^2\bp_0
  = 0,
  \qquad
  \U\sin\theta\,\bp_0
  - \Dyy(\theta)\,\pd_\yc\bp_0 = 0,
  \quad \yc=\ycw_\pm(\theta)
  \label{eq:bp0}
\end{equation}
with solution
\begin{equation}
  \bp_0(\theta,\yc)
  =
  \Q(\theta,\T)\,
  \ee^{\vth(\theta)\,\yc},
  \qquad
  \vth(\theta) \ldef \nofrac{\U\sin\theta}{\Dyy(\theta)},
  \label{eq:bp0sol_inv}
\end{equation}
where~$\Q(\theta,\T)$ is as-yet undetermined.

At order~$\iPer^1$, \cref{eq:pde_invdens} is
\begin{subequations}
\label{eq:order1}
\begin{alignat}{2}
  \U\sin\theta\,\pd_\yc\bp_1
  - \Dyy(\theta)\,\pd_\yc^2\bp_1
  &= \pd_\theta^2\bp_0 - \pd_\T\bp_0;
  \label{eq:order1pde}\\
  \U\sin\theta\,\bp_1
  - \Dyy(\theta)\,\pd_\yc\bp_1 &= -\ycw_\pm'(\theta)\,\pd_\theta\bp_0,
  \quad &\yc=\ycw_\pm(\theta). \label{eq:order1BC}
\end{alignat}
\end{subequations}
Integrate \cref{eq:order1pde} from~$\yc=\ycw_-$ to~$\ycw_+$ and use the
boundary conditions~\eqref{eq:order1BC} to get on the left
\begin{align*}
  \int_{\ycw_-(\theta)}^{\ycw_+(\theta)}
  (\U\sin\theta\,\pd_\yc\bp_1 - \Dyy\,\pd_\yc^2\bp_1)\dint\yc
  &=
  \l[
  \U\sin\theta\,\bp_1 - \Dyy(\theta)\,\pd_\yc\bp_1
  \r]_{\ycw_-(\theta)}^{\ycw_+(\theta)} \nonumber\\
  &=
  -\ycw_+'(\theta)\,\pd_\theta\bp_0(\theta,\ycw_+(\theta))
  + \ycw_-'(\theta)\,\pd_\theta\bp_0(\theta,\ycw_-(\theta)).
\end{align*}
On the right, the integral of the term~$\pd_\theta^2\bp_0$ is
\begin{equation*}
  \int_{\ycw_-(\theta)}^{\ycw_+(\theta)}
  \pd_\theta^2\bp_0\dint\yc
  =
  \pd_\theta
  \int_{\ycw_-(\theta)}^{\ycw_+(\theta)}
  \pd_\theta\bp_0\dint\yc
  -\ycw_+'(\theta)\,\pd_\theta\bp_0(\theta,\ycw_+(\theta))
  + \ycw_-'(\theta)\,\pd_\theta\bp_0(\theta,\ycw_-(\theta)).
\end{equation*}
Combining the last two equations, we obtain
\begin{equation}
  \pd_\T
  \int_{\ycw_-(\theta)}^{\ycw_+(\theta)}
  \bp_0\dint\yc
  =
  \pd_\theta
  \int_{\ycw_-(\theta)}^{\ycw_+(\theta)}
  \pd_\theta\bp_0\dint\yc
  .
  \label{eq:solvond_order1}
\end{equation}
We can then carry out the $\yc$ integral on the right
of~\eqref{eq:solvond_order1} after using \cref{eq:bp0sol_inv}, to get
\begin{subequations}
\begin{align}
  \int_{\ycw_-(\theta)}^{\ycw_+(\theta)}
  \l(
  \pd_\theta \Q
  +
  \vth'(\theta)\Q\,\yc
  \r)\ee^{\vth(\theta)\,\yc}\dint\yc
  &=
  \w(\theta)\,\pd_\theta \Q
  -
  \nu(\theta) \Q
\end{align}
\end{subequations}
where we defined the weight
\begin{subequations}
\begin{align}
  \w(\theta)
  &=
  \int_{\ycw_-(\theta)}^{\ycw_+(\theta)}
  \ee^{\vth(\theta)\,\yc}
  \dint\yc \nonumber\\
  &=
  \l(
  \ee^{\vth(\theta)\,\ycw_+(\theta)}
  -
  \ee^{\vth(\theta)\,\ycw_-(\theta)}
  \r)/\vth(\theta)
  ,
  \label{eq:w}
\intertext{and the drift}
  \nu(\theta)
  &=
  -
  \vth'(\theta)
  \int_{\ycw_-(\theta)}^{\ycw_+(\theta)}
  \yc\,\ee^{\vth(\theta)\,\yc}
  \dint\yc \nonumber\\
  &=
  -
  \frac{\vth'(\theta)}{\vth(\theta)}\l(
  \l[\yc\,\ee^{\vth(\theta)\,\yc}\r]_{\ycw_-(\theta)}^{\ycw_+(\theta)}
  -
  \int_{\ycw_-(\theta)}^{\ycw_+(\theta)}
  \ee^{\vth(\theta)\,\yc}
  \dint\yc
  \r) \nonumber\\
  &=
  \frac{\vth'(\theta)}{\vth(\theta)}\l(\w(\theta)
  -
  \ee^{\vth(\theta)\,\ycw_+(\theta)}\,\ycw_+(\theta)
  +
  \ee^{\vth(\theta)\,\ycw_-(\theta)}\,\ycw_-(\theta)
  \r).
  \label{eq:nu}
\end{align}
\label{eq:wnu}%
\end{subequations}
Note that~$\w(\theta) > 0$ if~$\ycw_+(\theta) > \ycw_-(\theta)$, and
$\w(\theta) = 0$ if and only if~$\ycw_+(\theta) = \ycw_-(\theta)$.  Thus,
$\w(\theta)$ only vanishes when the domain ``pinches off,'' as described in
\cref{sec:closed}.  Despite the apparent singularity, the weight~$\w$ is
nonsingular when~$\vth$ is small:
\begin{equation}
  \w(\theta) \sim \ycw_+(\theta) - \ycw_-(\theta),
  \qquad
  \vth \rightarrow 0.
\end{equation}

Another convenient form for the drift~$\nu$ is
\begin{equation}
  \nu(\theta)
  =
  \w(\theta)\,
  \frac{\vth'(\theta)}{\vth(\theta)}
  \l[1
  -
  \frac{\vth(\theta)}{2\sinh\Dyc(\theta)}
  \l(
  \ee^{\Dyc(\theta)}\,\ycw_+(\theta)
  -
  \ee^{-\Dyc(\theta)}\,\ycw_-(\theta)
  \r)
  \r]
\end{equation}
with
\begin{equation}
  \Dyc(\theta) \ldef
  \tfrac12\vth(\theta)\,(\ycw_+(\theta) - \ycw_-(\theta)).
  \label{eq:Dyc}
\end{equation}
The function~$\nu$ appears singular as~$\Dyc\rightarrow0$, but the limit
exists:
\begin{equation}
  \frac{\nu(\theta)}{\w(\theta)}
  \sim
  -\tfrac12\vth'(\theta)(\ycw_+(\theta) + \ycw_-(\theta)),
  \qquad \Dyc \rightarrow 0.
  \label{eq:nuw_small_Delta}
\end{equation}
This expression is valid whether~$\Dyc$ vanishes owing to~$\vth(\theta)=0$
or~$\ycw_+(\theta) = \ycw_-(\theta)$.

Doing the~$\yc$ integral on the left of~\eqref{eq:solvond_order1}, we finally
obtain the reduced equation
\begin{equation}
  \w(\theta)\,\pd_\T \Q
  + \pd_\theta(\nu(\theta) \Q
  -
  \w(\theta)\,\pd_\theta \Q) = 0.
  \label{eq:Qeq}
\end{equation}
The reduced equation is a (1+1)-dimensional drift-diffusion PDE that captures
the time-evolution of the marginal probability density
\begin{equation}
  \P(\theta,\T)
  =
  \int_{\ycw_-(\theta)}^{\ycw_+(\theta)}
  \bp_0(\theta,\yc,\T)\dint\yc
  =
  \w(\theta)\,\Q(\theta,\T).
  \label{eq:P}
\end{equation}
The weight function~$\w(\theta)$ and drift~$\nu(\theta)$ encode the effect of
the shape of the configuration space.

We can transform~\eqref{eq:Qeq} into an equation for~$\P$:
\begin{equation}
  \pd_\T \P
  +
  \pd_\theta(
  \mu(\theta)\, \P - \pd_\theta \P
  )
  =
  0,
  \label{eq:Peq}
\end{equation}
with
\begin{equation}
  \w(\theta)\,\mu(\theta)
  \ldef
  \nu(\theta) + \w'(\theta)
  =
  \ee^{\vth(\theta)\,\ycw_+(\theta)}\ycw_+'(\theta)
  - \ee^{\vth(\theta)\,\ycw_-(\theta)}\ycw_-'(\theta).
  \label{eq:mu}
\end{equation}
An explicit form for~$\mu$ in terms of~$\Dyc$ in~\eqref{eq:Dyc} is
\begin{equation}
  \mu(\theta)
  =
  \frac{\vth(\theta)}
  {2\sinh\Dyc(\theta)}\l(
  \ee^{\Dyc(\theta)}\,\ycw_+'(\theta)
  - \ee^{-\Dyc(\theta)}\,\ycw_-'(\theta)\r).
  \label{eq:munice}
\end{equation}
Though \cref{eq:Peq} is slightly nicer than \cref{eq:Qeq}, it has the
disadvantage that it requires a derivative~$\w'(\theta)$ in~$\mu(\theta)$,
which can cause problems for nonsmooth swimmer shapes.

\begin{figure}
  \begin{center}
    \includegraphics[height=.4\textheight]{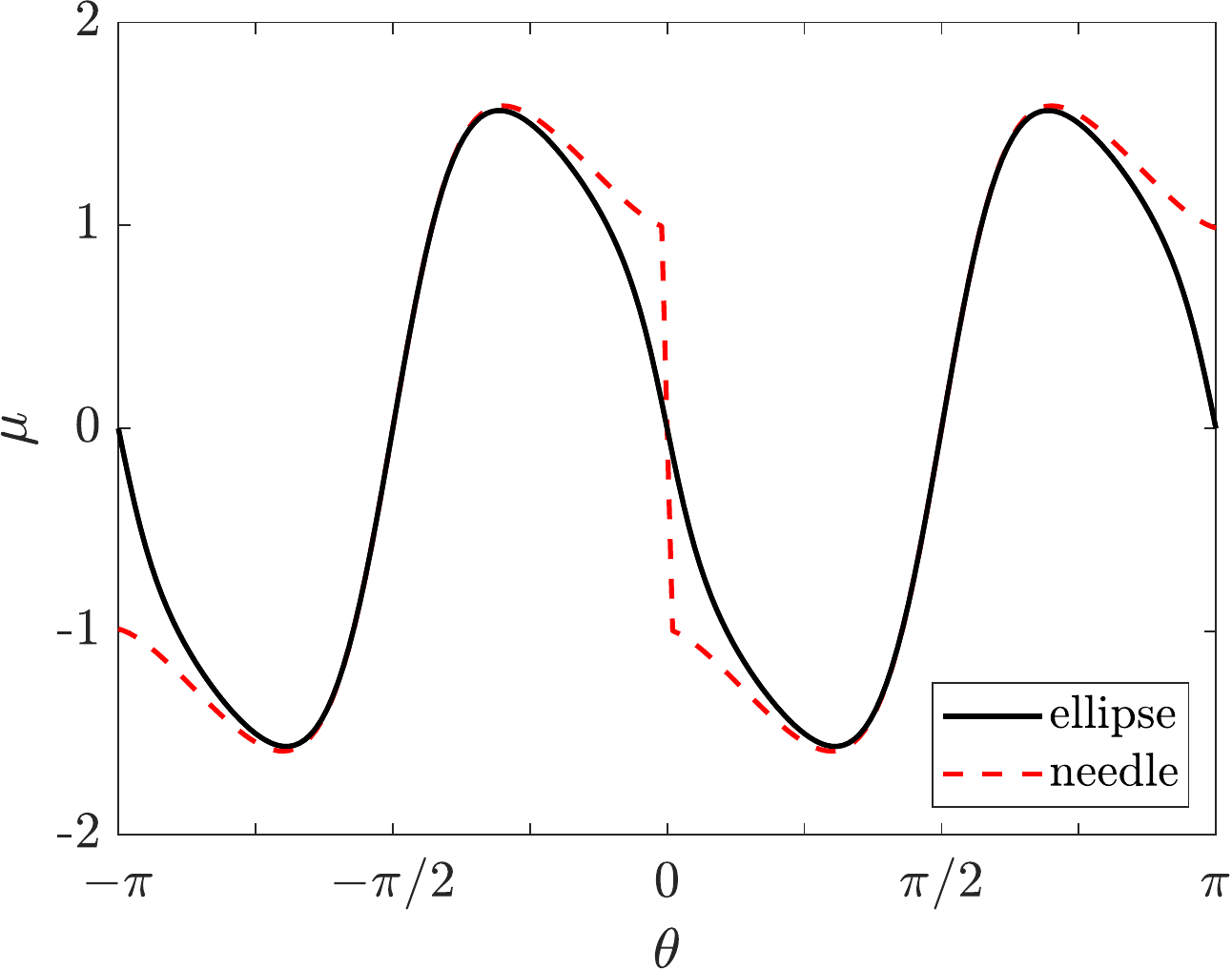}
  \end{center}
  \caption{The angular drift~$\mu_{\mathrm{ellipse}}(\theta)$
    (\cref{eq:mu_ellipse}) in a channel of width~$L=1.2$ for an ellipse
    with~$\aaa=1/2$, $\bbb=1/8$, $\U=\DX=\DY=1$, $\Xcrot=-1/4$.  The dashed
    line is~$\mu_{\mathrm{needle}}(\theta)$ ($\bbb=0$, \cref{eq:mu_needle}).}
  \label{fig:channel_mu_ellipse}
\end{figure}

\begin{Example}[$\mu(\theta)$ for elliptical and needle swimmers]%
  \label{ex:mu}
  For the elliptical swimmer described by \cref{eq:ycstar_ellipse_ecc}, we
  have
  \begin{equation}
    \mu_{\mathrm{ellipse}}(\theta)
    =
    -\tfrac12\ecc^2\,
    \frac{\aaa\vth(\theta)\sin2\theta}
    {\sqrt{1 - \ecc^2\cos^2\theta}}\coth\Dyc(\theta)
    + \Xcrot\,\vth(\theta)\cos\theta
    \label{eq:mu_ellipse}
  \end{equation}
  with~$\Dyc(\theta) = \tfrac12\vth(\theta)(L - 2\aaa\sqrt{1 -
    \ecc^2\cos^2\theta})$.  Note that~$\mu$ vanishes when~$\ecc=0$ (circular
  or point swimmer).  The needle is the limit of~\eqref{eq:mu_ellipse}
  as~$\ecc\rightarrow1$:
  \begin{equation}
    \mu_{\mathrm{needle}}(\theta)
    =
    -\aaa\vth(\theta)\cos\theta\sgn\theta\coth\Dyc(\theta)
    + \Xcrot\,\vth(\theta)\cos\theta
    .
    \label{eq:mu_needle}
  \end{equation}
  These are plotted in \cref{fig:channel_mu_ellipse}.  Note
  that~$\mu_{\mathrm{needle}}(\theta)$ is discontinuous at~$\theta=0$, due to
  the singular derivative of~$\w$ in~\eqref{eq:mu}.
\end{Example}

\section{Invariant density}
\label{sec:inv}

A natural quantity to compute from the reduced equation \eqref{eq:Qeq} is the
\emph{invariant density} for the swimmer.  This is the time-independent
solution~$\Q(\theta,\T) = \Qinv(\theta)$ to~\eqref{eq:Qeq}:
\begin{equation}
  \frac{d}{d\theta}\,(\nu(\theta)\, \Qinv(\theta)
  -
  \w(\theta)\,\Qinv'(\theta)) = 0.
  \label{eq:Qinv}
\end{equation}
$\Qinv$ is unique for a periodic domain~$\Omega$ or a single
component~$\Omega_i$; see \cref{sec:channelgeom}.  Note that~$\Qinv$ (and
hence the invariant density) is independent of~$\Drot$ at leading order.

To find the invariant density, we first integrate \cref{eq:Qinv} once to get
\begin{equation}
  \nu(\theta)\, \Qinv(\theta)
  -
  \w(\theta)\, \Qinv'(\theta)
  =
  \cc_2\,.
  \label{eq:Qinv_oneint}
\end{equation}
The solution to~\eqref{eq:Qinv_oneint} then can be written
\begin{equation}
  \Qinv(\theta)
  =
  \cc_1\,\l(1 - \cc_2\,F(\theta)\r)\,\ee^{\Phi(\theta)}
  \label{eq:Qinvsol}
\end{equation}
where
\begin{equation}
  \Phi(\theta)
  \ldef
  \int_{\thetaLint}^{\theta} \frac{\nu(\vartheta)}{\w(\vartheta)}\dint\vartheta,
  \qquad
  F(\theta)
  \ldef
  \int_{\thetaL}^\theta
  \frac{\dint\vartheta}{\cc_1\w(\vartheta)\,\ee^{\Phi(\vartheta)}}
  \label{eq:PhiF}
\end{equation}
and~$\thetaLint$ is the leftmost domain limit ($\thetaL=-\pi$ for~$\Omega$
and~$\thetaL=\thetaL_i$ for~$\Omega_i$; see \cref{sec:channelgeom}).  The
integrand in~\eqref{eq:PhiF} appears singular as~$\Dyc\rightarrow0$, but the
limit exists as we saw in~\cref{eq:nuw_small_Delta}.

Next we need to determine the constants~$\cc_1$ and~$\cc_2$.  Normalization
of~$\Pinv \ldef \w\Qinv$ determines~$\cc_1$, but~$\cc_2$ depends on whether we
have an open or closed channel configuration space (\cref{sec:channelgeom}).
We treat these two cases separately.

\subsection{Open channel}
\label{sec:open_Pinv}

For the open channel configuration space as described in \cref{sec:open},
$\w(\theta)$ and~$\nu(\theta)$ are $2\pi$-periodic.  The boundary condition on
$\Qinv(\theta)$ is that it be periodic as well.  Choosing~$\thetaLint=-\pi$ in
\cref{eq:PhiF}, we have
\begin{equation}
  \Qinv(-\pi)
  =
  \cc_1
  =
  \Qinv(\pi)
  =
  \cc_1\,\l(1
  -
  \cc_2\,
  F(\pi)\r)\,\ee^{\Phi(\pi)}.
  \label{eq:c1c2condperiodic}
\end{equation}
We solve for~$\cc_2$ in \cref{eq:c1c2condperiodic} to obtain
\begin{equation}
  \cc_2
  =
  \l(1 - \ee^{-\Phi(\pi)}\r) / F(\pi)
  \label{eq:c2sol}
\end{equation}
and
\begin{equation}
  \Qinv(\theta)
  =
  \cc_1\,\ee^{\Phi(\theta)}
  \l(
  1
  -
  \l(1 - \ee^{-\Phi(\pi)}\r)
  {F(\theta)}/{F(\pi)}
  \r).
  \label{eq:Qinvsol2}
\end{equation}
The constant~$\cc_1$ is chosen to enforce the normalization
of~$\Pinv = \w\Qinv$:
\begin{equation}
  \int_{-\pi}^\pi\int_{\ycw_-(\theta)}^{\ycw_+(\theta)}
  \bp_0(\theta,\yc)\dint\yc\dint\theta
  =
  \int_{-\pi}^\pi
  \Pinv(\theta)\dint\theta = 1.
\end{equation}

If~$\Phi(\theta)$ happens to be $2\pi$-periodic, then we have~$\Phi(\pi)=0$,
so~$\cc_2=0$ and
\begin{equation}
  \Qinv(\theta)
  =
  \cc_1\,\ee^{\Phi(\theta)},
  \qquad
  \text{($\Phi(\theta)$ $2\pi$-periodic)}.
  \label{eq:Qinv_open_Phi_periodic}
\end{equation}
The invariant probability density in this case satisfies \emph{detailed
  balance}~\cite{Pavliotis}.  In fact~$\Phi(\theta)$ is periodic for the very
important case of a left-right symmetric swimmer, for then we
have~$\ycw_+(\theta) = -\ycw_-(-\theta)$, which follows from
symmetries~\eqref{eq:LRsym} and~\eqref{eq:channelsym}.  This leads
to~$\Dyc(-\theta) = -\Dyc(\theta)$ and the integrand of \cref{eq:PhiF} is odd
in~$\theta$.  Choosing~$\thetaLint=-\pi$ then gives~$\Phi(-\pi)=\Phi(\pi)=0$,
\ie, $\Phi$ is periodic.

\begin{figure}
  \begin{center}
    \subcaptionbox{\label{fig:channel_invdens_open}}{
      \includegraphics[height=.26\textheight]{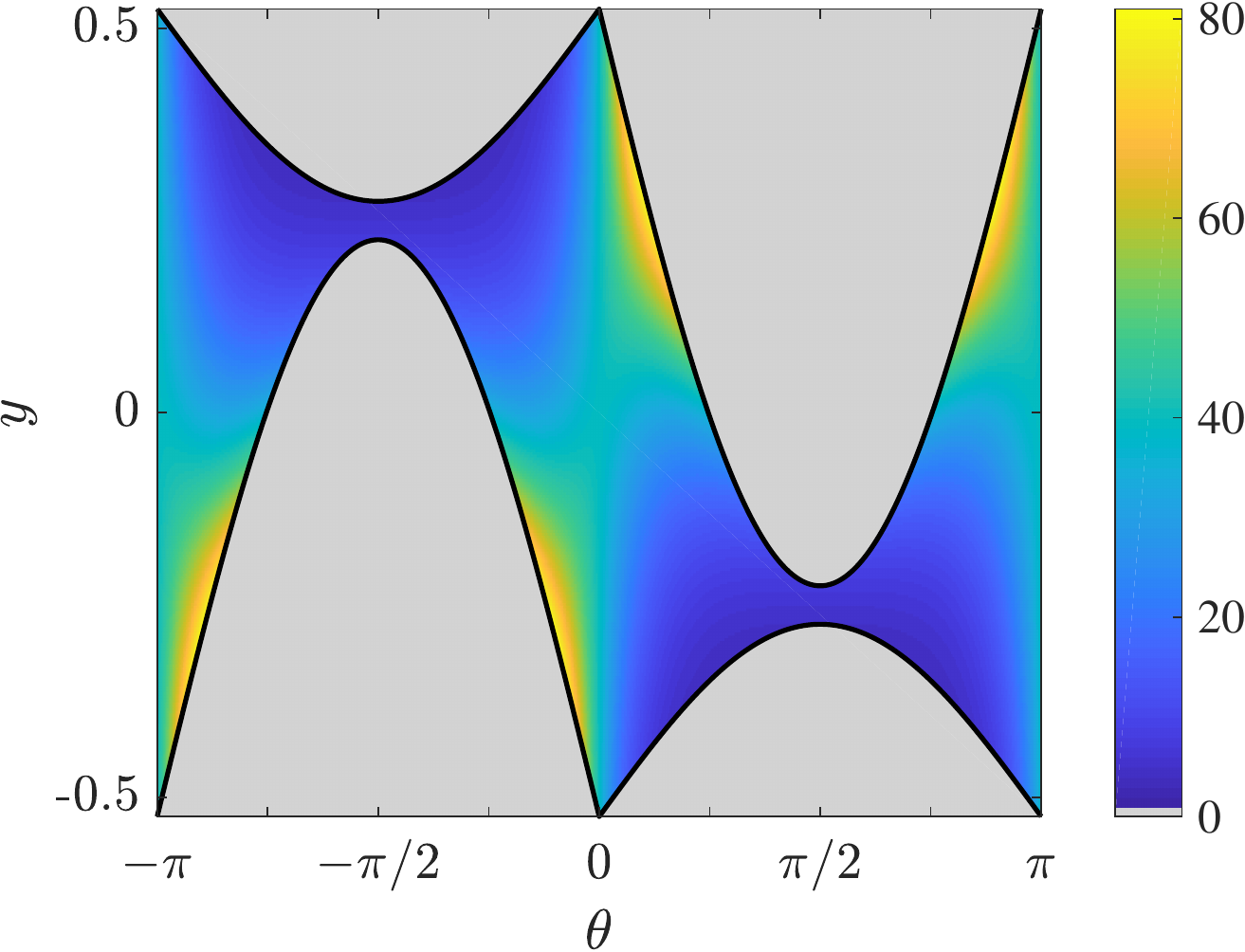}}
    \hspace{.05\textwidth}
    \subcaptionbox{\label{fig:channel_invdens_closed}}{
      \includegraphics[height=.26\textheight]{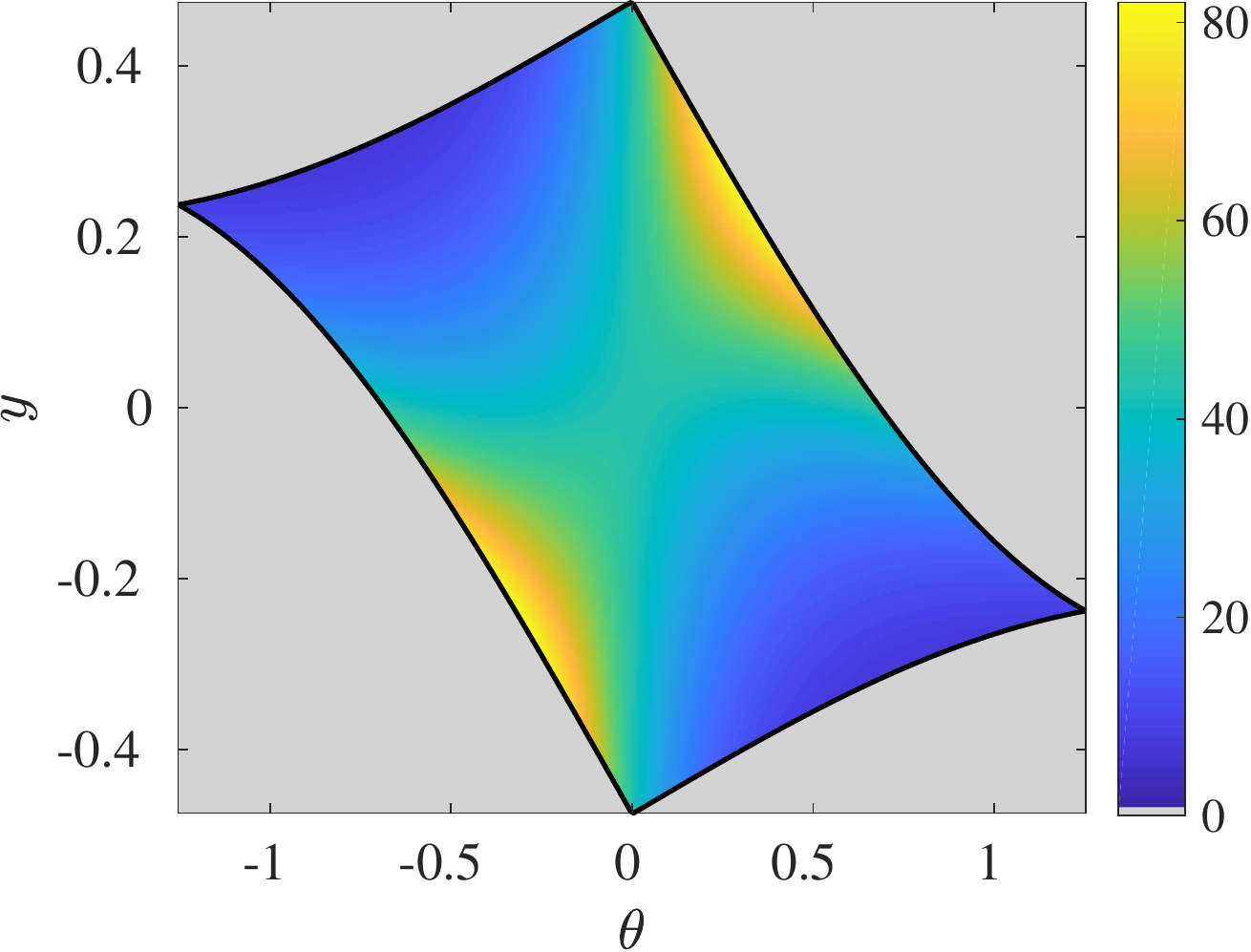}}
  \end{center}
  \caption{Invariant density~$\bp_0 = \Qinv(\theta)\,\ee^{\vth(\theta)\,\yc}$
    for~$\U=1$ and~$\DX=\DY=0.1$ for the needle in
    \cref{fig:swimmer_walldist_needle_Xrot} of length~$\ell=2\aaa=1$ in (a) an
    open channel of width~$L = 1.05$; (b) a closed channel of
    width~$L = 0.95$, for the domain~$\Omega_1$ in
    \cref{fig:channel_conf_closed}.}
  \label{fig:channel_invdens}
\end{figure}

\begin{figure}
  \begin{center}
    \subcaptionbox{\label{fig:channel_Pinv}}{
      \includegraphics[height=.26\textheight]{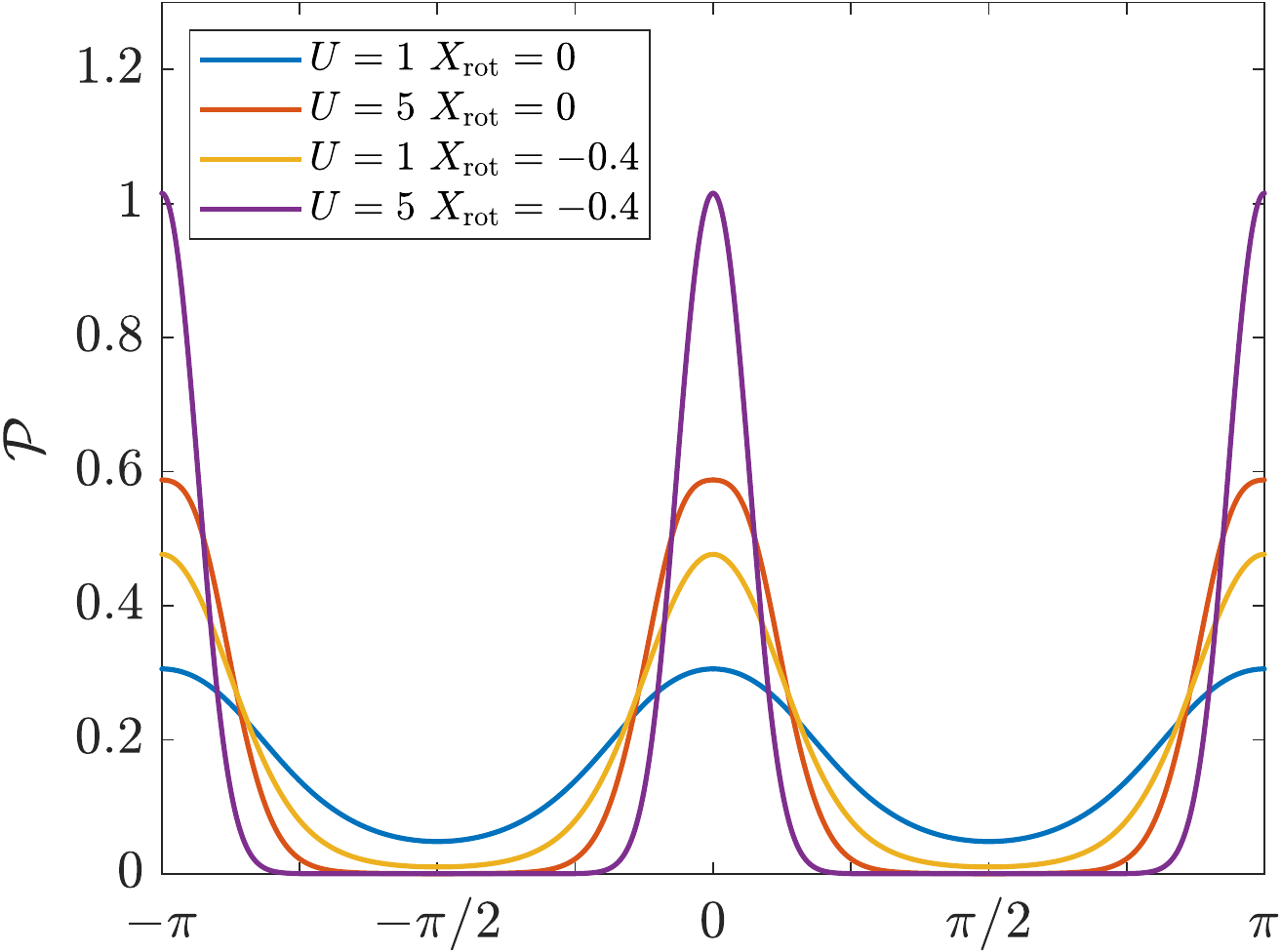}}
    \hspace{.05\textwidth}
    \subcaptionbox{\label{fig:channel_invPinv}}{
      \includegraphics[height=.26\textheight]{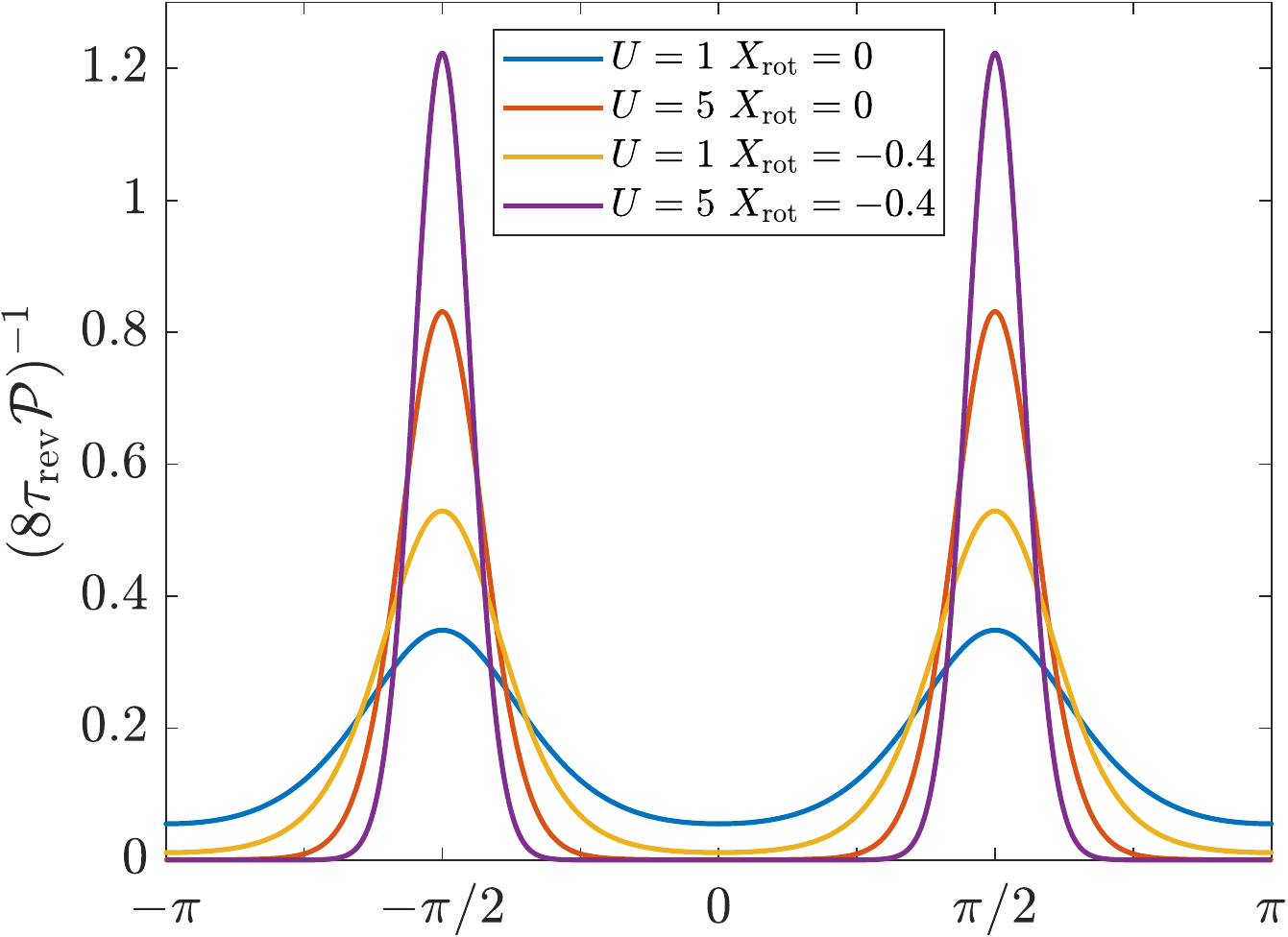}}
  \end{center}
  \caption{For an ellipse with~$2\aaa=\bbb=1$, $\DX=\DY=0.1$, $\Drot=0.01$,
    $\U=1$, in a channel of width~$L=1.2$: (a) Marginal invariant probability
    density~$\Pinv(\theta)$; (b) $1/\Pinv$, normalized to unit area (see
    \cref{eq:taurevsimp} for definition of~$\taurev$).}
  \label{fig:channel_Pinv_invPinv}
\end{figure}

From~$\Qinv$, we can reconstruct the full invariant density
from~\cref{eq:bp0sol_inv} as
$\bp_0(\theta,\yc) = \Qinv(\theta)\,\ee^{\vth(\theta)\yc}$,
with~$\vth(\theta) = \nofrac{U\sin\theta}{\Dyy(\theta)}$.  The exponential
term reflects the accumulation near both walls, as observed in experiments and
simulations.  The thickness of the boundary layer is
$\nofrac{\Dyy}{U\sin\theta}$, which agrees qualitatively with the results for
a spherical swimmer in \cite{Ezhilan2015,Elgeti2013}.  A typical invariant
density in an open-channel configuration is shown in
\cref{fig:channel_invdens_open} for a needle swimmer.  The marginal invariant
probability density~$\Pinv(\theta)$ is plotted in \cref{fig:channel_Pinv} for
elliptical swimmers with different velocities~$\U$ and centers of
rotation~$\Xcrot$.  From \cref{fig:channel_invdens}, the invariant marginal
density in $\yc$ peaks near both walls, but not exactly at the walls, in
accordance with the simulations in the appendix of \cite{Ezhilan2015}.


What is the meaning of nonzero~$\cc_2$?  It represents an average rotational
drift of the needle's stochastic angle~$\theta(\t)$.  To see this, note that
in the equilibrium state, we have the expectation
\begin{equation}
  \E\mu(\theta(\T))
  =
  \int_{-\pi}^\pi \mu(\theta)\,\Pinv(\theta)\dint\theta
  =
  \int_{-\pi}^\pi (\Pinv'(\theta) + \cc_2)\dint\theta
  =
  2\pi \cc_2
  \label{eq:Emu}
\end{equation}
since~$\mu\,\Pinv - \Pinv' = \cc_2$ and~$\Pinv(\theta)$ is periodic.  Hence,
the average rate of angular rotation of the swimmer is~$\omega = 2\pi \cc_2$.
From~\eqref{eq:PhiF}, the periodic average of~$\mu(\theta)$ is
\begin{equation}
  \mub
  =
  \frac{1}{2\pi}
  \int_{-\pi}^\pi \mu(\theta)\dint\theta
  =
  \frac{1}{2\pi}
  \int_{-\pi}^\pi \frac{\nu(\theta)}{\w(\theta)}\dint\theta
  =
  \frac{\Phi(\pi)}{2\pi}
  =
  \frac{1}{2\pi}
  \log(1 - \cc_2 F(\pi))^{-1}
  \label{eq:mub}
\end{equation}
which is zero if and only if $\cc_2=0$ (\cref{eq:c2sol}).

\begin{Example}[invariant density for fast needle swimmer]%
  \label{ex:needle_invdens}
  It is in general quite challenging to get closed-form solutions for the
  invariant density of a swimmer, but it can be done in the large-$\U$ limit.
  From~\cref{eq:PhiF,eq:mu} we have
  \begin{equation}
    \Phi(\theta)
    =
    \int_{-\pi}^\theta \mu(\vartheta)\dint\vartheta
    -\log\w(\theta)
    + \text{const.},
  \end{equation}
  and so the leading-order invariant marginal
  density~$\Pinv=\cc_1\w\,\ee^\Phi$ for a left-right symmetric swimmer is
  \begin{equation}
    \Pinv(\theta)
    =
    \bar\cc_1\,\exp\l(\int_{-\pi}^\theta \mu(\vartheta)\dint\vartheta\r)
    \label{eq:Pinv_mu}
  \end{equation}
  with~$\bar\cc_1$ a normalization constant.  For large~$\U$, the
  constant~$\bar\cc_1$ can be determined by approximating the normalization
  integral using the maxima of~$\mu$.

  We illustrate this here for the needle swimmer of \cref{ex:needle,ex:mu}.
  For large~$\U$, we can approximate~$\coth\Dyc \approx \sgn(\theta)$
  for~$\mu = \mu_{\mathrm{needle}}$ in~\cref{eq:mu_needle}, and we have
  \begin{equation}
    \mu(\theta)
    \approx
    -
    \vth(\theta)\cos\theta\,
    \l(\aaa - \Xcrot\r),
    \qquad
    \U \rightarrow \infty,
    \label{eq:mu_needle_largeU}
  \end{equation}
  with~$\vth$ defined in \cref{eq:bp0sol_inv}, and~$\aaa = \ell/2$ the needle
  half-length.  Note that the channel width~$L$ does not appear
  in~\eqref{eq:mu_needle_largeU} at leading order in large~$\U$: the needle
  spends most of its time stuck to one of the walls, so the channel width is
  not important.  We can integrate~\eqref{eq:mu_needle_largeU} and use the
  result in~\eqref{eq:Pinv_mu} to find
  \begin{equation}
    \Pinv(\theta)
    =
    \bar\cc_1\exp\l(
    \beta\,
    \log\l(
    \alpha\sin^2\theta + \cos^2\theta
    \r)^{1/(1-\alpha)}\r),
    \qquad
    \label{eq:Pinv_needle0}
  \end{equation}
  with
  \begin{equation}
    \alpha \ldef \nofrac{\DX}{\DY},\qquad
    \beta \ldef \nofrac{\U(\aaa -\Xcrot)}{2\DY} \gg 1.
    \label{eq:alphabeta}
  \end{equation}
  We can see that ``large $\U$'' in nondimensional terms means large~$\beta$,
  which is a P\'eclet number that accounts for the position of the center of
  rotation: $\beta$ is maximized when the center of rotation is at the rear
  ($\Xcrot=-\aaa$), which is a pusher-like swimmer.  We can now use Laplace's
  method to find the normalization constant~$\bar\cc_1$.  The maxima of the
  argument of the exponential in~\eqref{eq:Pinv_needle0} correspond to the
  zeros of~$\mu$ at~$\theta=0$ and~$\pi$ (with~$-\aaa \le \Xcrot < \aaa$).  We
  thus find
  \begin{equation}
    \Pinv(\theta)
    =
    \sqrt{\frac{\beta}{4\pi}}\,
    \l(
    \alpha\sin^2\theta + \cos^2\theta
    \r)^{\beta/(1 - \alpha)},
    \qquad
    \beta \gg 1.
    \label{eq:Pinv_needle}
  \end{equation}
  In the limit~$\alpha=1$ (equal diffusivities),
  \cref{eq:Pinv_needle} simplifies to
  \begin{equation}
    \Pinv(\theta)
    =
    \sqrt{\frac{\beta}{4\pi}}\,\ee^{-\beta\sin^2\theta},
    \qquad
    \beta \gg 1.
    \label{eq:Pinv_needle_alpha1}
  \end{equation}
  Note that the limit~$\alpha\rightarrow0$ in~\eqref{eq:Pinv_needle} is
  well-defined and
  gives~$\Pinv(\theta) = \sqrt{\nofrac{\beta}{4\pi}}\,
  \lvert\cos\theta\rvert^{2\beta}$.  However, the
  limit~$\alpha \rightarrow \infty$ gives an improperly normalized density,
  indicating that the limits~$\beta$, $\alpha\rightarrow\infty$ do not
  commute.  (The quartic term in a Taylor series expansion of the $\log$
  in~\eqref{eq:Pinv_needle0} has coefficient proportional to~$\alpha$, and
  cannot be neglected when applying Laplace's method.)

  In \cref{fig:channel_Pinv_invPinv_largeU_needle} we compare a numerical
  solution for~$\Pinv$ to the large-$\U$ form~\eqref{eq:Pinv_needle}.
  \begin{figure}
    \begin{center}
      \subcaptionbox{\label{fig:channel_Pinv_largeU_needle}}{
        \includegraphics[height=.26\textheight]{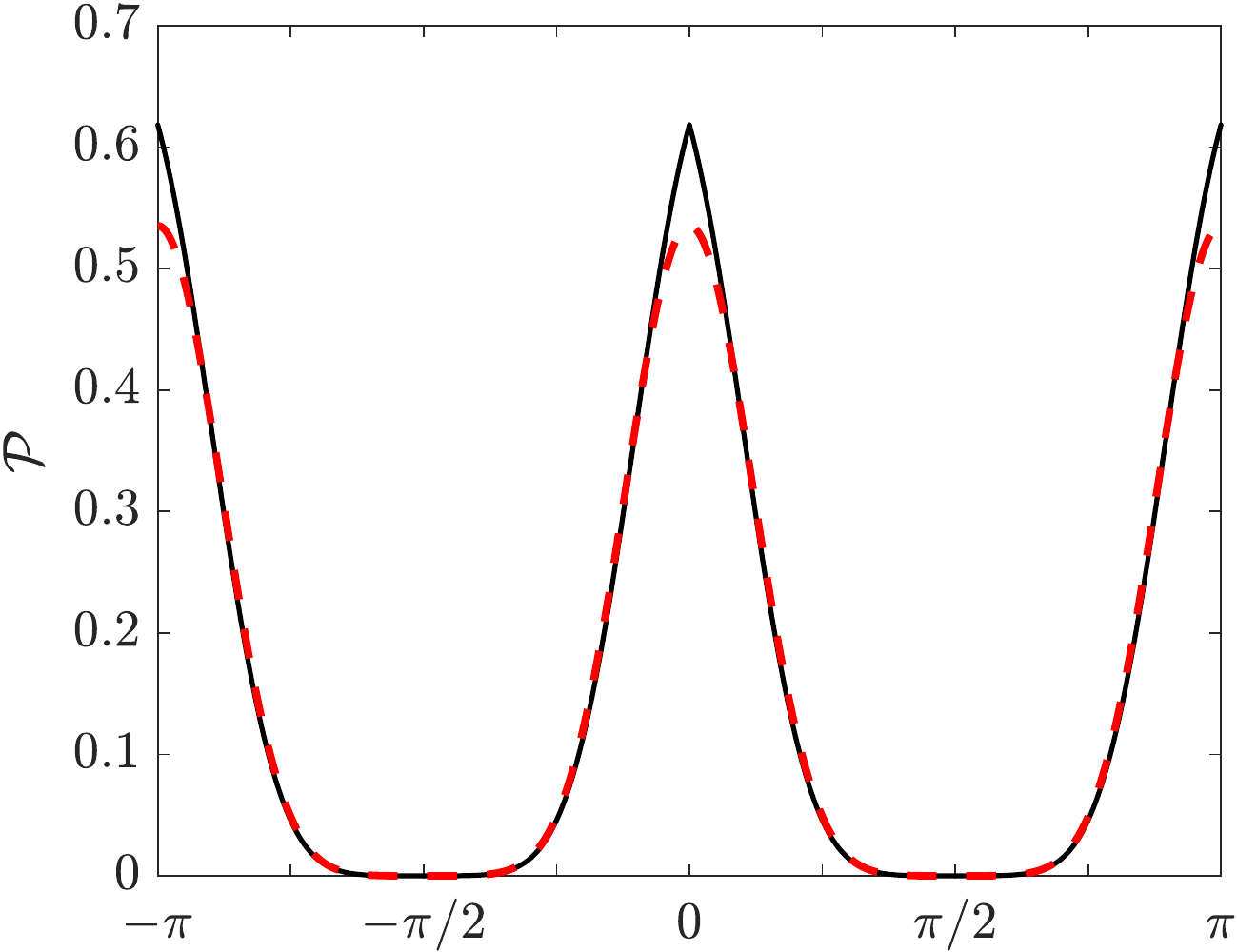}}
      \hspace{.05\textwidth}
      \subcaptionbox{\label{fig:channel_invPinv_largeU_needle}}{
        \includegraphics[height=.26\textheight]{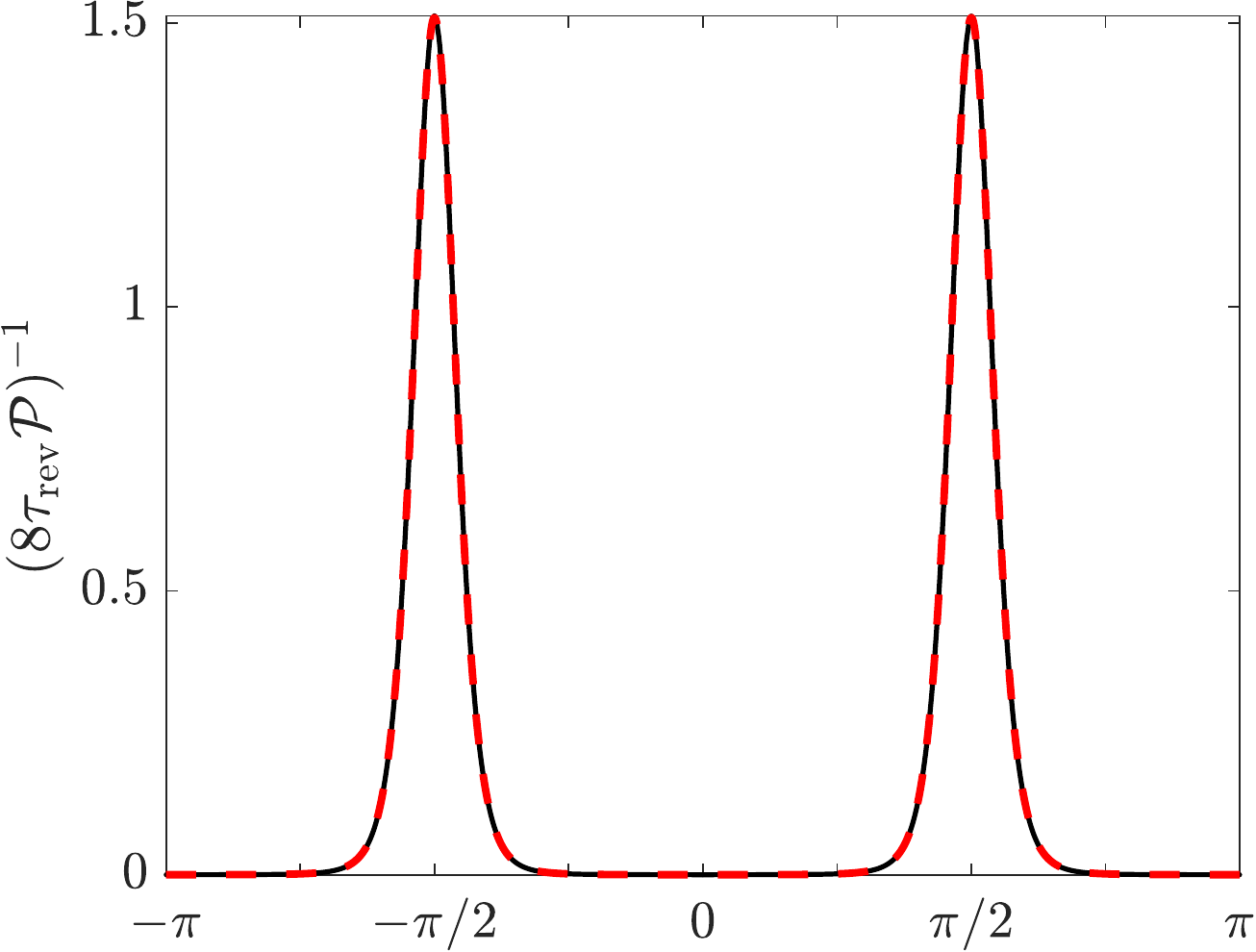}}
    \end{center}
    \caption{For a needle with~$\ell=1$, $\U=8$, $\DX=0.1$, $\DY=1$,
      $\Drot=0.01$, in a channel of width~$L=1.2$ ($\beta=3.6$): (a) Marginal
      invariant probability density~$\Pinv(\theta)$ (solid) and large-$\U$
      form~\eqref{eq:Pinv_needle} (dashed).  Notice the discrepancy
      near~$\theta=0,\pi$, which goes away for large~$\U$. (b) $1/\Pinv$,
      normalized to unit area, for the same parameters.}
    \label{fig:channel_Pinv_invPinv_largeU_needle}
  \end{figure}
  There is some discrepancy near~$\theta=0,\pi$, which comes from the
  approximation~\eqref{eq:mu_needle_largeU} breaking down near those points,
  but the difference vanishes as~$\U$ gets larger.  We also plot~$1/\Pinv$,
  which we shall need in \cref{sec:MERT}, for which the approximation is
  uniformly much better.

\end{Example}

\begin{Example}[invariant density for fast circular swimmer]%
  \label{ex:circ_invdens}
  When the swimmer is perfectly circular, putting~$\ecc=0$ in
  \cref{eq:mu_ellipse} gives~$\mu(\theta) = \Xcrot\,\vth(\theta)\cos\theta$,
  which is the same as~\eqref{eq:mu_needle_largeU} with~$\aaa=0$ in the
  previous example, except that here this expression is exact.  The invariant
  density thus has the form~\eqref{eq:Pinv_needle0},
  with~$\beta = \nofrac{-\U\Xcrot}{2\DY}$, which can have either sign.
  For~$\beta>0$ and large we
  recover~\cref{eq:Pinv_needle,eq:Pinv_needle_alpha1} --- the circular swimmer
  tends to align parallel to the wall.  For~$-\beta>0$ and large the maxima
  of~$\int\mu\dint\theta$ switch from~$\{0,\pi\}$ to~$\pm\pi/2$, and we get
  instead of \cref{eq:Pinv_needle}:
  \begin{equation}
    \Pinv(\theta)
    =
    \sqrt{\frac{\lvert\beta\rvert}{4\pi\alpha}}\,
    \l(
    \sin^2\theta + \alpha^{-1}\cos^2\theta
    \r)^{-\lvert\beta\rvert/(1 - \alpha)},
    \qquad
    -\beta \gg 1,
    \label{eq:Pinv_circ}
  \end{equation}
  which for~$\alpha=1$ simplifies to
  \begin{equation}
    \Pinv(\theta)
    =
    \sqrt{\frac{\lvert\beta\rvert}{4\pi}}\,\ee^{-\lvert\beta\rvert\cos^2\theta},
    \qquad
    -\beta \gg 1.
    \label{eq:Pinv_circ_alpha1}
  \end{equation}
  Comparing the latter to~\eqref{eq:Pinv_needle_alpha1}, we can see that
  puller-like circular swimmers ($\Xcrot>0$) collect at~$\theta=\pm\pi/2$,
  swimming towards the wall, rather than aligning parallel to the wall.  The
  limit~$\alpha\rightarrow0$ in~\ref{eq:Pinv_circ} gives~$\Pinv(\theta) =
  \tfrac12\l(\delta(\theta - \pi/2) + \delta(\theta + \pi/2)\r)$.
\end{Example}

\subsection{Closed channel}
\label{sec:closed_Pinv}

For a closed channel configuration space, as described in \cref{sec:closed},
the domain is given by~$\Omega_i$ in \cref{eq:Omega_i}, for some fixed~$i$.
Now the boundary condition is that there be no net flux in the~$\theta$
direction, so the constant~$\cc_2=0$ in \cref{eq:Qinv_oneint}.  The solution
for the invariant density is thus
\begin{equation}
  \Qinv(\theta)
  =
  \cc_1\,\ee^{\Phi(\theta)},
  \qquad
  \thetaL_i \le \theta \le \thetaR_i,
  \label{eq:Qinv_closed}
\end{equation}
with~$\cc_1$ obtained by the normalization condition for~$\Pinv = \w\Qinv$:
\begin{equation}
  \int_{\thetaL_i}^{\thetaR_i}
  \int_{\ycw_-(\theta)}^{\ycw_+(\theta)}
  \bp_0(\theta,\yc)\dint\yc\dint\theta
  =
  \int_{\thetaL_i}^{\thetaR_i}
  \Pinv(\theta)\dint\theta = 1.
  \label{eq:Qinv_norm_closed}
\end{equation}
A typical invariant density in a closed-channel configuration is shown in
\cref{fig:channel_invdens_closed} for a needle swimmer.

\section{Mean exit and mean reversal times}
\label{sec:MERT}

A standard problem in drift-diffusion processes is to compute the \emph{mean
  exit time} (MET) of a particle to some exit, also called a first-passage
time
\cite{Redner,Holcman2014,Ward1993,Kurella2015,Grebenkov2016,Marcotte2018}.
Associated with the reduced drift-diffusion equation \eqref{eq:Peq} is a
reduced equation for the mean exit time~$\tau(\theta)$:
\begin{subequations}
  \begin{align}
    \mu(\theta)\, \tau' + \tau'' &= -1,
    \qquad \thetaL < \theta < \thetaR; \label{eq:reducedMETPDE}\\
    \tau(\thetaL) = \tau(\thetaR) &= 0.
  \end{align}
  \label{eq:reducedMET}%
\end{subequations}
The left-hand side of~\eqref{eq:reducedMETPDE} is the adjoint of the linear
operator in~\cref{eq:Peq} \cite{Redner, Holcman2014, Kurella2015}.  The
solution to \Cref{eq:reducedMET} gives the expected time~$\tau$ for a particle
starting at~$\theta$ (for any~$\yc$) to reach an `exit' at~$\theta=\thetaL$
or~$\theta=\thetaR$.  Note that since~$\T=\Drot\t$ in \cref{eq:Peq} the
dimensional MET is~$\tau/\Drot$, which goes to infinity
as~$\Drot\rightarrow0$, \ie, the exit cannot be reached if~$\Drot=0$.  The
word exit here is interpreted loosely: the MET merely signifies the first time
a swimmer's orientation achieves the value~$\thetaL$ or~$\thetaR$, starting
from some value~$\theta$.

\subsection{Solving the mean exit time equation}
\label{sec:METsol}

To solve \cref{eq:reducedMET}, define~$\wtau = \w\tau'$ which satisfies
\begin{equation}
  (\nu/\w)\, \wtau + \wtau' = -\w,
  \qquad
  \wtau = \w\tau',
  \label{eq:wtau}%
\end{equation}
where~$\w(\theta)$ and~$\nu(\theta)$ are defined in~\eqref{eq:wnu}.  Use the
integrating factor~$\cct_1\,\ee^{\Phi(\theta)}$ from \cref{eq:PhiF} to get
\begin{equation}
  \tau'(\theta)
  =
  -\frac{1}{\Pinvt(\theta)}
  \l(
  \Gt(\theta) - \Cc
  \r),
  \qquad
  \Gt(\theta) \ldef
  \int_{\thetaLint}^{\theta} \Pinvt(\vartheta)\dint\vartheta,
  \label{eq:taup}
\end{equation}
where~$\Pinvt(\theta) = \cct_1\,\w\,\ee^{\Phi}$ is analogous to the invariant
density for a closed channel~\eqref{eq:Qinv_closed} and~$\Cc$ is an
integration constant.  We choose~$\cct_1$ so that the normalization
condition~\eqref{eq:Qinv_norm_closed} is satisfied: $\Gt(\thetaRint) = 1$;
hence, $\Gt(\theta)$ is the equilibrium probability of finding the swimmer
between~$\thetaL$ and~$\theta$, \emph{if} the channel were closed.  The mean
exit time~$\tau$ has a unique maximum at~$\theta=\theta_*$
with~$\Cc = \Gt(\theta_*)$.

Now integrate~\cref{eq:taup}:
\begin{equation}
  \tau(\theta)
  =
  \int_{\thetaL}^{\theta}
  \frac{1}{\Pinvt(\vartheta)}
  \l(
  \Cc - \Gt(\vartheta)\r)
  \dint\vartheta,
  \label{eq:tausolfull}
\end{equation}
which satisfies the left boundary condition~$\tau(\thetaL)=0$.  The right
boundary condition then requires~$\tau(\thetaR) = 0$, which fixes the
integration constant
\begin{equation}
  \Cc
  =
  \int_{\thetaL}^{\thetaR}
  \frac{\Gt(\vartheta)}{\Pinvt(\vartheta)}\dint\vartheta
  \Bigr\slash
  \int_{\thetaL}^{\thetaR}
  \frac{\dint\vartheta}{\Pinvt(\vartheta)}.
  \label{eq:Cc}
\end{equation}
\Cref{eq:tausolfull,eq:Cc} give the mean exit time for a swimmer starting
at~$\theta$ and exiting at either~$\thetaL$ or~$\thetaR$.  We now focus on a
particular version of this mean exit time with a more natural interpretation
--- the mean reversal time.

\subsection{Mean reversal time}
\label{sec:MRT}

The \emph{mean reversal time}~$\taurev$ (or turnaround
time~\cite{Holcman2014}) is the expected time for a swimmer initially oriented
with~$\theta=0$ to reverse direction to~$\theta=\pm\pi$.  It can be obtained
from \cref{eq:tausolfull} by setting~$-\thetaL=\thetaR=\pi$ and~$\theta=0$:
\begin{equation}
  \taurev
  =
  \tau(0)
  =
  \int_{-\pi}^0
  \frac{1}{\Pinvt(\vartheta)}
  \l(
  \Cc - \Gt(\vartheta)\r)
  \dint\vartheta.
  \label{eq:taurevfull}
\end{equation}
In \cref{apx:mrt_calc} we show how the constant~$\Cc$ can be eliminated to
obtain
\begin{equation}
  \taurev
  =
  \frac{\Gt(0)}{1 + \ee^{\pi\mub}}
  \int_{-\pi}^0\frac{\dint\vartheta}{\Pinvt(\vartheta)}
  +
  \tanh(\pi\mub/2)
  \int_{-\pi}^0
  \frac{\Gt(\vartheta)}{\Pinvt(\vartheta)}
  \dint\vartheta
  \label{eq:taurev}
\end{equation}
where~$\mub$ is defined in~\eqref{eq:mub}.  For a left-right symmetric
swimmer, $\Pinvt=\Pinv$, $\mub = 0$, and~$\Gt(0)=1/2$ by symmetry, so we
obtain the compact expression
\begin{equation}
  \taurev
  =
  \tfrac14
  \int_0^\pi
  \frac{\dint\theta}{\Pinv(\theta)}.
  \label{eq:taurevsimp}
\end{equation}

\begin{Example}[mean reversal time with diffusion only]%
  \label{ex:mrt_diff}
  In the absence of swimming ($\U=0$), we have~$\nu = 0$ from~\cref{eq:nu},
  and~$\Pinvt = \Pinv = \cc_1\w$,
  with~$\cc_1^{-1} = \int_{-\pi}^\pi \w \dint\theta$.  Hence,
  \cref{eq:taurevsimp} is
  \begin{equation}
    \taurev
    =
    \frac{1}{2}\,
    \l(\int_0^\pi \w(\theta) \dint\theta\r)
    \l(\int_0^\pi
    \frac{\dint\theta}{\w(\theta)}\r).
    \label{eq:taurevdiff}
  \end{equation}
  For a needle with wall distance function~\eqref{eq:ycstar_needle}, we get
  from \cref{eq:ycw_mp}
  \begin{equation}
    \w(\theta)
    =
    \ycw_+(\theta) - \ycw_-(\theta)
    =
    L - \ell\,\lvert\sin\theta\rvert
  \end{equation}
  where~$\Xcrot$ drops out: the center of rotation is immaterial in the
  absence of swimming.  We can then easily compute the
  integral~\eqref{eq:taurevdiff} to obtain
  \begin{equation}
    \taurev
    =
    \frac{(\pi - 2\lambda)(\pi - \arccos\lambda)}{\sqrt{1 - \lambda^2}},
    \qquad
    \lambda \ldef \ell/L < 1.
    \label{eq:taurevdiff_needle}
  \end{equation}
  The `narrow exit' limit corresponds to~$\lambda = 1 - \delta$, with~$\delta$
  small.
  \begin{equation}
    \taurev
    =
    \frac{\pi(\pi - 2)}{\sqrt{2\delta}}
    +
    \Order{\delta^0},
    \qquad
    \delta \ll 1.
  \end{equation}
  This is similar but not identical to Holcman \& Schuss's result
  \cite[Eq.~(5.13)]{Holcman2014}:
  \begin{equation}
    \taurev
    =
    \frac{\pi(\pi-2)}{\sqrt{2\delta}}\,
    \sqrt{\frac{\DY}{2L^2\Drot}}
    \label{eq:taurevdiff_HS}
  \end{equation}
  valid as~$\delta\rightarrow0$, but otherwise unconstrained.  In
  \cref{fig:channel_revtime_needle}, a comparison to a finite-element
  numerical solution of the full PDE (\ie, without using the reduced
  equation) shows excellent agreement with our small-$\Drot$
  form~\eqref{eq:taurevdiff_needle}, but less so with the
  form~\eqref{eq:taurevdiff_HS}.  Possibly there is a parameter regime
  where~\eqref{eq:taurevdiff_HS} shows better agreement.

  \begin{figure}
    \begin{center}
      \includegraphics[width=.6\textwidth]{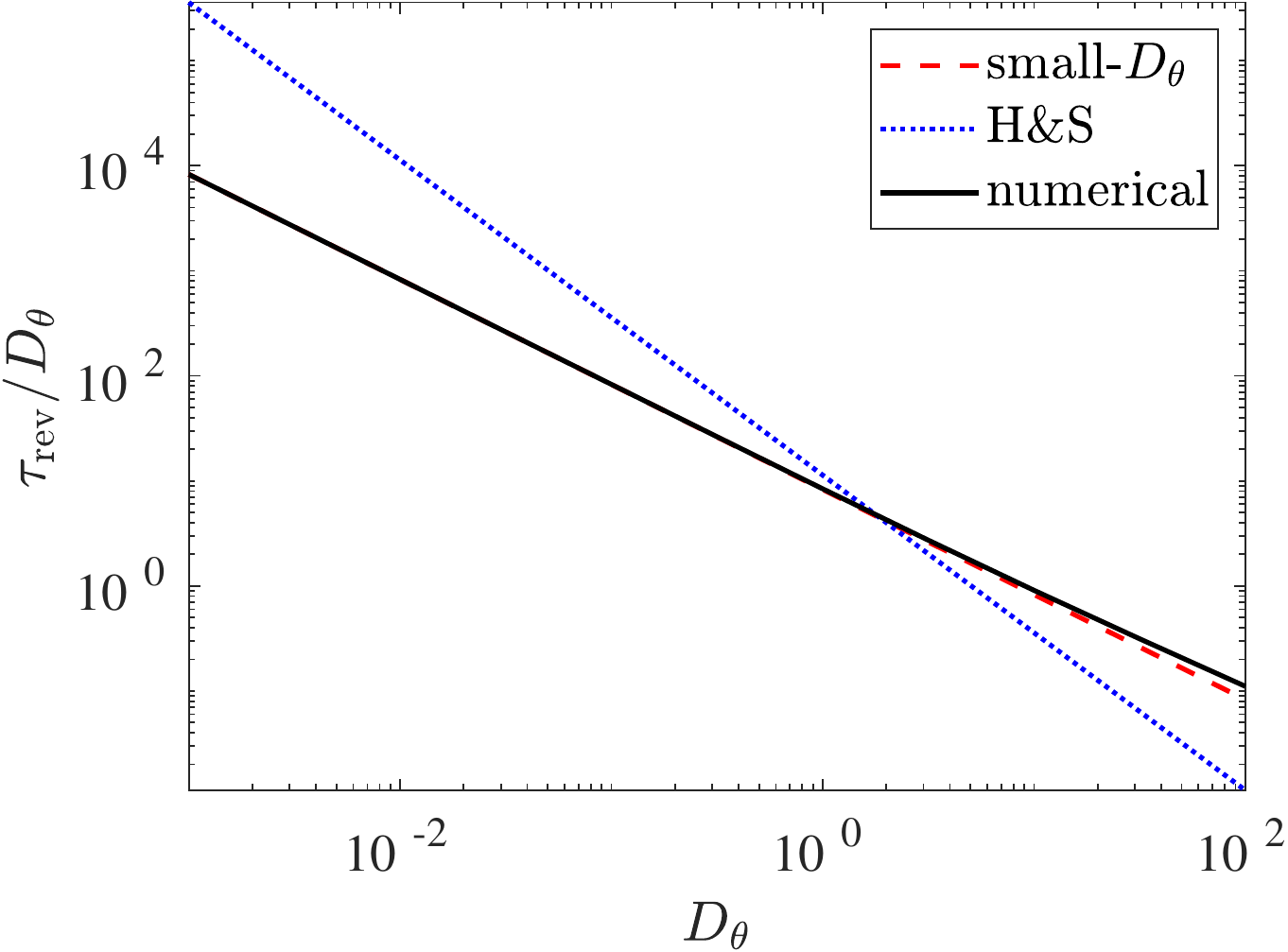}
    \end{center}
    \caption{Mean reversal time for a needle as a function of~$\Drot$,
      for~$\lambda=.9$, $\DY=1$.  The dashed line
      is~\eqref{eq:taurevdiff_needle}, which is technically valid for
      small~$\Drot$ but applies over a wide range.  The dotted line is
      from~\cite{Holcman2014}.  The solid line is from a finite-element
      simulation of the full PDE.}
    \label{fig:channel_revtime_needle}
  \end{figure}

  \begin{figure}
    \begin{center}
      \includegraphics[width=.6\textwidth]{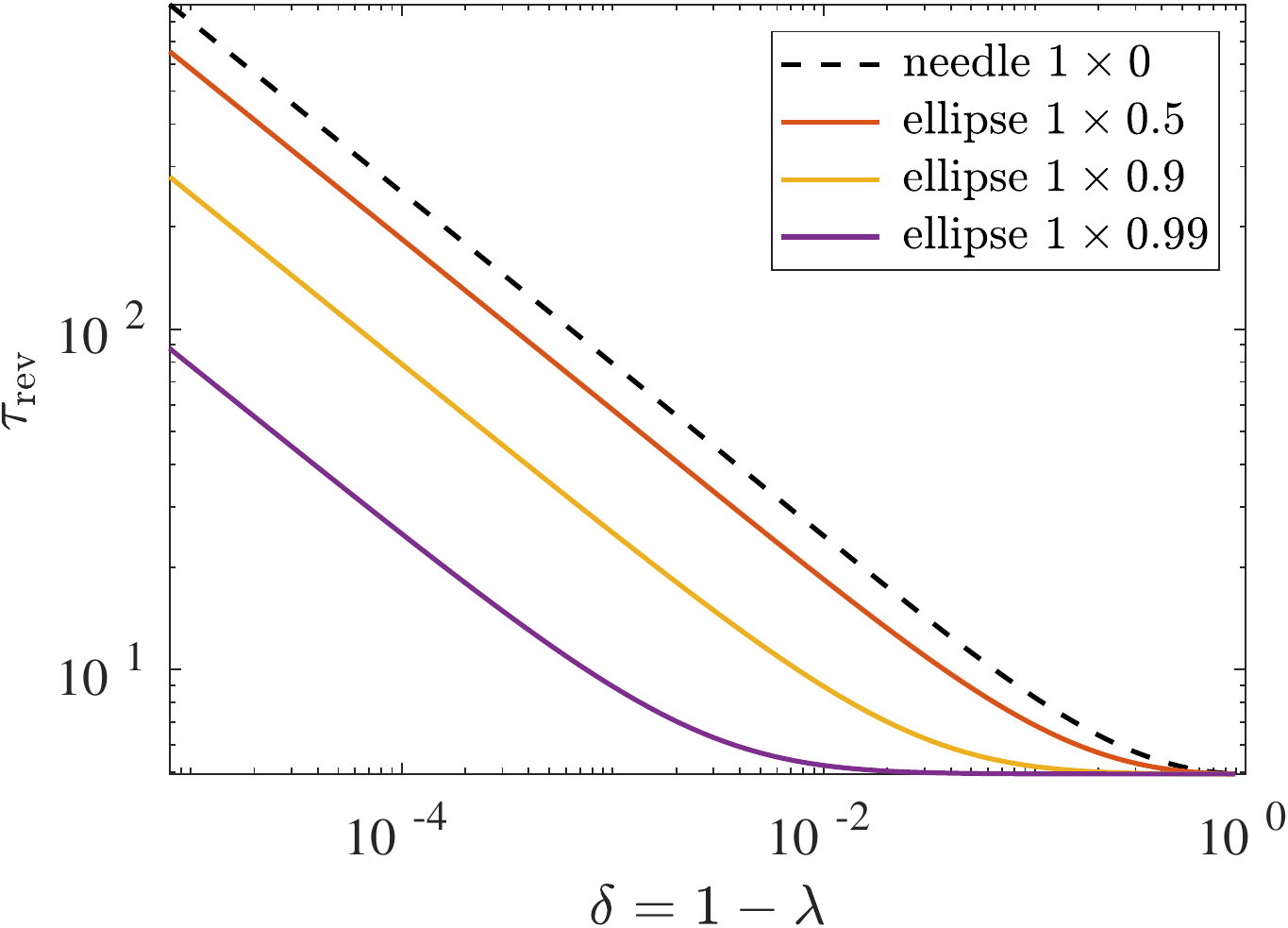}
    \end{center}
    \caption{Mean reversal time~$\taurev$ for a needle and different
      ellipses, as a function of gap size~$\delta$.  An ellipse reverses more
      rapidly as it becomes more circular, since it spends less time aligning
      with the wall.}
    \label{fig:channel_revtime_ellipse}
  \end{figure}

  For an elliptical shape~\eqref{eq:ycstar_ellipse_ecc} with~$\U=0$, we have
  \begin{equation}
    \w(\theta) = L\,(1 - \lambda\sqrt{1 - \ecc^2\cos^2\theta}).
  \end{equation}
  Then
  \begin{equation}
    L^{-1}\int_0^\pi\w(\theta)\dint\theta
    =
    \pi - 2\lambda\EE(\ecc),
    \label{eq:well_int1}
  \end{equation}
  where~$\EE$ is a complete elliptic integrals of the second kind.
  We also have
  \begin{equation}
    \int_0^\pi\frac{L\dint\theta}{\w(\theta)}
    =
    \frac{\pi}{\sqrt{(1 - \lambda^2)(1 - (1 - \ecc^2)\lambda^2)}}
    - \frac{2}{\lambda}\EK(\ecc)
    + \frac{2}{\lambda (1 - \lambda^2)}\,
    \EPi\l(\frac{\ecc^2}{1 - \lambda^{-2}}\,\middle|\,\ecc\r)
    \label{eq:well_int2}
  \end{equation}
  where~$\EK$ and~$\EPi$ are the complete elliptic integrals of the first and
  third kind.  Together the product of~\cref{eq:well_int1,eq:well_int2}
  into~\eqref{eq:taurevdiff} give the reversal time for an ellipse.  The mean
  reversal time for a needle is compared to different ellipses in
  \cref{fig:channel_revtime_ellipse}.
\end{Example}

\subsection{Asymptotics of mean reversal time}
\label{sec:METasym}

The integrand in \cref{eq:taurevsimp} is the inverse
of~$\Pinv(\theta) = \cc_1 \w(\theta)\,\ee^{\Phi(\theta)}$.  The integral
itself will thus typically be dominated by the minima of~$\Pinv$, which
correspond to values of~$\theta$ that the swimmer finds difficult to cross
when trying to reverse.  This could be because~$\delta = 1-\ell/L$ is small:
this is the \emph{narrow escape} problem discussed by Holcman \&
Schuss~\cite{Holcman2014} (see \cref{ex:mrt_diff}).  However, for a swimmer
the long reversal time is usually due to a swimmer `sticking' to the top or
bottom wall for long times.

To approximate the integral in~\eqref{eq:taurevsimp}, we look for
minima~$\theta_*$ of~$\Pinv(\theta)$, where we can approximate
\begin{equation}
  \Pinv(\theta)
  =
  \Pinv(\theta_*)\,
  \ee^{C_*\xi^2 + \Order{\xi^3}},
  \qquad
  \xi = \theta - \theta_*.
  \label{eq:Pstarexp}
\end{equation}
This gives us the approximation
\begin{equation}
  \taurev
  \approx
  \frac{1}{4\Pinv(\theta_*)}
  \int_{-\infty}^\infty \ee^{-C_*\xi^2}\dint\vartheta
  =
  \frac{1}{4\Pinv(\theta_*)}\,\sqrt{\frac{\pi}{C_*}}
  .
  \label{eq:taurev_star_approx}
\end{equation}
This should be multiplied by the number of minima with value~$\Pinv(\theta_*)$
in~$0 \le \theta < \pi$, should there be more than one.  We can easily obtain
a more accurate, if messier, approximation by retaining higher-order terms
in~\eqref{eq:Pstarexp}.

\begin{Example}[mean reversal time for a fast needle swimmer]%
  \label{ex:mrt_fast_needle}
  An ideal example for the asymptotic approximation of~$\taurev$ is the fast
  needle swimmer of \cref{ex:needle_invdens}.
  \Cref{fig:channel_invPinv_largeU_needle} clearly shows strong peaks
  in~$1/\Pinv$ at~$\theta_* = \pm\pi/2$ even for this modest value of~$\U$.
  Hence, we are justified in expanding around the minimum
  of~\eqref{eq:Pinv_needle} at~$\theta_*=\pi/2$, which gives
  \begin{equation}
    \Pinv(\theta_*)
    =
    \sqrt{\frac{\beta}{4\pi}}\,\alpha^{\beta/(1-\alpha)},
    \qquad
    C_* = \beta/\alpha.
  \end{equation}
  Recall that~$\alpha = \DX/\DY$ is the ratio of diffusivities, and $\beta$ is
  a large P\'eclet number defined in \cref{eq:alphabeta}.  Inserting these in
  \cref{eq:taurev_star_approx}, we find
  \begin{equation}
    \taurev
    \approx
    \frac{\pi}{2\beta}\,\alpha^{\frac12 - \frac{\beta}{1-\alpha}},
    \label{eq:taurev_needle_largeU}
  \end{equation}
  or in the case~$\alpha=1$:
  \begin{equation}
    \taurev
    \approx
    \frac{\pi}{2\beta}\,\ee^{\beta},
    \qquad
    \alpha = 1.
    \label{eq:taurev_needle_largeU_alpha1}
  \end{equation}
  This is exponential in the P\'eclet number~$\beta$, so that the reversal
  time can become extremely long.  Note also that the mean reversal time is
  independent of the channel width~$L$ in this limit.  For the parameters in
  \cref{fig:channel_Pinv_invPinv_largeU_needle} ($\beta=3.6$), the numerical
  mean reversal time is~$\taurev \approx 1.54\times 10^3$, whereas the
  approximation~\eqref{eq:taurev_needle_largeU} gives~$1.38 \times 10^3$.  The
  approximation is thus reasonably good even for a modest P\'eclet
  number~$\beta$.
\end{Example}

\section{Effective diffusion along the channel}
\label{sec:effdiff}

For the open channel configuration, as a microswimmer travels down the
channel, it will occasionally reverse direction.  For long times, we expect
these reversals to lead to an effective diffusion process on large scales.
One way to capture this limit exactly is to derive an effective diffusion
equation using a homogenization approach~\cite{Childress1989, McCarty1988,
  Sagues1986, Majda1999}.  We proceed to do so for our microswimmer in a
channel, and find the effective diffusivity in the same reduced limit as in
\cref{sec:reduced} (small~$\Drot$).

\subsection{Homogenized equation}
\label{sec:homog}

Rewrite the Fokker--Planck \cref{eq:FP} as
\begin{equation}
  \pd_\t \p
  + \pd_\xc(\Uxc\,\p)
  + \pd_\yc(\Uyc\,\p)
  =
  \pd_\xc^2(\Dxx\,\p)
  + 2\pd_\xc\pd_\yc(\Dxy\,\p)
  + \pd_\yc^2(\Dyy\,\p)
  + \pd_\theta^2(\Drot\,\p)
  \label{eq:Fp_exp}
\end{equation}
where~$\Uv = (\Uxc,\Uyc)$.  In this section we only assume that~$\Dt$,
$\Drot$, and~$\Uv$ do not depend on~$\xc$.  (The derivation could easily be
extended to allow~$\xc$ dependence.)  Later (\cref{sec:Deff_smallDr}) we will
specialize to the forms in~\cref{eq:uD}, which are functions of~$\theta$ only.

We will homogenize in the~$\xc$ direction only, since the swimmer is confined
between walls in the~$\yc$ direction and the~$\theta$ direction is periodic.
We introduce a large scale~$\xc/\eta$ and long time~$\t/\eta^2$, where~$\eta$
is a small expansion parameter.  After rescaling~$\t \rightarrow \thom/\eta^2$
and~$\xc \rightarrow \xchom/\eta$, \cref{eq:Fp_exp} becomes
\begin{subequations}
\begin{equation}
  \Lc\,\p
  =
  - \eta\,\pd_\xchom(\Uxc\,\p)
  + 2\eta\,\pd_\xchom\pd_\yc(\Dxy\,\p)
  - \eta^2\pd_\thom \p
  + \eta^2\,\pd_\xchom^2(\Dxx\,\p)
  \label{eq:FP_Lc_eq}
\end{equation}
where we defined the linear operator
\begin{equation}
  \Lc\,p
  \ldef
  \pd_\yc(\Uyc\,\p)
  - \pd_\yc^2(\Dyy\,\p)
  - \pd_\theta^2(\Drot\,\p).
  \label{eq:Lc}
\end{equation}
The no-flux boundary conditions for~\eqref{eq:FP_Lc_eq} are
\begin{equation}
  \l(\Uyc\,\p - \pd_\yc(\Dyy\,\p) - \eta\,\pd_\xchom(\Dxy\,\p)\r)
  + \ycw_\pm'(\theta)\,\pd_\theta(\Drot\,\p)
  =
  0,
  \qquad
  \yc = \ycw_\pm(\theta).
  \label{eq:FP_Lc_BC}
\end{equation}%
\label{eq:FP_Lc}
\end{subequations}

We expand the probability density~$\p$ as a regular series in~$\eta$:
\begin{equation}
  \p = \p_0 + \eta\,\p_1 + \eta^2\,\p_2 + \ldots.
  \label{eq:petaexp}
\end{equation}
Define the cell integral of~$f(\theta,\yc)$ as
\begin{equation}
  \avg{f}
  \ldef
  \int_{-\pi}^\pi
  \int_{\ycw_-(\theta)}^{\ycw_+(\theta)} f(\theta,\yc) \dint\yc\dint\theta.
  \label{eq:avg}
\end{equation}
To ensure uniqueness at each order, we enforce the cell-integrated
probability~$\avg{\p} = \avg{\p_0}$, so that~$\avg{\p_i} = 0$ for~$i>0$.

Now we collect powers of~$\eta$ in~\cref{eq:FP_Lc} with the
expansion~\eqref{eq:petaexp}.  At leading order in~$\eta$ we have
from~\cref{eq:FP_Lc_eq}
\begin{subequations}
\begin{equation}
  \Lc\,\p_0 = 0
  \label{eq:Lcp0_eq}
\end{equation}
with boundary conditions from \cref{eq:FP_Lc_BC}:
\begin{equation}
  \Uyc\,\p_0 - \pd_\yc(\Dyy\,\p_0)
  + \ycw_\pm'(\theta)\,\pd_\theta(\Drot\,\p_0)
  =
  0,
  \qquad
  \yc = \ycw_\pm(\theta).
  \label{eq:Lcp0_BCs}
\end{equation}
\label{eq:Lcp0}%
\end{subequations}
The operator~$\Lc$ only involves~$(\theta,\yc)$, so we can
solve~\cref{eq:Lcp0} with
\begin{equation}
  \p_0 = \phom(\xchom,\thom)\,\Pinvz(\theta,\yc),
  \qquad
  \Lc\Pinvz = 0,
  \quad
  \avg{\Pinvz} = 1,
  \label{eq:Pinvz}
\end{equation}
where~$\Pinvz(\theta,\yc)$ is the cell-normalized $\xc$-independent invariant
density for~\eqref{eq:Fp_exp}, and so~$\avg{\p_0} = \phom(\xchom,\thom)$.

At order~$\eta^1$, \cref{eq:FP_Lc_eq} gives
\begin{align}
  \Lc\,\p_1
  &=
  - \pd_\xchom(\Uxc\,\p_0)
  + 2\pd_\xchom\pd_\yc(\Dxy\,\p_0) \nonumber\\
  &=
  \l(
  - \Uxc\,\Pinvz
  + 2\pd_\yc(\Dxy\,\Pinvz)
  \r)\,\pd_\xchom\phom.
\end{align}
We can solve this by letting
\begin{equation}
  \p_1 = \chi\,\pd_\xchom\phom
\end{equation}
where~$\chi(\theta,\yc)$ satisfies the \emph{cell problem}
\begin{subequations}
\label{eq:cell}
\begin{align}
  \Lc\,\chi
  &=
  - \Uxc\,\Pinvz
  + 2\pd_\yc(\Dxy\,\Pinvz); \\
  0 &= \Uyc\,\chi - \pd_\yc(\Dyy\,\chi)
  + \ycw_\pm'(\theta)\,\pd_\theta(\Drot\,\chi)
  - \Dxy\,\Pinvz,
  \qquad
  \yc = \ycw_\pm(\theta),
\end{align}
\end{subequations}
with~$\avg{\chi}=0$.  The solvability condition for the cell
problem~\eqref{eq:cell} demands
\begin{equation}
  \avg{\Uxc\,\Pinvz} = \avg{\pd_\yc(\Dxy\,\Pinvz)}.
  \label{eq:noratchet}
\end{equation}
In our case, the left and right sides of~\eqref{eq:noratchet} vanish
separately after using the channel
symmetry~$\Pinvz(\theta+\pi,\yc) = \Pinvz(\theta,-\yc)$.  On the left, we have
$\Uxc(\theta) = \U\cos\theta$, so $\Uxc(\theta+\pi) = -\Uxc(\theta)$ and the
integral must vanish.  On the right, we
have~$\Dxy(\theta) = \tfrac12(\DX-\DY)\sin2\theta$ from~\eqref{eq:uD},
so~$\Dxy(\theta+\pi) = \Dxy(\theta)$
and~$\pd_\yc\Pinvz(\theta+\pi,\yc) = -(\pd_\yc\Pinvz)(\theta,-\yc)$ and the
integral again vanishes.  There is thus no `ratchet effect' to cause a net
drift, since there is no breaking of the left-right symmetry of the channel.
Having patterned walls as in Yariv and Schnitzer \cite{Yariv2014} would
possibly cause such a drift.

At order~$\eta^2$, \cref{eq:FP_Lc} gives
\begin{subequations}
\begin{align}
  \Lc\,\p_2
  &=
  - \pd_\xchom(\Uxc\,\p_1)
  + 2\pd_\xchom\pd_\yc(\Dxy\,\p_1)
  - \pd_\thom \p_0
  + \pd_\xchom^2(\Dxx\,\p_0)
  \\
  0 &=
  \l(\Uyc\,\p_2 - \pd_\yc(\Dyy\,\p_2) - \pd_\xchom(\Dxy\,\p_1)\r)
  + \ycw_\pm'(\theta)\,\pd_\theta(\Drot\,\p_2),
  \qquad
  \yc = \ycw_\pm(\theta).
\end{align}
\end{subequations}
The solvability condition then yields the effective heat equation
\begin{equation}
  \pd_\thom \phom = \Deff\,\pd_\xchom^2\phom
  \label{eq:phom_heat}
\end{equation}
where the effective diffusivity in~$\xchom$ is
\begin{equation}
  \Deff
  =
  \avg{\Dxx\,\Pinvz}
  - \avg{\Uxc\,\chi}
  + \avg{\pd_\yc(\Dxy\chi)}.
  \label{eq:Deff}
\end{equation}
The solvability condition~\eqref{eq:noratchet} implies that~$\chi$ only
affects the effective diffusivity up to an additive multiple of~$\Pinvz$.

\subsection{Reduced equation limit}
\label{sec:Deff_smallDr}

We solve the cell problem~\eqref{eq:cell} in the same small-$\Drot$ limit as
in \cref{sec:reduced}.  Anticipating that the effective diffusivity should
diverge as~$\iPer=\Drot$ becomes smaller, we expand
\begin{equation}
  \chi = \iPer^{-1}\,\chi_0 + \chi_1 + \iPer\,\chi_2 + \ldots.
  \label{eq:chi_iPer_exp}
\end{equation}
The leading-order~$\iPer^{-1}$ cell problem~\eqref{eq:cell} is the the same
as~\eqref{eq:bp0}, with solution
\begin{equation}
  \chi_0(\theta,\yc)
  =
  \Chi(\theta)\,
  \ee^{\vth(\theta)\yc}.
  \label{eq:chi0sol}
\end{equation}
At next order~$\iPer^0$ we have the PDE and boundary conditions
\begin{subequations}
\label{eq:chi1}
\begin{alignat}{2}
  \U\sin\theta\,\pd_\yc\chi_1
  - \Dyy(\theta)\,\pd_\yc^2\chi_1
  &=
  \pd_\theta^2\chi_0
  -
  \l(\U\cos\theta
  - 2\vth\Dxy\r)\,\Qinv(\theta)\,\ee^{\vth(\theta)\yc};
  \label{eq:chi1pde}\\
  \U\sin\theta\,\chi_1
  - \Dyy(\theta)\,\pd_\yc\chi_1 &= -\ycw_\pm'(\theta)\,\pd_\theta\chi_0
  + \Dxy\,\Qinv(\theta)\,\ee^{\vth(\theta)\yc},
  \qquad \yc=\ycw_\pm(\theta), \label{eq:chi1BC}
\end{alignat}
\end{subequations}
where we used the invariant
density~$\bp_0 = \Qinv(\theta)\,\ee^{\vth(\theta)\yc}$.
Integrate~\eqref{eq:chi1pde} from~$\yc=\ycw_-$ to~$\ycw_+$ and use the
boundary conditions~\eqref{eq:chi1BC}:
\begin{equation}
  \pd_\theta(\nu\Chi - \w\,\pd_\theta\Chi)
  =
  -\Xi\,\Pinv,
  \label{eq:Chi}
\end{equation}
where~$\Pinv = \w\Qinv$, and
\begin{equation}
  \Xi(\theta)
  \ldef
  \U\cos\theta - \vth\Dxy
  =
  \frac{\U\cos\theta}{\cos^2\theta + \alpha\sin^2\theta}
  \label{eq:Xi}
\end{equation}
with~$\alpha = \DX/\DY$ as in~\eqref{eq:Pinv_needle}.  We
integrate~\cref{eq:Chi} once:
\begin{equation}
  \nu\Chi - \w\,\pd_\theta\Chi
  =
  \dd
  -
  H(\theta),
  \qquad
  H(\theta) \ldef
  \int_{-\pi}^\theta\Xi(\vartheta)\,\Pinv(\vartheta)\dint\vartheta,
  \label{eq:Chiint}
\end{equation}
with~$\dd$ a constant of integration.  The solvability
condition~\eqref{eq:noratchet} ensures that~$H(\theta)$ is a periodic function
of~$\theta$.  Next use the integrating factor~$\ee^{\Phi(\theta)}$
from~\eqref{eq:PhiF}, with~$\thetaL=-\pi$:
\begin{equation}
  \w\,\ee^{\Phi}\,\pd_\theta(\ee^{-\Phi}\Chi)
  =
  H(\theta) - \dd.
\end{equation}
We integrate again and find
\begin{equation}
  \ee^{-\Phi(\theta)}\,\Chi(\theta) - \Chi(-\pi)
  =
  \int_{-\pi}^\theta\frac{H(\vartheta)}{\w(\vartheta)\,\ee^{\Phi(\vartheta)}}
  \dint\vartheta
  -
  \cc_1\dd\,F(\theta)
\end{equation}
where we used~$F(\theta)$ from~\cref{eq:PhiF}.  By rearranging and
using~\cref{eq:Qinvsol2}, we find after introducing new constants
\begin{equation}
  \Chi(\theta)
  =
  \cct_1\,\ee^{\Phi(\theta)}
  \l(
  \int_{-\pi}^\theta\frac{H(\vartheta)}{\Pinvt(\vartheta)}
  \dint\vartheta
  -
  \dd_2 F(\theta)
  \r)
  +
  \dd_1\Qinv(\theta)
\end{equation}
where we used~$\Pinvt = \cct_1\w\,\ee^{\Phi}$ as in \cref{sec:METsol}.  The
constant~$\dd_1$ is adjusted to satisfy~$\avg{\chi}=0$ and is immaterial to
the effective diffusivity.  The constant~$\dd_2$ is used to enforce
periodicity of~$\Chi$ and can be eliminated to obtain
\begin{equation}
  \Chi(\theta)
  =
  \cct_1\,\ee^{\Phi(\theta)}
  \l(
  \int_{-\pi}^\theta\frac{H(\vartheta)}{\Pinvt(\vartheta)}
  \dint\vartheta
  -
  \frac{F(\theta)}{F(\pi)}
  \int_{-\pi}^\pi\frac{H(\vartheta)}{\Pinvt(\vartheta)}
  \dint\vartheta
  \r)
  +
  \dd_1\Qinv(\theta).
  \label{eq:Chisol}
\end{equation}
In this reduced limit, the effective diffusivity~\cref{eq:Deff} is
\begin{equation}
  \Deff
  =
  \E\Dxx
  +
  \Denh,
  \qquad
  \Denh
  \ldef
  -
  \int_{-\pi}^\pi
  \Xi\,\w\Chi
  \dint\theta,
  \label{eq:Deff_reduced}
\end{equation}
to leading order in~$\Drot$.  The expected value~$\E\Dxx$
is taken over the invariant marginal density~$\Pinv(\theta)$, and~$\Denh$ is
the `enhanced' part of the diffusivity.

\subsection{Bounding~\texorpdfstring{$\Deff$}{Deff} by~\texorpdfstring{$\taurev$}{taurev} for a left-right symmetric swimmer}
\label{sec:relate_effdiff_MRT}

Since the expressions are getting a bit complicated, for the sake of brevity
we will focus on the left-right symmetric case for this section.  In that case
the term involving~$F(\theta)$ vanishes in~\eqref{eq:Chisol},
and~$\cct_1\,\ee^{\Phi}$ becomes~$\Qinv$:
\begin{equation}
  \Chi(\theta)
  =
  \Qinv(\theta)
  \int_{-\pi}^\theta\frac{H(\vartheta)}{\Pinv(\vartheta)}
  \dint\vartheta
  +
  \dd_1\Qinv(\theta).
  \label{eq:Chi_LRsym}
\end{equation}
After restoring in \cref{eq:Chi_LRsym} the definition of~$H$ from
\cref{eq:Chiint}, it can be shown that the enhanced diffusivity~$\Denh$
in~\cref{eq:Deff_reduced} can be written as
\begin{equation}
  \Denh
  =
  4\int_0^{\pi/2}\Xi(\theta)\,\Pinv(\theta)\int_0^\theta
  \Xi(\theta')\,\Pinv(\theta')
  \int_\theta^{\pi - \theta} \frac{\dint\theta''}{\Pinv(\theta'')}
  \dint\theta'\dint\theta.
  \label{eq:intXiChi_taurev0}
\end{equation}
To derive this we used the
symmetries~$\Xi(\theta + \pi) = \Xi(\pi - \theta) = -\Xi(\theta)$,
$\Pinv(\theta + \pi) = \Pinv(\pi - \theta) = \Pinv(\theta)$, the latter
holding for a left-right-symmetric swimmer.  Rearranging the innermost
integral in~\eqref{eq:intXiChi_taurev} gives us
\begin{equation}
  \Denh
  =
  \tfrac12\taurev
  (\E\lvert\Xi\rvert)^2
  -
  8\int_0^{\pi/2}\Xi(\theta)\,\Pinv(\theta)\int_0^\theta
  \Xi(\theta')\,\Pinv(\theta)'
  \int_0^\theta \frac{\dint\theta''}{\Pinv(\theta'')}
  \dint\theta'\dint\theta,
  \label{eq:intXiChi_taurev}
\end{equation}
where the mean reversal time~$\taurev$ was defined in \cref{eq:taurevsimp}.
The expected value in~\eqref{eq:intXiChi_taurev} is taken over the invariant
marginal density~$\Pinv(\theta)$, as in \cref{eq:Deff_reduced}.

\begin{figure}
  \begin{center}
    \includegraphics[width=.7\textwidth]{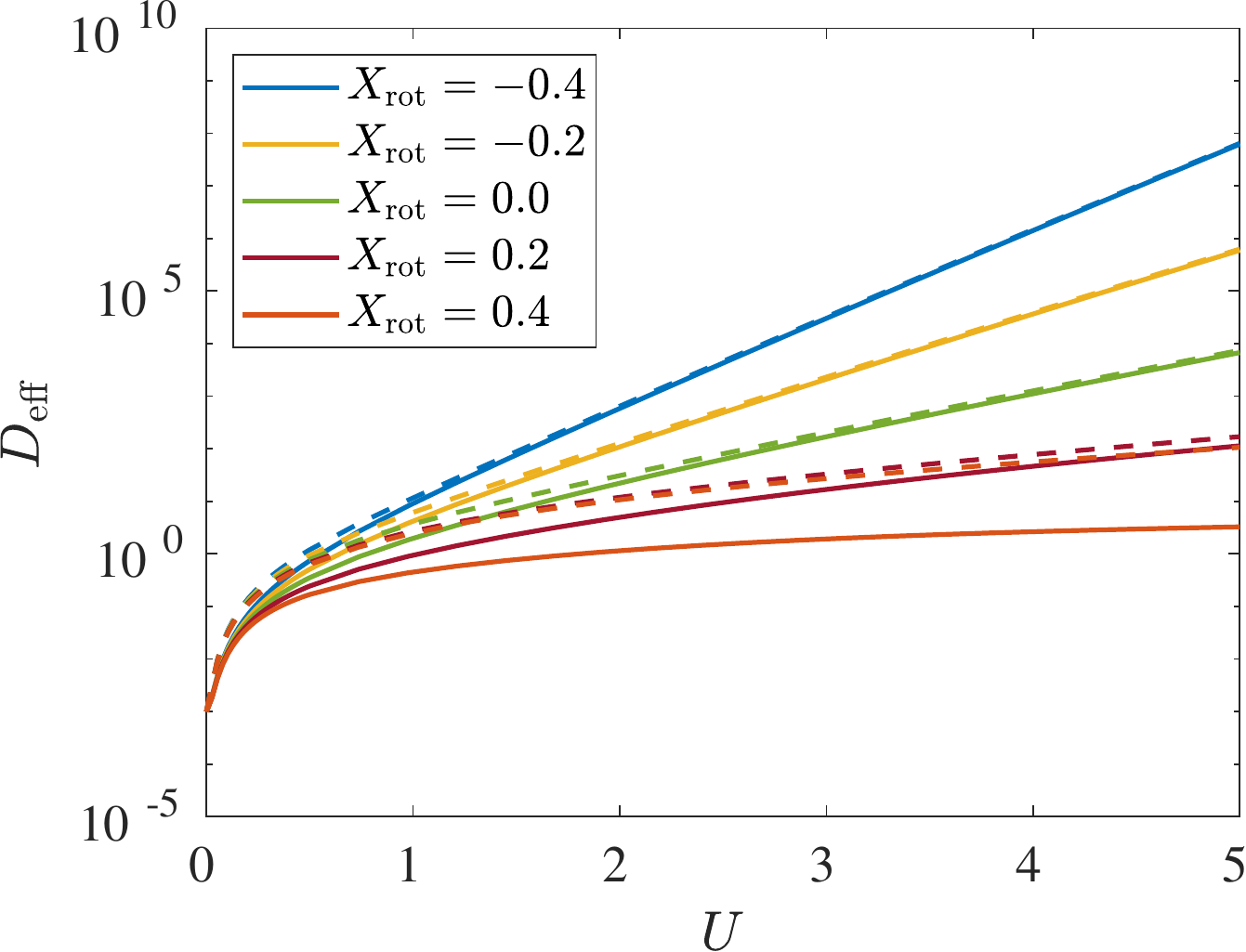}
  \end{center}
  \caption{Effective diffusivity from \cref{eq:Deff_reduced} for an ellipse
    with~$\aaa=1$, $\bbb=1/2$, $\DX=\DY=0.1$, $\Drot=0.01$, in a channel of
    width~$L=1.2$.  The dashed lines are the bound~$\tfrac12\U^2\taurev$
    from~\eqref{eq:Denh_bound}, with~$\alpha=1$ and~$\taurev$ given by
    \cref{eq:taurevsimp}.  The decrease in exponential rate as the center of
    rotation~$\Xcrot$ is moved forward is similar to the asymptotic form
    of~$\taurev$ for a needle, \cref{eq:taurev_needle_largeU_alpha1},
    with~$\beta$ given by~\eqref{eq:alphabeta}.}
  \label{fig:channel_effdiff}
\end{figure}

Since~$\Xi(\theta)$ defined by~\cref{eq:Xi} is non-negative in~$[0,\pi/2]$,
\cref{eq:intXiChi_taurev} immediately gives us the bound
\begin{subequations}
\begin{equation}
  \Denh
  \le
  \tfrac12\taurev(\E\lvert\Xi\rvert)^2
  \le
  \tfrac12\taurev\,\max_\theta \Xi^2(\theta),
  \label{eq:Denh_bound0}
\end{equation}
with
\begin{equation}
  \max_{0 \le \theta \le \pi/2} \Xi(\theta)
  =
  \begin{cases}
    \U, \qquad & \alpha \ge 1/2; \\
    \nofrac{\U}{\sqrt{4\alpha(1 - \alpha)}}, \qquad &\alpha < 1/2.
  \end{cases}
  \label{eq:Xi_bound}
\end{equation}%
\label{eq:Denh_bound}%
\end{subequations}
This useful bound allows us to estimate~$\Denh$ from~$\taurev$, which is
simpler to compute (\cref{eq:taurevsimp}).  The estimate tends to improve the
longer the swimmer spends aligned with the walls (\cref{fig:channel_effdiff}).

\section{Discussion}
\label{sec:discussion}

We used both theory and numerics to analyze the dynamics of finite-sized
swimmers in a channel.  The shape of swimmers is embedded in their
configuration space, which is defined even for nonsmooth swimmers via their
convex hull.  Shape enters solely in the boundary conditions to the
Fokker--Planck equation.  We then derived a reduced equation in the small
$\Drot$ limit for both open and closed channels, from which we computed the
invariant density.  For open channel geometry, we calculated the mean reversal
time.  We used homogenization techniques and solved the cell problem to find
the effective diffusivity.  The shape of a swimmer is encoded in the
rotational drift term~$\mu$ in the reduced equation~\eqref{eq:Peq}, and
appears explicitly in the integral solution for the invariant density, mean
reversal time, and effective diffusivity.  The integral of $\mu$ vanishes for
a left-right symmetric swimmer, and many of our expressions then greatly
simplify.  In particular the mean angular drift vanishes for such a symmetric
swimmer.

A particular novelty in our work is to explicitly allow the position of the
center of rotation, $\Xcrot$, to vary.  The sign of $\Xcrot$ affects the
configuration space, as shown in \cref{fig:swimmer_walldist}, and changes a
swimmers' tendency to align with walls.  When a swimmer's center of rotation
is behind its geometrical center, it tends to align with a wall; in the
opposite case, it tends to stay perpendicular to the wall even for a needle or
circular swimmer (\cref{ex:needle_invdens,ex:circ_invdens}).

In our work we focused on two-dimensional swimmers for simplicity and ease of
presentation.  However, generalizing the formalism to three-dimensional
axisymmetric swimmers is straightforward~\cite{Nitsche1990}.  In that case,
the wall distance function~$\yc_*$ is essentially the same as in 2D, and all
that changes is that the~$\pd_\theta^2$ operator must be replaced by the 3D
surface Laplacian.  The dynamics for 3D axisymmetric swimmers will thus be
similar to 2D left-right symmetric swimmers, with some quantitative changes.
In particular, following almost the same calculations, we found the invariant
density is given by a formula similar to \cref{eq:Qinv_closed}, and when
applied to spherical swimmers with~$\Xcrot=0$ it agrees with the results in
\cite{Ezhilan2015,Elgeti2013}.  A Fully 3D swimmer that is not axisymmetric
will require considerably more work to implement: the configuration space has
an extra dimension due to the extra degrees of freedom.  Of course, this will
potentially also make the dynamics richer, and will allow for interesting
effects such as chirality.

Another challenging direction is to add hydrodynamic interactions with
boundaries, in a manner similar to the work mentioned in the introduction.
This is in principle straightforward, though the equations will be harder to
solve.  Here, we left out these interactions in order to emphasize steric
effects and keep the equations simple, permitting analytic solutions.  It is
also possible to include variable swimmer shapes, which would allow the
inclusion of flagella, by letting the configuration space itself be
time-dependent.

\section*{Acknowledgments}

Some of the initial work on this project was carried out with Jacob Gloe.  The
authors are grateful to Saverio Spagnolie for many discussions.

\ifthenelse{\isundefined{\arXiv}}
{%
\section*{Declaration of interest}

The authors report no conflict of interest.
}{}

\bibliographystyle{amsplain}
\bibliography{needle}

\appendix

\section{Derivation of the mean reversal time}
\label{apx:mrt_calc}

In this \namecref{apx:mrt_calc} we show how to from the reversal time formula
\cref{eq:taurevfull} to the more explicit form~\eqref{eq:taurev} by
eliminating the constant~$\Cc$.  First note that the numerator of \cref{eq:Cc}
is
\begin{equation}
  \int_{-\pi}^\pi
  \frac{\Gt(\vartheta)}{\Pinvt(\vartheta)}\dint\vartheta
  =
  \int_{-\pi}^0
  \frac{\Gt(\vartheta)}{\Pinvt(\vartheta)}\dint\vartheta
  +
  \int_{-\pi}^0
  \frac{\Gt(\vartheta + \pi)}{\Pinvt(\vartheta + \pi)}\dint\vartheta.
  \label{eq:Ccnum0}
\end{equation}
We wish to relate~$\Pinvt(\theta + \pi)$
to~$\Pinvt(\theta) = \cct_1\,\w(\theta)\,\ee^{\Phi(\theta)}$.  We
have~$\w(\theta + \pi) = \w(\theta)$.  The change in~$\Phi$ over a period is,
from \cref{eq:PhiF},
\begin{equation}
  \Phi(\theta + \pi) - \Phi(\theta)
  =
  \int_0^\pi \mu(\vartheta)\dint\vartheta
  =
  \pi\mub,
  \label{eq:Phipi}
\end{equation}
since~$\mu(\theta)$ is~$\pi$-periodic.  Here~$\mub$ is the
period-averaged~$\mu(\theta)$ from \cref{eq:mub}; $\Phi(\theta)$ is only
$\pi$-periodic when~$\mub=0$, or equivalently~$\cc_2=0$.  It follows from
\cref{eq:Phipi} that $\Phi$ must have the form
\begin{equation}
  \Phi(\theta)
  =
  \mub\,\theta + \Phit(\theta),
  \label{eq:Phit}
\end{equation}
where~$\Phit(\theta)$ is $\pi$-periodic.  We conclude
from~$\Pinvt(\theta) = \cct_1\,\w(\theta)\,\ee^{\Phi(\theta)}$
that~$\Pinvt(\theta + \pi) = \ee^{\pi\mub}\,\Pinvt(\theta)$.  Similarly,
for~$-\pi \le \theta \le 0$,
\begin{align}
  \Gt(\theta + \pi)
  &=
  \int_{-\pi}^{\theta + \pi}
  \Pinvt(\vartheta)\dint\vartheta\nonumber\\
  &=
  \int_{-\pi}^0
  \Pinvt(\vartheta)\dint\vartheta
  +
  \int_0^{\theta + \pi}
  \Pinvt(\vartheta)\dint\vartheta
  \nonumber\\
  &=
  \Gt(0) + \ee^{\pi\mub}\,\Gt(\theta).
\end{align}
Returning to \cref{eq:Ccnum0}, we obtain
\begin{align}
  \int_{-\pi}^\pi
  \frac{\Gt(\vartheta)}{\Pinvt(\vartheta)}\dint\vartheta
  &=
  \int_{-\pi}^0
  \frac{\Gt(\vartheta)}{\Pinvt(\vartheta)}\dint\vartheta
  +
  \int_{-\pi}^0
  \frac{\Gt(0) + \ee^{\pi\mub}\,\Gt(\theta)}
  {\ee^{\pi\mub}\,\Pinvt(\vartheta)}\dint\vartheta \nonumber\\
  &=
  2
  \int_{-\pi}^0
  \frac{\Gt(\vartheta)}{\Pinvt(\vartheta)}\dint\vartheta
  +
  \Gt(0)\,\ee^{-\pi\mub}
  \int_{-\pi}^0\frac{\dint\vartheta}{\Pinvt(\vartheta)}.
  \label{eq:Ccnum}
\end{align}
The denominator of~$\Cc$ in~\eqref{eq:Cc} is
\begin{equation}
  \int_{-\pi}^\pi
  \frac{\dint\vartheta}{\Pinvt(\vartheta)}
  =
  \int_{-\pi}^0
  \frac{\dint\vartheta}{\Pinvt(\vartheta)}
  +
  \int_{-\pi}^0
  \frac{\dint\vartheta}{\Pinvt(\vartheta + \pi)}
  =
  (1 + \ee^{-\pi\mub})
  \int_{-\pi}^0
  \frac{\dint\vartheta}{\Pinvt(\vartheta)}.
  \label{eq:Ccdenom}
\end{equation}
Together~\eqref{eq:Ccnum} and~\eqref{eq:Ccdenom} give
\begin{equation}
  \Cc
  \int_{-\pi}^0
  \frac{\dint\vartheta}{\Pinvt(\vartheta)}
  =
  \frac{1}{1 + \ee^{-\pi\mub}}
  \l(
  2
  \int_{-\pi}^0
  \frac{\Gt(\vartheta)}{\Pinvt(\vartheta)}\dint\vartheta
  +
  \Gt(0)\,\ee^{-\pi\mub}
  \int_{-\pi}^0\frac{\dint\vartheta}{\Pinvt(\vartheta)}
  \r).
\end{equation}
We use this in the reversal time~\eqref{eq:taurevfull}:
\begin{align}
  \taurev
  &=
  \Cc
  \int_{-\pi}^0
  \frac{\dint\vartheta}{\Pinvt(\vartheta)}
  -
  \int_{-\pi}^0
  \frac{\Gt(\vartheta)}{\Pinvt(\vartheta)}
  \dint\vartheta \nonumber\\
  &=
  \frac{1}{1 + \ee^{-\pi\mub}}
  \l(
  2
  \int_{-\pi}^0
  \frac{\Gt(\vartheta)}{\Pinvt(\vartheta)}\dint\vartheta
  +
  \Gt(0)\,\ee^{-\pi\mub}
  \int_{-\pi}^0\frac{\dint\vartheta}{\Pinvt(\vartheta)}
  \r)
  -
  \int_{-\pi}^0
  \frac{\Gt(\vartheta)}{\Pinvt(\vartheta)}
  \dint\vartheta \nonumber\\
  &=
  \frac{1}{1 + \ee^{-\pi\mub}}
  \l(
  \Gt(0)\,\ee^{-\pi\mub}
  \int_{-\pi}^0\frac{\dint\vartheta}{\Pinvt(\vartheta)}
  +
  (1 - \ee^{-\pi\mub})
  \int_{-\pi}^0
  \frac{\Gt(\vartheta)}{\Pinvt(\vartheta)}
  \dint\vartheta
  \r)
  \nonumber\\
  &=
  \frac{\Gt(0)\,\ee^{-\pi\mub}}{1 + \ee^{-\pi\mub}}
  \int_{-\pi}^0\frac{\dint\vartheta}{\Pinvt(\vartheta)}
  +
  \frac{1 - \ee^{-\pi\mub}}{1 + \ee^{-\pi\mub}}
  \int_{-\pi}^0
  \frac{\Gt(\vartheta)}{\Pinvt(\vartheta)}
  \dint\vartheta,
\end{align}
which is \cref{eq:taurev}.

\clearpage

\section{Notation}
\label{apx:notation}

\begin{table}[!htbp]
  \caption{Notation used in the paper, \cref{sec:conf}.}
  \label{tab:notation1}
  \begin{center}
    {\renewcommand{\arraystretch}{1.2}
    \begin{tabular}{ll}
      \hline\hline
      symbol & description \\[2pt]
      \hline
      \multicolumn{2}{l}{shape parameters (\cref{sec:walldist})} \\
      \hline
      $\rv = (\xc,\yc)$ & coordinates of point in lab (spatial) frame; \\
      & \quad also center of rotation of swimmer \\
      $\Rv = (\Xc,\Yc)$ & coordinates of point in swimmer's frame; \\
      & \quad $(\Xc,\Yc) = (0,0)$ is center of rotation \\
      $\Rvb(\varphi) = (\Xcb(\varphi),\Ycb(\varphi))$
      & boundary in swimmer's frame \\
      $\rvb(\varphi) = (\xcb(\varphi),\ycb(\varphi))$
      & boundary in lab frame \\
      $\theta$ & swimming direction w.r.t. $\xc$ axis \\
      $\Qrot_\theta$ & rotation matrix \\
      $W$ & point of contact between swimmer and wall \\
      $\tvb$ & tangent to swimmer's boundary \\
      $\yc_*(\theta)$ & wall distance function \\
      $\Xcrot$ & center of rotation offset ($>0$ toward front) \\
      $\ell$ & maximum swimmer diameter \\
      $\aaa$, $\bbb$ & ellipse semi-axis $\parallel$ and $\perp$ to swimming
      direction \\
      $\ecc$ & ellipse eccentricity \\
      \hline
      \multicolumn{2}{l}{channel geometry (\cref{sec:channelgeom})} \\
      \hline
      $L$ & channel width \\
      $\ycw_\pm(\theta)$ & range of $\yc$ in channel: $\ycw_-(\theta) \le \yc
      \le \ycw_+(\theta)$ \\
      $\Omega$ & open channel configuration space of~$(\xc,\yc,\theta)$ \\
      $\Omega_i$ & closed channel configuration space component \\
      $\thetaL_i$, $\thetaR_i$ &
      $\theta$ left and right limits of $\Omega_i$ \\
      \hline\hline
    \end{tabular}
    }
  \end{center}
\end{table}

\begin{table}
  \caption{Notation used in the paper,
    \cref{sec:stochmod,sec:reduced,sec:inv}.}
  \label{tab:notation2}
  \begin{center}
    {\renewcommand{\arraystretch}{1.2}
    \begin{tabular}{ll}
      \hline\hline
      symbol & description \\[2pt]
      \hline
      \multicolumn{2}{l}{stochastic model (\cref{sec:stochmod})} \\
      \hline
      $\t$ & time \\
      $\U$ & swimming speed \\
      $W_i(\t)$ & Brownian motion \\
      $\DX$, $\DY$ & diffusivity $\parallel$ and $\perp$ to swimming
      direction \\
      $\Drot$ & rotational diffusivity \\
      $\p(\xc,\yc,\theta,\t)$ & probability density of swimmers \\
      $\Uv = (\Uxc,\Uyc)$ & swimming velocity vector \\
      $\Dt$, $\Dxx$, $\Dyy$, $\Dxy$ &
      spatial diffusivity tensor and components \\
      $\xuv$, $\yuv$, $\thetauv$ & unit vectors \\
      $\nv$ & normal vector to boundary \\
      $\bp$, $\bfv$, $\bnv$ & $\xc$-independent quantities for infinite
      channel \\
      \hline
      \multicolumn{2}{l}{reduced equation (\cref{sec:reduced})} \\
      \hline
      $\iPer = \Drot$ & small expansion parameter \\
      $\bp_i$ & probability density at order~$\iPer^i$ \\
      $\T = \iPer\t$ & dimensionless slow time \\
      $\Q$ & solution at order $\eps^0$ \eqref{eq:bp0sol_inv} \\
      $\vth = \U\sin\theta / \Dyy$ & \eqref{eq:bp0sol_inv} \\
      $\w$ & channel weight function \eqref{eq:w} \\
      $\nu$ & channel drift for~$\Q$ \eqref{eq:nu} \\
      $\mu = (\nu + \w')/w$ & channel drift for~$\P$ \eqref{eq:mu} \\
      $\Dyc = \tfrac12\vth\,(\ycw_+ - \ycw_-)$ & \eqref{eq:Dyc} \\
      $\P=\w\Q$ & $\theta$ marginal density at order $\eps^0$ \eqref{eq:P} \\
      \hline
      \multicolumn{2}{l}{invariant density (\cref{sec:inv})} \\
      \hline
      $\Qinv$, $\Pinv$ & marginal invariant density for $\Q$, $\P$ \\
      $\bp_0 = \Qinv\,\ee^{\vth\yc}$ & invariant density \\
      $\cc_1$, $\cc_2$ & integration constants \eqref{eq:Qinvsol} \\
      $\Phi$, $F$ & integrating factor terms \eqref{eq:PhiF} \\
      $\E$ & expected value over invariant marginal density~$\Pinv(\theta)$ \\
      $\omega = \E\mu = 2\pi\cc_2$ & angular rotation rate of swimmmer \\
      $\mub$ & $\theta$-averaged drift \eqref{eq:mub} \\
      $\alpha$ & diffusivity ratio $\nofrac{\DX}{\DY}$ \\
      $\beta$ & center-offset P\'eclet number $\nofrac{\U(\aaa
        -\Xcrot)}{2\DY}$ \\
      \hline\hline
    \end{tabular}
    }
  \end{center}
\end{table}

\begin{table}
  \caption{Notation used in the paper, \cref{sec:MERT,sec:effdiff}.}
  \label{tab:notation3}
  \begin{center}
    {\renewcommand{\arraystretch}{1.2}
    \begin{tabular}{ll}
      \hline\hline
      symbol & description \\[2pt]
      \hline
      \multicolumn{2}{l}{mean exit time (\cref{sec:MERT,apx:mrt_calc})} \\
      \hline
      $\tau(\theta)$ & dimensionless mean exit time to~$\thetaL$ or~$\thetaR$
      \\
      $\wtau = \w\tau'$ & \eqref{eq:wtau} \\
      $\Pinvt = \cct_1\,\w\,\ee^{\Phi}$ &
      closed-channel invariant density \eqref{eq:taup} \\
      $\Cc$ & integration constant \\
      $\Gt$ & \eqref{eq:taup} \\
      $\taurev$ & dimensionless mean reversal time
      \eqref{eq:taurev} \\
      $\lambda = \ell/L$, $\delta = 1 - \lambda$
      & narrow exit parameters \eqref{eq:taurevdiff_needle} \\
      $\Phit$ & periodic part of $\Phi$ \eqref{eq:Phit} \\
      \hline
      \multicolumn{2}{l}{homogenized equation (\cref{sec:homog})} \\
      \hline
      $\eta$ & scale separation parameter \eqref{eq:FP_Lc_eq} \\
      $\p_i$ & probability density at order~$\eta^i$ \\
      $\Lc$ & linear operator \eqref{eq:FP_Lc_eq} \\
      $\avg{\cdot}$ & cell integral \eqref{eq:avg} \\
      $\phom$ & large-scale probability density \eqref{eq:phom_heat} \\
      $\Pinvz$ & cell-normalized invariant density \eqref{eq:Pinvz} \\
      $\chi$ & solution to cell problem~\eqref{eq:cell} \\
      $\Deff$ & effective diffusivity \eqref{eq:Deff} \\
      $\Denh$ & enhanced diffusivity \eqref{eq:Deff_reduced} \\
      \hline
      \multicolumn{2}{l}{reduced equation (\cref{sec:Deff_smallDr,%
          sec:relate_effdiff_MRT})} \\
      \hline
      $\chi_i$ & $\chi$ at order~$\iPer^i$ \eqref{eq:chi_iPer_exp} \\
      $\Chi$ & solution at order $\eps^0$ \eqref{eq:chi0sol} \\
      $\Xi$ & right-hand side of $\Chi$ equation~\eqref{eq:Chi} \\
      $H$ & \eqref{eq:Chiint} \\
      $\dd_i$ & integration constants \\
      \hline\hline
    \end{tabular}
    }
  \end{center}
\end{table}

\end{document}